\pdfoutput=1
\documentclass{tcibook}
\usepackage{fancyhea}
\usepackage{work}
\usepackage{bm}
\usepackage{graphicx}
\usepackage{hyperref}
\hypersetup{colorlinks=true}
\usepackage{paralist}
\usepackage{color}
\usepackage{chngcntr}
\counterwithout{figure}{chapter}

\begin{document}

\def\bibname{References}

\bibliographystyle{utphys}

\raggedbottom

\pagenumbering{roman}

\parindent=0pt
\parskip=8pt
\setlength{\evensidemargin}{0pt}
\setlength{\oddsidemargin}{0pt}
\setlength{\marginparsep}{0.0in}
\setlength{\marginparwidth}{0.0in}
\marginparpush=0pt

\pagenumbering{arabic}

\renewcommand{\chapname}{chap:intro_}
\renewcommand{\chapterdir}{.}
\renewcommand{\arraystretch}{1.25}
\addtolength{\arraycolsep}{-3pt}

%%%%%% WIMP Dark Matter Direct Detection Chapter  %%%%%%%%%%%%%%%%

\chapter*{Snowmass CF1 Summary: WIMP Dark Matter Direct Detection}
\renewcommand*\thesection{\arabic{section}}

%\begin{center}
%\textbf{SNOWMASS 2013}

%\end{center}

%%%%%%%%%%%%%%%%%%%%%%%%%%%%%%%%%%%%%%%%%%%%%%%%%%%%%%%%%%%
%%%%%%%%%%%%%%%%%%%%%%%%%%%%%%%%%%%%%%%%%%%%%%%%%%%%%%%%%%%
%%%%%%%%%%%%%%%%%%%%%%%%%%%%%%%%%%%%%%%%%%%%%%%%%%%%%%%%%%%
%%%%%%%%%%%%%%%%%%%%%%%%%%%%%%%%%%%%%%%%%%%%%%%%%%%%%%%%%%%
\begin{center}\begin{boldmath}
\textbf{Convenors: P. Cushman, C. Galbiati, D. N. McKinsey, H. Robertson, and T. M. P. Tait}
\begin{center}
D.~Bauer,
A.~Borgland,
B.~Cabrera,
F.~Calaprice,
J.~Cooley,
P.~Cushman,
T.~Empl,
R.~Essig,
E.~Figueroa-Feliciano,
R.~Gaitskell,
C.~Galbiati,
S.~Golwala,
J.~Hall,
R.~Hill,
A.~Hime,
E.~Hoppe,
L.~Hsu,
E.~Hungerford,
R.~Jacobsen,
M.~Kelsey,
R.~F.~Lang,
W.~H.~Lippincott,
B.~Loer,
S.~Luitz,
V.~Mandic,
J.~Mardon,
J.~Maricic,
R.~Maruyama,
D.~N.~McKinsey,
R.~Mahapatra,
H.~Nelson,
J.~Orrell,
K.~Palladino,
E.~Pantic,
R.~Partridge,
H.~Robertson,
A.~Ryd,
T.~Saab,
B.~Sadoulet,
R.~Schnee,
W.~Shepherd,
A.~Sonnenschein,
P.~Sorensen,
M.~Szydagis,
T.~M.~P.~Tait,
T.~Volansky,
M.~Witherell,
D.~Wright,
K.~Zurek.
\end{center} 
%Conveners are also listed separately in authorlist.tex

\end{boldmath}\end{center}

%%%%%%%%%%%%%%%%%%%%%%%%%%%%%%%%%%%%%%%%%%%%%%%%%%%%%%%%%%%
%%%%%%%%%%%%%%%%%%%%%%%%%%%%%%%%%%%%%%%%%%%%%%%%%%%%%%%%%%%
%%%%%%%%%%%%%%%%%%%%%%%%%%%%%%%%%%%%%%%%%%%%%%%%%%%%%%%%%%%
%%%%%%%%%%%%%%%%%%%%%%%%%%%%%%%%%%%%%%%%%%%%%%%%%%%%%%%%%%%

%%%%%%%%%%%%%%%%%%%%%%%%%%%%%%%%%%%%%%%%%%%%%%%%%%%%%%%%%%%
\section{Executive Summary}\label{sec:cf1:exec}

%%%%%%%%%%%%%%%%%%%%%%%%%%%%%%%%%%%%%%%%%%%%%%%%%%%%%%%%%%%
\textbf{\emph{Dark matter exists}}

It is now generally accepted in the scientific community that roughly 85\% of the matter in the universe is in a form that neither emits nor absorbs electromagnetic radiation.  Multiple lines of evidence from cosmic microwave background probes, measurements of cluster and galaxy rotations, strong and weak lensing and big bang nucleosynthesis all point toward a model containing cold dark matter particles  as the best explanation for the universe we see.  Alternative theories involving modifications to Einstein's theory of gravity have not been able to explain the
observations across all scales.

%%%%%%%%%%%%%%%%%%%%%%%%%%%%%%%%%%%%%%%%%%%%%%%%%%%%%%%%%%%
\textbf{\emph{WIMPs are an excellent candidate for the dark matter}}

Weakly Interacting Massive Particles (WIMPs) represent a class of dark matter particles that froze out of thermal equilibrium in the early universe with a relic density that matches observation.  This coincidence of scales - the relic density and the weak force interaction scale - provides a compelling rationale for WIMPs as particle dark matter.  Many particle physics theories beyond the Standard Model provide natural candidates for WIMPs, but there is a huge range in the possible WIMP masses (1\,GeV to 100\,TeV) and interaction cross sections with normal matter (10$^{-40}$ to 10$^{-50}$\,cm$^2$).  It is expected that WIMPs would interact with normal matter by elastic scattering with nuclei \cite{Goodman:1984dc}, requiring detection of nuclear recoil energies in the 1-100 \,keV range.  These low energies and cross sections represent an enormous experimental challenge, especially in the face of daunting backgrounds from electron recoil interactions and from neutrons that mimic the nuclear recoil signature of WIMPs.  Direct detection describes an experimental program that is designed to identify the interaction of WIMPs with normal matter.

%%%%%%%%%%%%%%%%%%%%%%%%%%%%%%%%%%%%%%%%%%%%%%%%%%%%%%%%%%%
\textbf{\emph{Discovery of WIMPs may come at any time}}

\begin{figure}[h!]%Figure 1
\begin{center}
\includegraphics[width=0.6\textwidth]{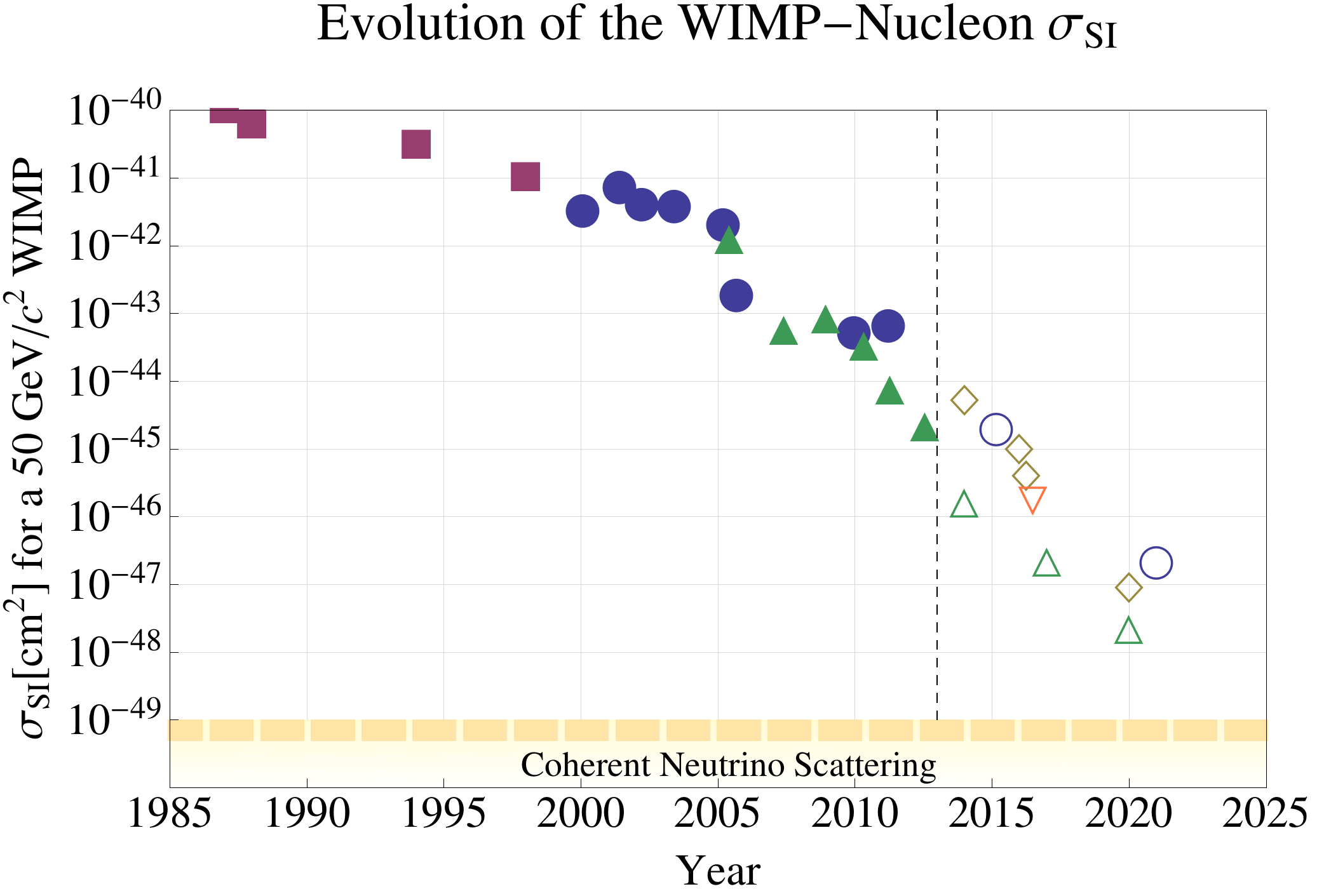}
\caption{\label{fig:50GeV}History and projected evolution with time of spin-independent WIMP-nucleon cross section limits for a 50\,GeV WIMP.  The shapes correspond to technologies: cryogenic solid state (blue circles), crystal detectors (purple squares), liquid argon (brown diamonds), liquid xenon (green triangles), and threshold detectors (orange inverted triangle). Below the yellow dashed line, WIMP sensitivity is limited by coherent neutrino-nucleus scattering.}
\end{center}
\end{figure}

Direct detection experiments have made tremendous progress in the last three decades, with sensitivity to WIMPs doubling roughly every 18~months, as seen in  Fig.~\ref{fig:50GeV}.
This rapid progress has been driven by remarkable innovations in detector technologies that have provided extraordinary active rejection of normal matter backgrounds.  A comprehensive program to model and reduce backgrounds, using a combination of material screening, radiopure passive shielding and active veto detectors, has resulted in projected background levels of $\sim$1\,event/ton of target mass/year.  Innovations in all of these areas are continuing, and promise to increase the rate of progress in the next two decades.  Ultimately, direct detection experiments will start to see signals from coherent scattering of solar, atmospheric and diffuse supernova neutrinos.  Although interesting in their own right, these neutrino signals will eventually require background subtraction or directional capability in WIMP direct detection detectors to separate them from the dark matter signals. 

%%%%%%%%%%%%%%%%%%%%%%%%%%%%%%%%%%%%%%%%%%%%%%%%%%%%%%%%%%%
\begin{center}

\textbf{A Roadmap for Direct Detection}

\textbf{\emph{Discovery}}

Search for WIMPS over a wide mass range (1 GeV to 100 TeV), with at least an order 
of magnitude improvement in sensitivity in each generation, until we encounter 
the coherent neutrino scattering signal that will arise from solar, atmospheric 
and supernova neutrinos

\textbf{\emph{Confirmation}}

Check any evidence for WIMP signals using experiments with complementary technologies, 
and also with an experiment using the original target material, but having better 
sensitivity

\textbf{\emph{Study}}

If a signal is confirmed, study it with multiple technologies in order to extract 
maximal information about WIMP properties

\textbf{\emph{R\&D}}

Maintain a robust detector R\&D program on technologies that can enable discovery, 
confirmation and study of WIMPs.

\end{center}

This comprehensive direct detection program carries the potential for an extraordinary discovery and subsequent understanding of particles that may constitute most of the matter in our universe.

%%%%%%%%%%%%%%%%%%%%%%%%%%%%%%%%%%%%%%%%%%%%%%%%%%%%%%%%%%%
\textbf{\emph{The US has a well-defined and leading role in direct detection experiments}}

The US has a clear leadership role in the field of direct dark matter detection experiments, with most major collaborations having major involvement of US groups.  In order to maintain this leadership role and to reduce the risk inherent in pushing novel technologies to their limits, a variety of US-led direct search experiments is required.  To maximize the science reach of these experiments, any proposed new direct detection experiment should demonstrate that it meets at least one of the following two criteria:

\begin{compactitem}
\item Provide at least an order of magnitude improvement in cross section sensitivity for some range of WIMP masses and interaction types.
\item Demonstrate the capability to confirm or deny an indication of a WIMP signal from another experiment.
\end{compactitem}

\textbf{\emph{Direct detection will provide complementary information about dark
matter}}

\begin{figure}[h!] %Figure 2
\begin{center}
\includegraphics[width=0.32\textwidth]{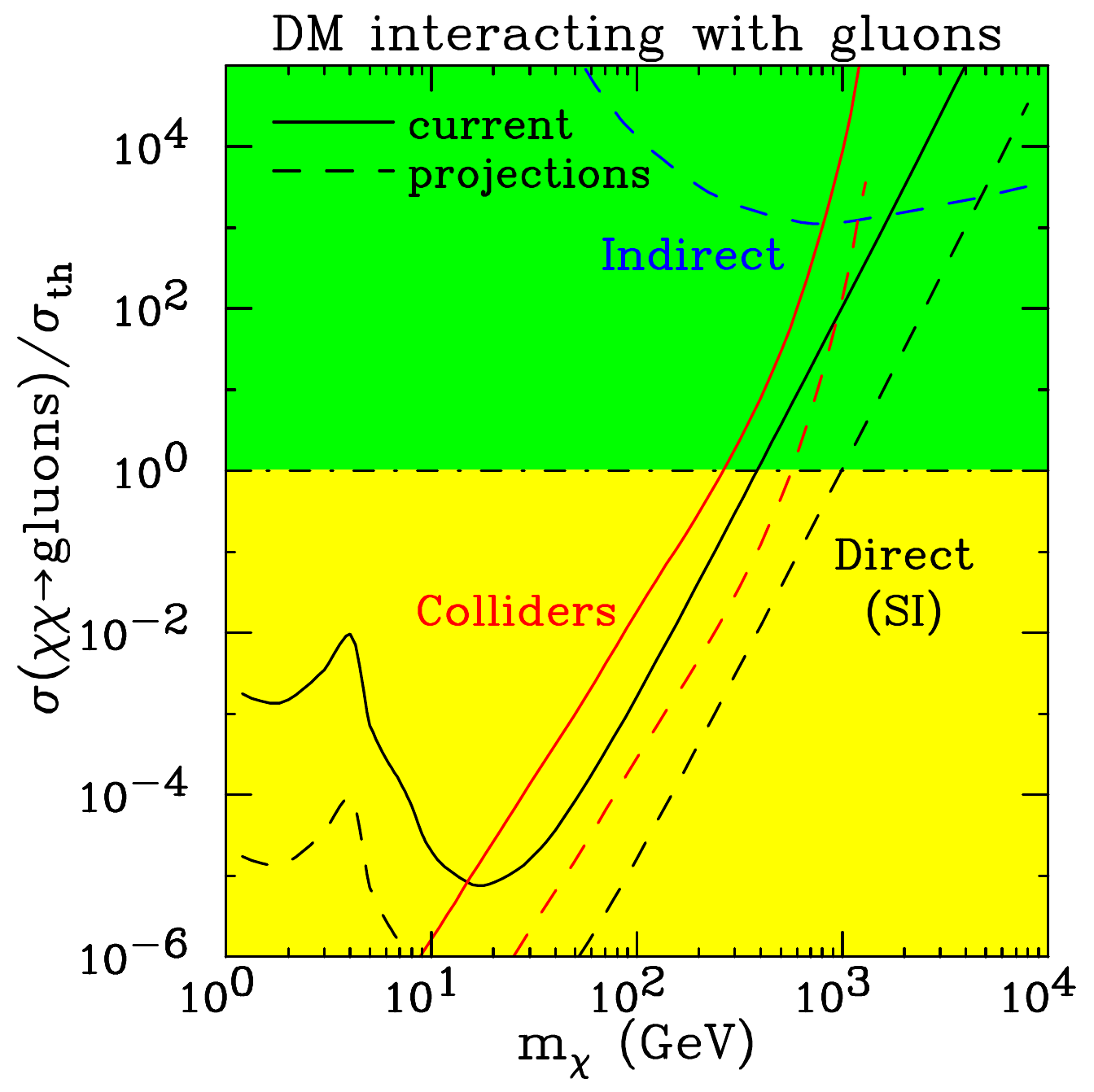}
\includegraphics[width=0.32\textwidth]{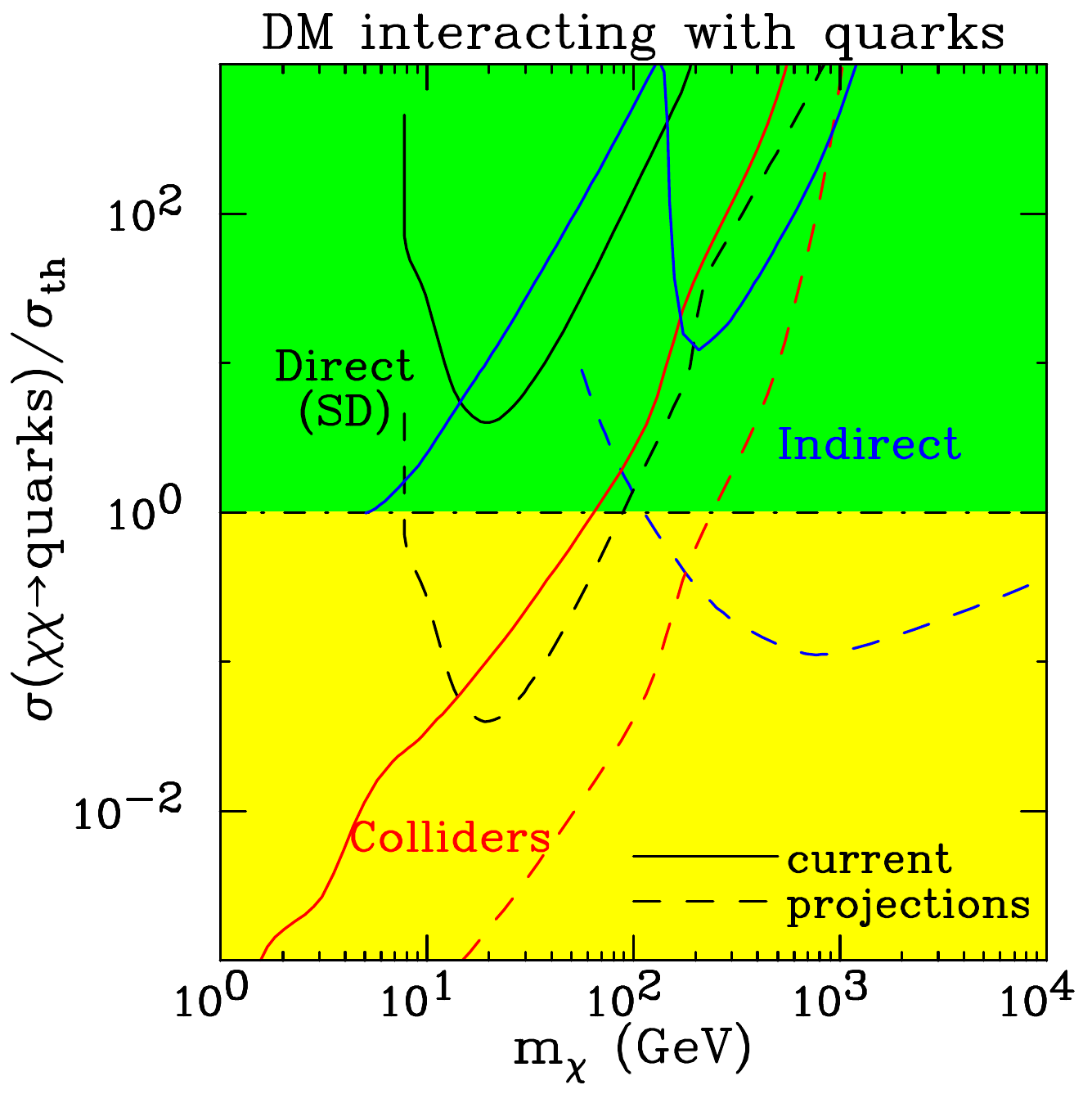}
\includegraphics[width=0.32\textwidth]{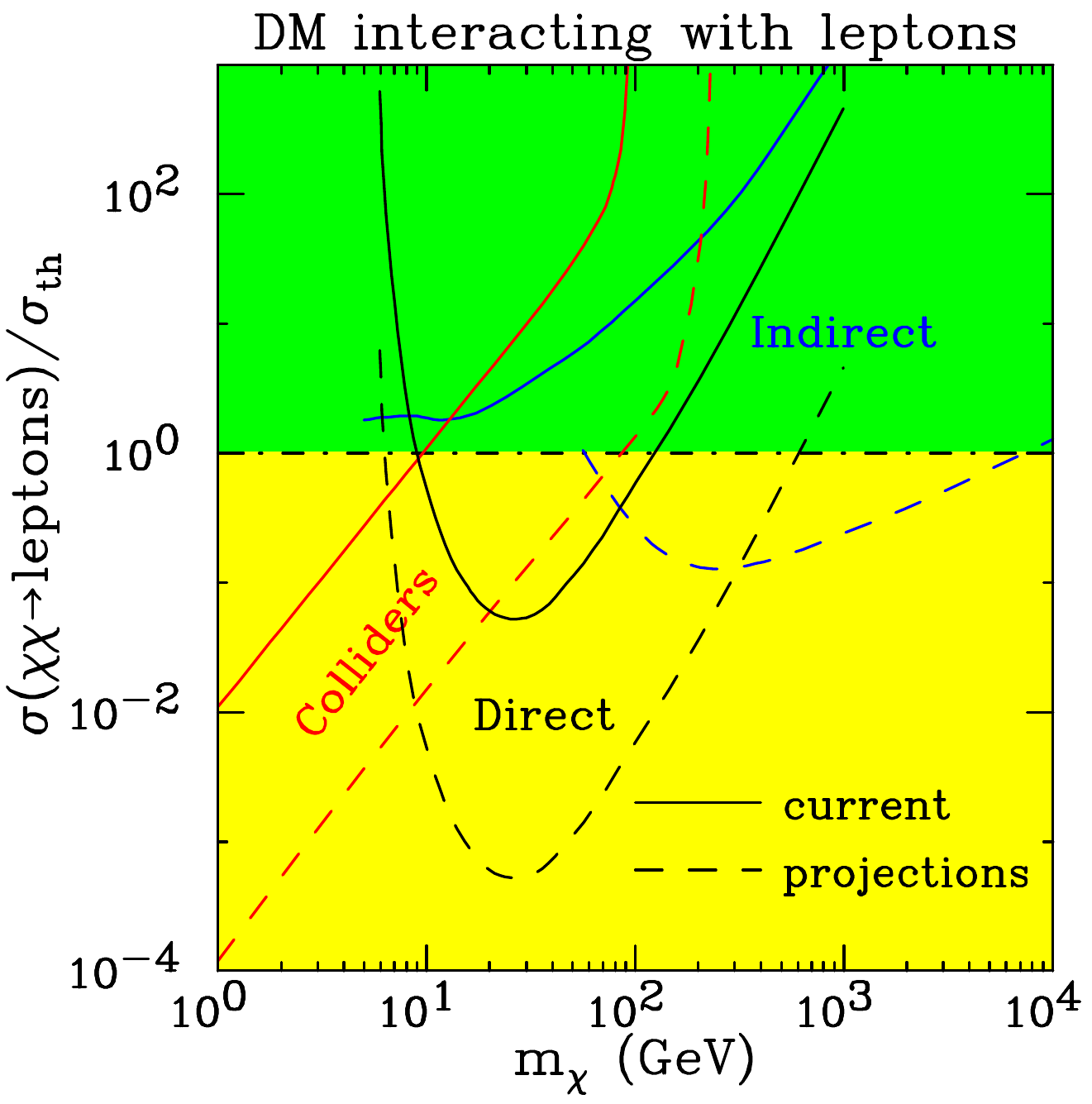}
\caption{\label{fig:complementarity}Dark matter discovery prospects in the (m$_\chi$, $\sigma$/$\sigma_{\rm th}$) plane for current and future direct detection, indirect detection, and particle colliders for dark matter coupling to gluons, quarks, and leptons, as indicated.  See Ref.~\cite{complementarity} and references cited therein for a detailed description.}
\end{center}
\end{figure}

A confirmed signal from direct detection experiments would prove that WIMPs exist and that they come from dark matter in our galaxy.  Studying the signal with several experimental targets would provide a measure of the WIMP mass, the form of the interactions with normal matter and even astrophysical information about the distribution of dark matter in our galaxy. This information is complementary to that which can be obtained from particle colliders or indirect detection of dark matter, as shown in Fig.~\ref{fig:complementarity}
from the CF4 complementarity report~\cite{complementarity}.

%%%%%%%%%%%%%%%%%%%%%%%%%%%%%%%%%%%%%%%%%%%%%%%%%%%%%%%%%%%
\newpage
\section{Introduction}\label{sec:cf1:intro}

Deciphering the nature of dark matter is one of the primary goals of particle physics for the next decade.  Astronomical evidence of many types, including cosmic microwave background measurements, cluster and galaxy rotation curves, lensing studies and spectacular observations of galaxy cluster collisions, all point towards the existence of cold dark matter particles.  Cosmological simulations based on the Cold Dark Matter (CDM) model have been remarkably successful at predicting the actual structures we see in the universe.  Alternative explanations involving modification of Einstein's theory of general relativity have not been able to explain this large body of evidence across all scales.

Weakly Interacting Massive Particles (WIMPs) are strong candidates to explain dark matter, because of a simple mechanism for the production of the correct thermal relic abundance of dark matter in the early Universe.  If WIMPs exist, they should be detectable through their scattering on atomic nuclei on Earth, by production at particle colliders or through detection of their annihilation radiation in our galaxy and its satellites.  The first of these methods, ``direct detection'', involves the construction of deep underground particle detectors to directly register the interactions of through-going dark matter particles

\begin{figure}[h!]%Figure 3
\begin{center}
\includegraphics[width=0.5\textwidth]{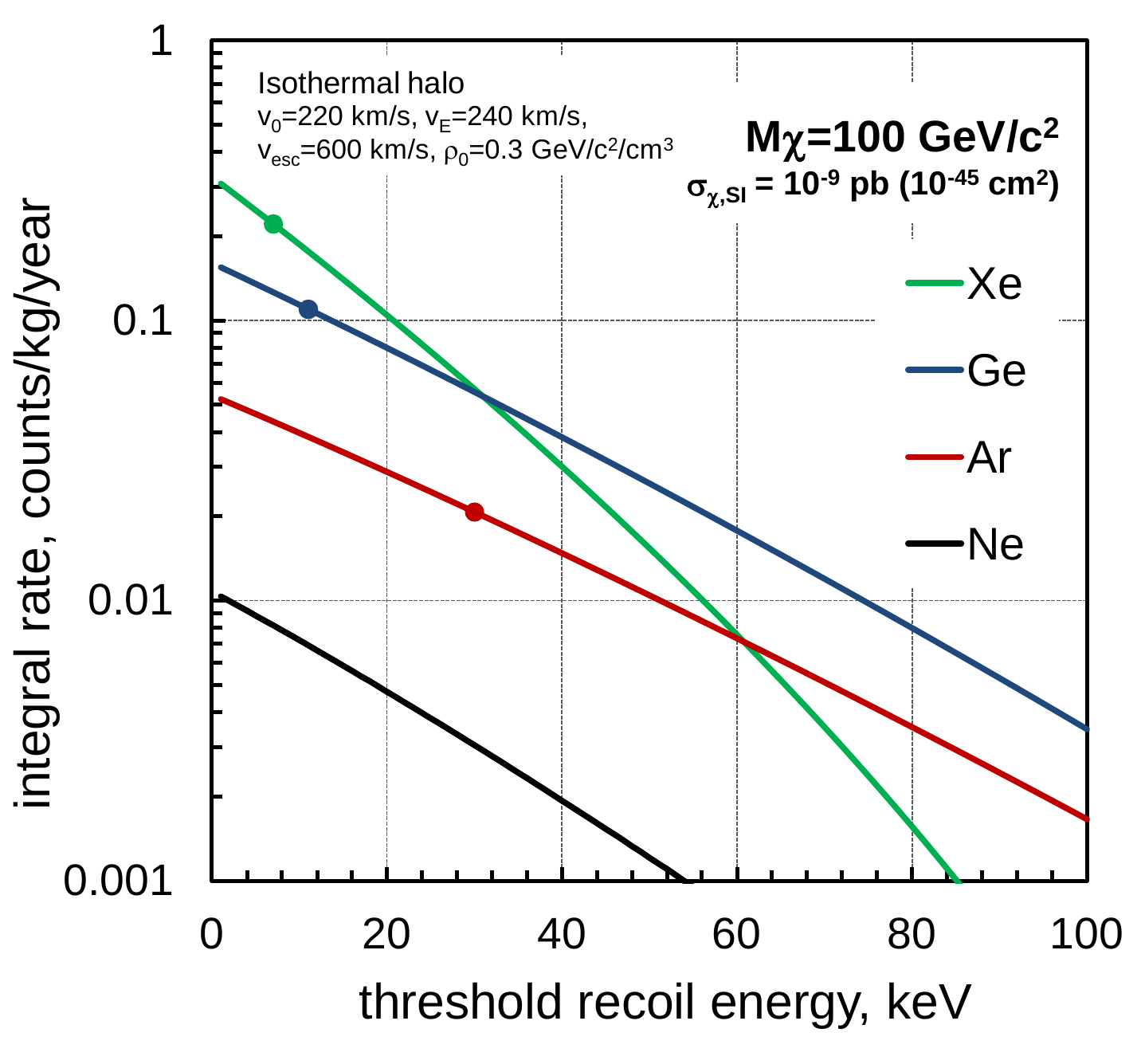}
\caption{\label{fig:rates-vs-a}Predicted integral spectra for WIMP elastic scattering for Xe, Ge, Ar and Ne (in order of decreasing rate at zero threshold), assuming perfect energy resolution~\cite{chepel}.  Dark matter rates are for a 100\,GeV WIMP with 10$^{-45}$\,cm$^2$ interaction cross section per nucleon, calculated with the halo parameters shown; the markers indicate typical WIMP-search thresholds for each target.}
\end{center}
\end{figure}

The energy scale for WIMP scattering on nuclei is determined by the gravitational binding energy of our galaxy.  Typical energy spectra for a 100\,GeV WIMP interacting with various targets are shown in Fig.~\ref{fig:rates-vs-a}.
The shapes of these spectra do not, in general, depend on the underlying particle physics model; astrophysical uncertainties are believed to play only a small role.  N-body simulations of galactic halos do show a departure on small scales from the standard smooth isothermal model, but the effect of micro-halos on direct detection experiments has been shown to be minimal~\cite{schneider}.  However, the expected WIMP-nucleon total interaction rate is highly dependent on particle physics models and subject to many orders of magnitude uncertainty.

In the non-relativistic limit, WIMP-nucleon couplings are usefully classified as ``spin-dependent'', when the sign of the scattering amplitude depends on the relative orientation of particle spins, or ``spin-independent'' when spin orientations do not affect the amplitude.  For spin-dependent interactions, the WIMP effectively couples to the net nuclear spin, due to cancellation between opposite spin pairs.  It will differ depending on whether the net nuclear spin is carried primarily by a residual neutron or proton.  For spin-independent couplings, if all nucleons couple to WIMPs in the same way, the total nuclear cross section is enhanced by the square of the atomic mass due to coherent summation over all the scattering centers in the nucleus.  This greatly increases event rates on heavy target nuclei relative to lighter nuclei.  Finally, in some models (so-called ``isospin dependent dark matter''), the proton and neutron contributions can be different in magnitude or sign, breaking the simple $A^2$ scaling.  As a result, information about the interaction type can be obtained by comparing results obtained with different target nuclei.

The field has progressed since the first experiments in the late 1980's by achieving sensitivity to progressively smaller WIMP-nucleon couplings. In the last decade, sensitivity has increased by three orders of magnitude and now probes cross sections as low as $\sim$10$^{-45}$\,cm$^2$.  The main driver of increased sensitivity is the development of a surprisingly diverse set of techniques for the elimination of background events from environmental radioactivity and cosmic rays.

\begin{figure}[h!]%Figure 4
\begin{center}
\includegraphics[width=0.8\textwidth]{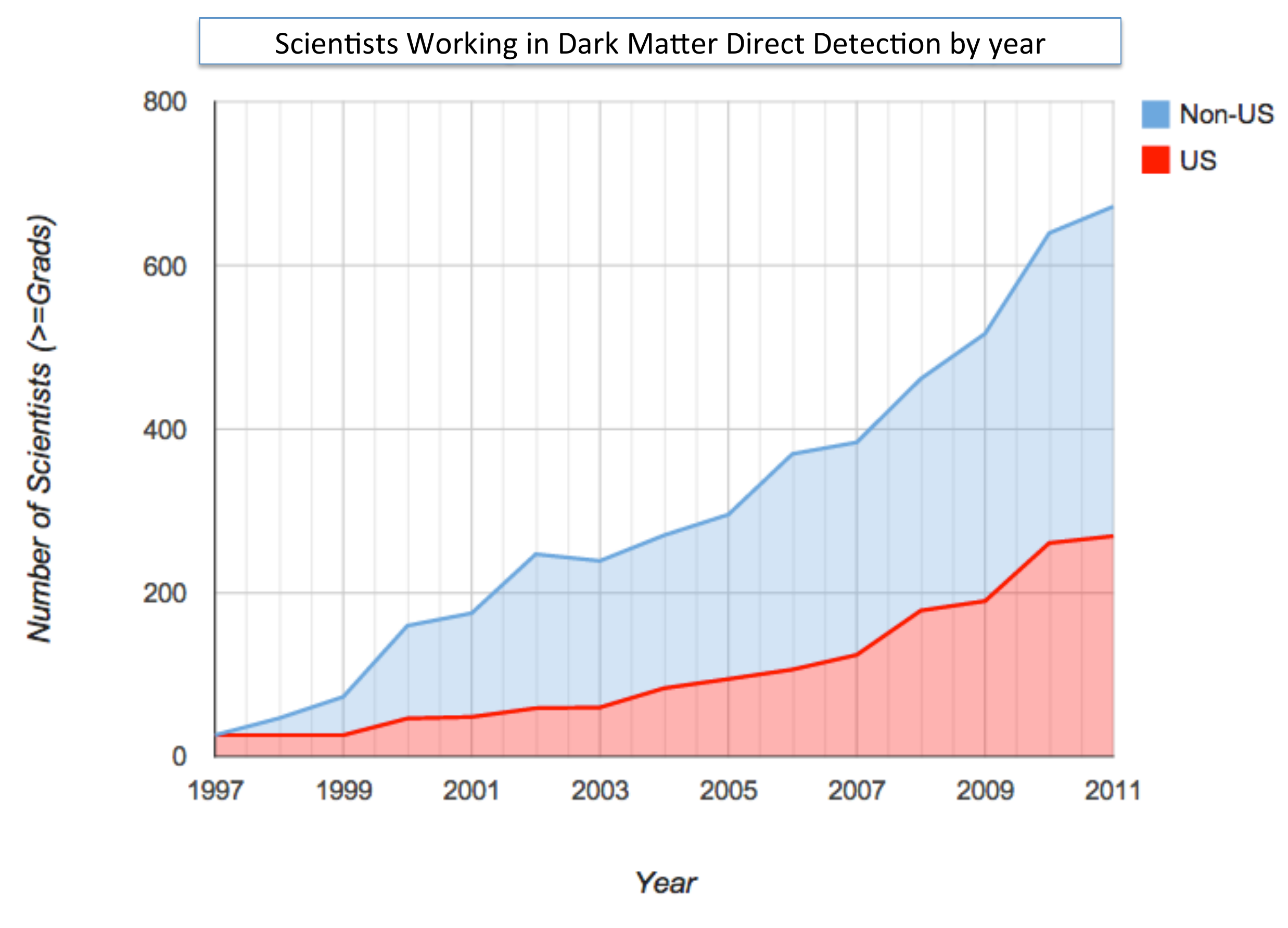}
\caption{\label{fig:demographics}Dark matter direct detection experiment demographics.}
\end{center}
\end{figure}

The enormous appeal of direct detection of dark matter to the worldwide particle physics community is shown in Fig.~\ref{fig:demographics}.
It is also highlighted by the large number of highly-trained graduate students produced by the field.  The rapid succession of experiments and their modest size leads to PhD theses that usually involve both hardware work as well as high-level analysis tasks. As such, students graduating from the field of direct detection are trained in a wide variety of experimental and analytical techniques. Our PhDs have an ideal preparation to tackle problems in broad areas of basic science, engineering, industry, and even the financial sectors.

In this paper, we discuss the context for direct detection experiments in the search for dark matter and describe briefly the current state of theoretical models for WIMPs.  A brief review of the technologies and experiments is presented, along with a discussion of facilities and instrumentation that enable such experiments, and a description of other physics that these experiments can do. We end with a discussion of how the field is likely to evolve over the next two decades, with a specific roadmap and criteria for new experiments.

The international dark matter program is expected to evolve from currently-running (G1) experiments to G2 experiments (defined as in R\&D or construction now), to G3 experiments which will eventually reach the irreducible neutrino background.  Down-selection and consolidation will occur at each stage, given the growing financial cost and manpower needs of these experiments.  The DOE has a formal down-selection process for one or more major G2 experiments.  Since substantial NSF contributions are also expected, XENON1T is considered to be a joint NSF/international US-led G2 experiment.  Additional G2 experiments may also move to construction in the coming year by either having relatively low overall cost or relatively low cost to DOE/NSF.  It is unclear when and how the U.S. funding agencies will select G3 experiments, but such a stage is on their planning horizon.  It is expected that only one or two U.S.-led G3 experiments at the \$100M range will be financially tenable.

%%%%%%%%%%%%%%%%%%%%%%%%%%%%%%%%%%%%%%%%%%%%%%%%%%%%%%%%%%%
\section{Dark Matter Direct Detection in Context}\label{sec:cf1:dmddic}

\begin{figure}[h!]%Figure 5
\begin{center}
\includegraphics[width=\textwidth]{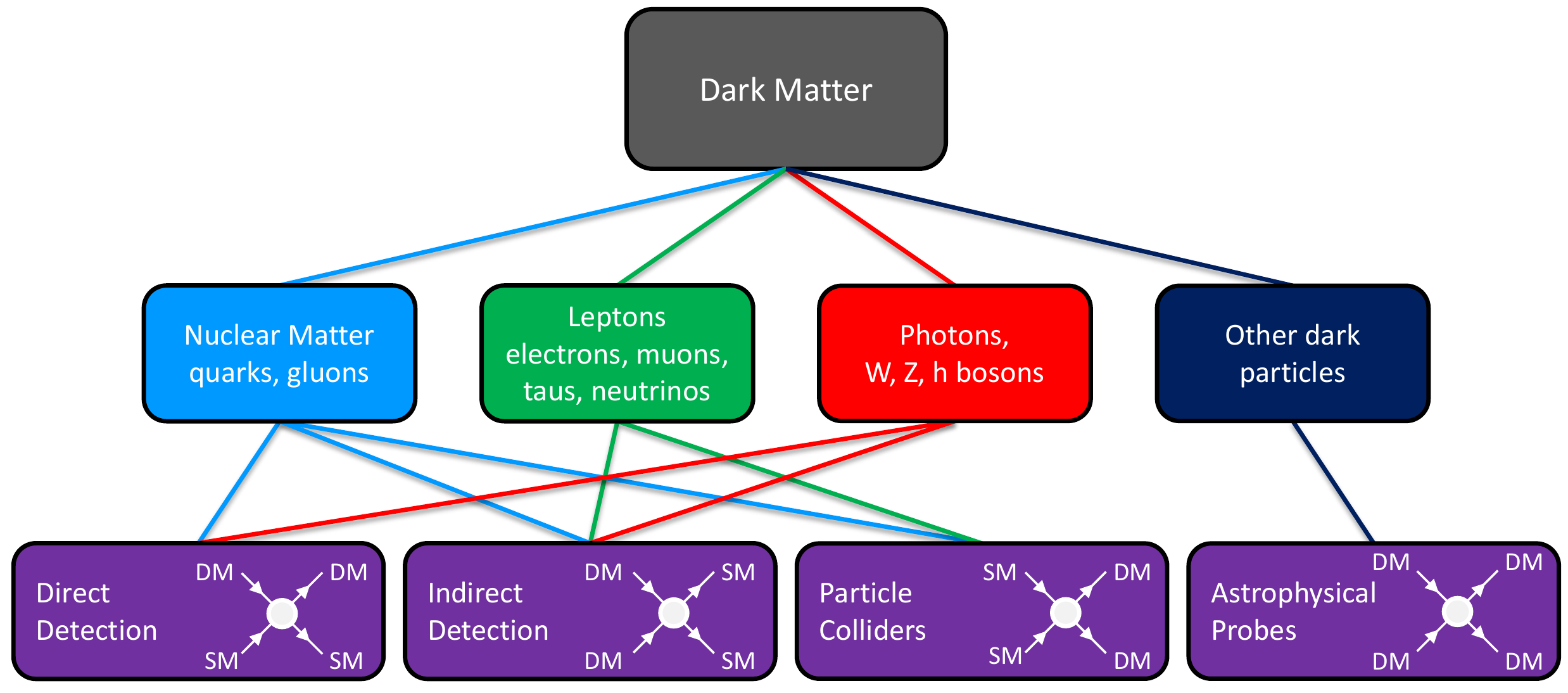}
\caption{\label{fig:interactions}Dark matter may have non-gravitational interactions with any of the known particles as well as other dark particles, and these interactions can be probed in several different ways.}
\end{center}
\end{figure}

Direct detection is only one method to search for dark matter.  Because dark matter can potentially interact with any of the known particles or, as in the case of hidden sector dark matter, another currently unknown particle (as shown in Fig.~\ref{fig:interactions}),
it is important to place direct detection in the larger context of dark matter research.  The Snowmass Cosmic Frontier Working Group CF4 has prepared a report~\cite{complementarity} exploring the complementarity of different methods of dark matter detection.  In this section, we extract a few of the key results relevant to direct detection.

Fig.~\ref{fig:complementarity} shows sensitivity plots for direct, indirect and collider experiments assuming a generic contact interaction involving gluons, quarks or leptons.  In these plots, the cross section shown on the $y$-axis has been normalized to the cross section required for a thermal relic to fully account for the amount of dark matter in the universe.  In other words, the observation of a dark matter candidate with $\sigma$ above $\sigma_{\rm th}$ (green shaded region) would not be able to account for all of the dark matter because too much dark matter would have annihilated during the evolution of the universe.  Such an observation would point to the existence of another dark matter species waiting to be discovered.  On the other hand, if an observed cross section were below $\sigma_{\rm th}$, the corresponding relic density would be too large and another annihilation channel would need to be observed.

These plots show how direct detection experiments have an advantage over the other methods in observing dark matter interacting with gluons or leptons.  In particular, we see that for these generic models, direct searches probe mass regions at the TeV scale that are beyond the reach of colliders.  On the other hand, at low masses, where dark matter cannot deposit as much energy in a direct detection experiment, colliders become much more competitive.

The contact interactions illustrated in Fig.~\ref{fig:interactions} represent the exchange of very heavy particles.  If instead, the dark matter interactions are mediated by a light particle, effective field theory of the kind illustrated in Fig.~\ref{fig:interactions} breaks down and the mediator has to be included in the calculation explicitly.  In these cases, colliders have a relative disadvantage to direct detection experiments because the cross section can be suppressed.

\begin{figure}[h!]%Figure 6
\begin{center}
\includegraphics[width=0.8\textwidth]{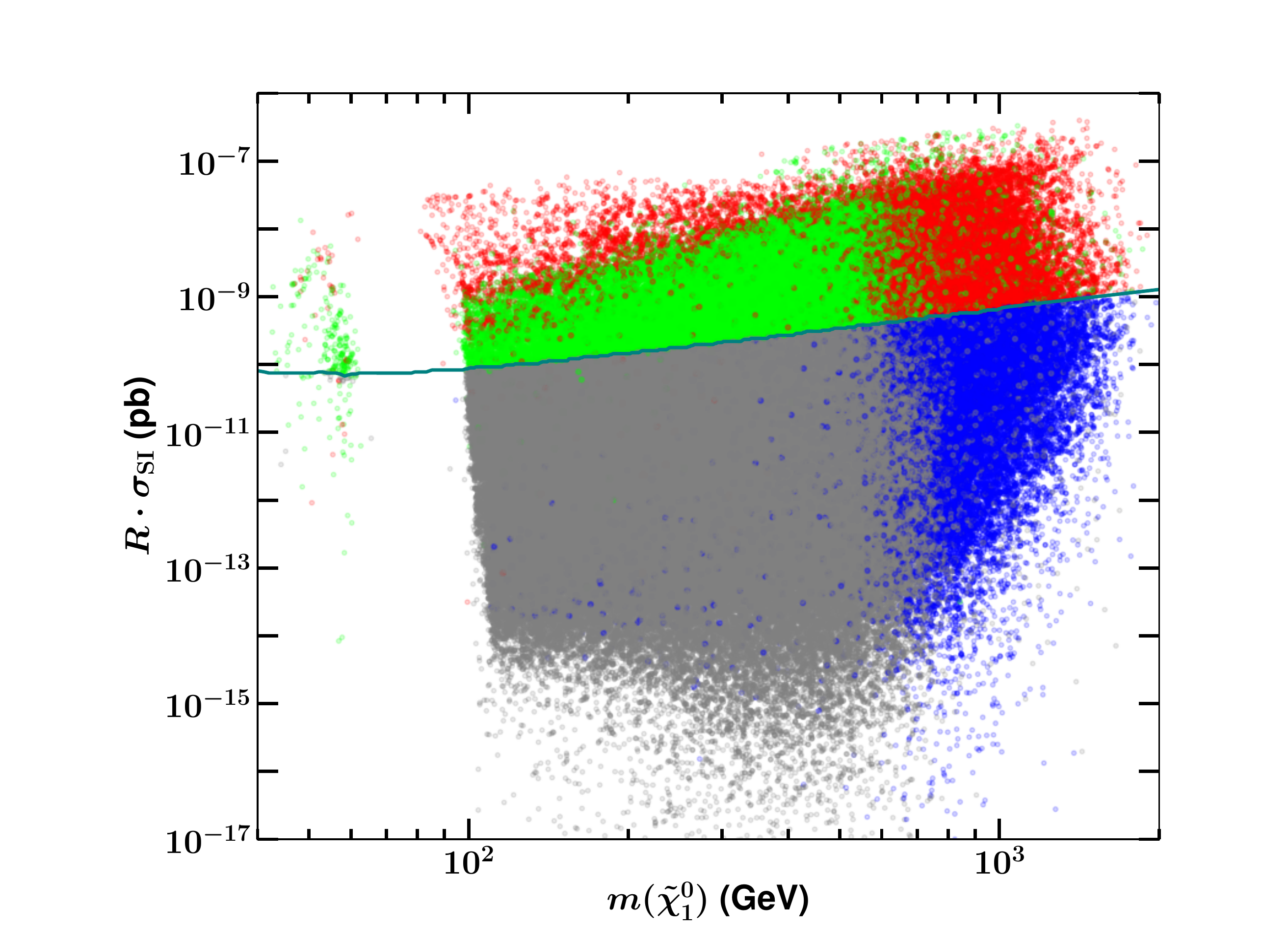}
\caption{\label{fig:scan} Results from a model-independent scan of the full parameter space in MSSM.  The models are divided into categories depending on whether dark matter can be discovered in future direct detection experiments (green points), indirect detection experiments (blue points) or both (red points). Purple points represent models that may be discovered at an upgraded LHC but escape detection via the other two methods.  See Ref.~\cite{complementarity} and references cited therein for a detailed description.}
\end{center}
\end{figure}

The complementarity of dark matter searches is also illustrated by examining specific theoretical models, for example supersymmetry.  Fig.~\ref{fig:scan}
shows sensitivity to a scan of the parameter space in MSSM, and highlights which models are, or can be, excluded by different approaches to dark matter detection.  Again, the complementarity of the methods is readily apparent.  The next section provides a more detailed description of dark matter models and the prospects for direct detection in each case.

%%%%%%%%%%%%%%%%%%%%%%%%%%%%%%%%%%%%%%%%%%%%%%%%%%%%%%%%%%%
\section{ Models of Dark Matter}\label{sec:cf1:modm}

While model-dependent, there is generically a connection between the rate of scattering with nuclei and the annihilation cross section which determines its relic abundance in the early Universe.  Already, direct detection puts important constraints on the parameter space of WIMP dark matter, and in the future it offers the possibility to cover the interesting regions entirely~\cite{complementarity}.

%%%%%%%%%%%%%%%%%%%%%%%%%%%%%%%%%%%%%%%%%%%%%%%%%%%%%%%%%%%
\subsection{Minimal Supersymmetric Standard Model (MSSM)}

The minimal supersymmetric extension of the Standard Model (MSSM) remains one of the most well-motivated theoretical frameworks for beyond the Standard Model physics, solving the gauge hierarchy problem and leading naturally to Grand Unification.  Under the assumption that R-parity is conserved, the lightest supersymmetric particle (LSP) is stable and serves as a compelling dark matter candidate.  The $\sim$100 free parameters of the MSSM are unwieldy and highly constrained by flavor and CP-violating observables, and thus are difficult to use to provide the guidance that experimental groups need to focus their search efforts.  However, constrained scenarios, where the number of free parameters is reduced considerably, do provide specific predictions for WIMP mass and interaction cross sections.  With the discovery of the Higgs, lack of positive signals from the LHC, and null results from direct detection experiments, the simplest and most highly constrained scenario (CMSSM) now resides in considerable tension with recent experimental data, and has now become largely disfavored.  Moderately constrained scenarios remain compelling even in the light of these experimental bounds.

As an example, the pMSSM is defined to have flavor-diagonal soft masses for the sfermions, which are degenerate for the first and second generations.  The $A$-terms are also assumed to be diagonal, and negligible for the first and second generations.  In addition, the remainder of the supersymmetry-breaking parameters are also defined to be real.  All told, this reduces the nominal 105 parameters down to 19 or 20.  Scans of the pMSSM space, with subsequent application of constraints from recent experimental data while allowing the relic density of the neutralino LSP to be below its observed value, would imply that dark matter is not composed of a single particle species~\cite{cahill}.  The remaining parameter space largely favors neutralino masses above 100\,GeV and up to several\,TeV.  The favored spin-independent cross sections range from 10$^{-43}$\,cm$^2$ down to well below 10$^{-50}$\,cm$^2$, though the generated models exhibit clustering at the upper end of this range.  Spin-dependent cross sections range from 10$^{-41}$\,cm$^2$ to below 10$^{-48}$\,cm$^2$.  In the spin-independent case, next generation dark matter experiments will probe deeply in this region.  They conclude that maximal coverage of the model set requires a combination of direct detection, indirect detection and collider searches.

Taking a different approach, the authors of Ref.~\cite{sanford} argue that obtaining a neutralino relic density which explains all of the dark matter leads to a characteristic spin-independent cross section which for a mass larger than 70\,GeV is expected to lie between 10$^{-45}$\,cm$^2$ and 4$\times$10$^{-45}$\,cm$^2$.  The current state of direct detection is close to covering this entire range.  If no discovery is made in this range, the generic conclusion will be that the MSSM as a theory of dark matter becomes increasingly fine-tuned.

The NMSSM is a simple extension of the MSSM which introduces a gauge singlet superfield whose vacuum expectation value explains the apparent correlation of the mu parameter with the supersymmetry-breaking masses.  These theories introduce new scalar particles as well as an additional neutralino (the singlino) which can impact the physics of the LSP.  A scan of parameter space~\cite{cao} finds models which saturate the relic density and have masses between a few and 20\,GeV, and spin-independent cross sections clustered around 10$^{-45}$\,cm$^2$.  More complicated extensions are also possible, and lead to a wide range of theories of dark matter.  One other particular extension is the observation~\cite{kumar} that for some ``WIMPless'' models of super-symmetry breaking, dark matter in a hidden sector can automatically inherit the correct relationship between mass and coupling.  This results in the correct relic abundance, even when the dark matter mass is smaller than is usual for dark matter in the MSSM.

%%%%%%%%%%%%%%%%%%%%%%%%%%%%%%%%%%%%%%%%%%%%%%%%%%%%%%%%%%%
\subsection{Model-independent approaches with effective theories}

Phenomenological sketches of dark matter which seek to describe some key properties, but do not present complete visions of Nature represent an approach that is distinct from the construction of specific models (such as the MSSM).  In the limits where either the mass of the mediating particle (contact interactions) or the mass of the WIMP (heavy-WIMP effective theory) is much larger than the weak scale ($\sim$100\,GeV), such theories become particularly simple.

In the case of heavy mediating particles, the symmetries of the Standard Model imply that a wide range of complete theories of dark matter will manifest themselves as a relatively limited set of non-renormalizable interactions at low energies, whose form is dictated by gauge and Lorentz invariance, and depend on the spin and electroweak charge of the dark matter~\cite{goodman,abe,bai}.  Since the effective theories have a limited number of parameters, one can study their predictions without the need to scan through multi-dimensional parameter spaces.  A wide range of behavior in direct detection experiments is nonetheless captured.  This includes cases with light ($<$10\,GeV) dark matter particles, dark matter that has either predominantly spin-independent or spin-dependent interactions with nuclei, and dark matter whose interactions with protons and neutrons are different (e.g. isospin-violating dark matter).

In the case of the heavy-WIMP effective theory~\cite{solon}, the discovery of a Standard Model-like Higgs boson and the hitherto absence of evidence for other new states indicates that if WIMPs comprise cosmological dark matter, they may plausibly be heavy compared to the electroweak scale, $M_\chi$$\gg$$M_{W}$,$M_{Z}$.  In this limit, the relic density suggests a mass scale for the dark matter of a few TeV, and the absolute cross section for a WIMP of given electroweak quantum numbers to scatter from a nucleon becomes computable in terms of only Standard Model parameters.  For example, a scalar iso-triplet of ``dark pions'' is a generic composite WIMP~\cite{hill}.  Direct detection of such a WIMP is similar to that of a supersymmetric wino, with the relation becoming exact in the heavy WIMP limit.   A pure electroweak SU(2) triplet scatters with per-nucleon cross section of 10$^{-47}$\,cm$^2$ .  Larger cross sections typically require the lightest WIMP to be a mixture of different electroweak multiplets, leading to cross sections in the range from 10$^{-44}$\,cm$^2$ to 10$^{-47}$\,cm$^2$, depending on the strength of a Higgs coupling parameter mediating interactions between the multiplets.

%%%%%%%%%%%%%%%%%%%%%%%%%%%%%%%%%%%%%%%%%%%%%%%%%%%%%%%%%%%
\subsection{Low-mass WIMPs and alternatives to the default thermal relic WIMP paradigm}

An interesting variation on the usual thermal relic WIMP paradigm is provided by the Asymmetric Dark Matter scenario, which posits a dark matter particle that is not its own anti-particle, and seeks a connection between the primordial dark matter asymmetry and the baryon asymmetry.  Details depend on the specific models, but generically if they relate the number density of baryons and WIMPs, one has the generic expectation that the dark matter has a natural mass scale of $m_{p }$$\times$$\rho_{DM}$/$\rho_{b}$$\sim$5\,GeV.  The scattering cross-sections are very model-dependent~\cite{lin}.  Nonetheless, important benchmarks can be reached, and ruled out.  For example, the need for thermal equilibrium in the early universe is suggestive of scattering with nucleons with a cross section above 10$^{-49}$\,cm$^2$.  Scattering off electrons when the particle mediating the scattering is heavier than the dark matter particle implies cross-sections larger than 10$^{-41}$\,cm$^2$ for 1\,GeV dark matter.

More structure in the dark matter sector gives rise to more predictive models for the scattering cross-section.  Since the dark matter in these models is typically a singlet under SM gauge interactions, the scattering off of nucleons or electrons often arises through a mixing with a SM particle such as the Higgs or a photon.  An example of one such class of hidden sector dark matter models is a model with a dark photon, where an explicit model in the context of ``Asymmetric Dark Matter'' was constructed in~\cite{cohen}, where it was found that the spin-independent cross section is bounded from below by 10$^{-41}$\,cm$^2$ for a dark matter mass of $\sim$5\,GeV.  Dark matter could also couple predominantly via a small magnetic dipole moment~\cite{sigurdson,fortin}, for which the relic density combined with current limits from direct detection are suggestive of masses below 10\,GeV.

Though recent accelerator results~\cite{atlas,cms}, CMB observations, and gamma ray observatories all place constraints on particular models of light dark matter, the models of low-mass WIMPs described above are not generically ruled out.  Accelerator results can place limits on light dark matter assumed a generalized operator framework that assumes contact interactions and heavy mediators, but which are not valid for the light mediators typically used in the above models.  Likewise, CMB constraints~\cite{giesen} place limits on energy injection during recombination delaying photon decoupling and increasing optical depth, but the above low-mass WIMP models do not inject significant energy or distort the CMB.  Recent results from Fermi/LAT place limits on light dark matter that preferentially decays into $b$$\bar{b}$ but do not generically rule out light dark matter models.

To date, the only claimed observation of dark matter is by the DAMA/LIBRA experiment.  DAMA/LIBRA observes a 9-$\sigma$ annual modulation, which can be interpreted as elastic scattering of a 10\,GeV WIMP.  This, along with observations of excess events at low recoil energy made by several other direct detection experiments (CoGeNT, CDMS-Si, and CRESST), has fueled a great deal of interest in WIMPs of mass $<$10\,GeV.  When interpreted as WIMP scattering, the data from these experiments yield cross sections that range from a few times 10$^{-40}$\,cm$^2$ to 10$^{-42}$\,cm$^2$ at masses that range from 5 to 20\,GeV.  Many theoretical models give rise to dark matter with masses in this range.  While provocative, the dark matter interpretations of these experiments under the standard elastic scattering scenario are not terribly consistent with each other, and are in tension with null results from other experiments, particularly XENON-100.

Generalized models have been proposed as attempts to explain these observations.  In particular, iso-spin violating dark matter (IVDM)~\cite{marfatia} represents a mild generalization which allows for the coupling between WIMPs and protons to differ from the coupling between WIMPs and neutrons (parameterized by the ratio of the couplings $f_{n}$/$f_{p}$).  In order to explore isospin-violating scenarios its necessary to have several direct detection experiments each with a different target nucleus.  Specific benchmarks for isospin violating scenarios are discussed in Ref.~\cite{feng}.

Despite the richness of the theoretical work, none of these models can provide a physical explanation that simultaneously accounts for all of the experimental observations without making the assumptions that one or more of the measurements is flawed.  Resolution of this confusing state of experimental data remains a high priority for the field.  It is also potentially within reach for the next generation of experiments.  Several technologies, which include point-contact Ge detectors, cryogenic Ge detectors, two-phase xenon detectors, bubble chambers and CCD-based searches, are designing the next generation experiments with the goal of pushing energy thresholds lower.  Such experiments are expected to improve sensitivities by an order of magnitude or more in the 1--10\,GeV range over the next 5--10 years.  In addition, isospin-violating scenarios strongly illustrate the need to have several direct detection experiments each with a different target nucleus.

%%%%%%%%%%%%%%%%%%%%%%%%%%%%%%%%%%%%%%%%%%%%%%%%%%%%%%%%%%%
\subsection{Direct Detection Methodology}

\begin{figure}[h!]%Figure 7
\begin{center}
\includegraphics[width=0.6\textwidth]{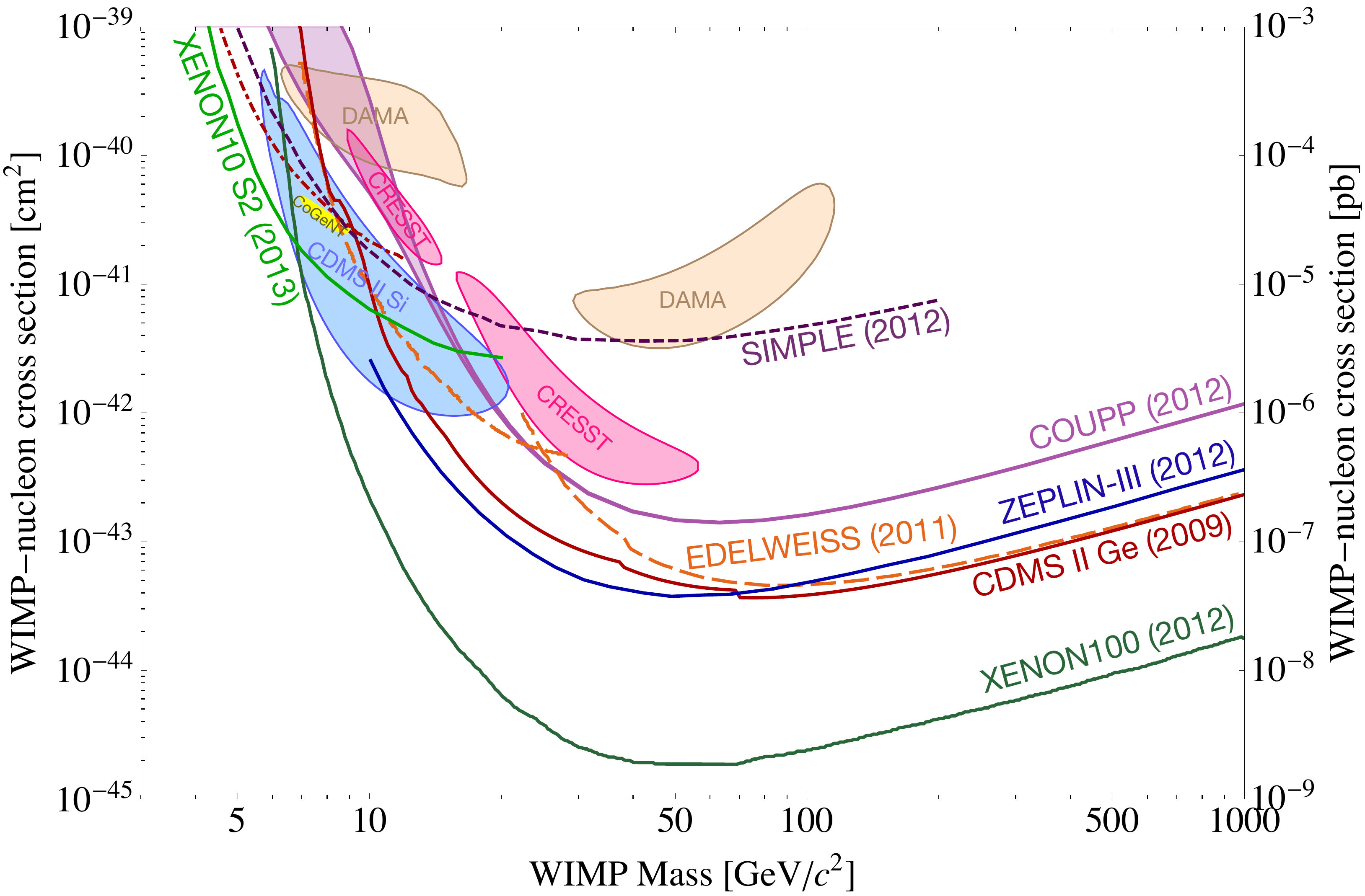}
\caption{\label{fig:SI}Spin-independent WIMP-nucleon cross section limits vs WIMP mass as of summer 2013. Experimental limits referenced \cite{Angle:2011th, Aprile:2012nq, Aprile:2011hi, Girard:2012zz, Behnke:2012ys, Aalseth:2012if, Angloher:2011uu, Bernabei:2010mq, Armengaud:2011cy, Akimov:2011tj, Agnese:2013rvf, Agnese:2013cvt, Ahmed:2009zw, Ahmed:2010wy}}
\end{center}
\end{figure}

\begin{figure}[h!]%Figure 8
\begin{center}
\includegraphics[width=0.6\textwidth]{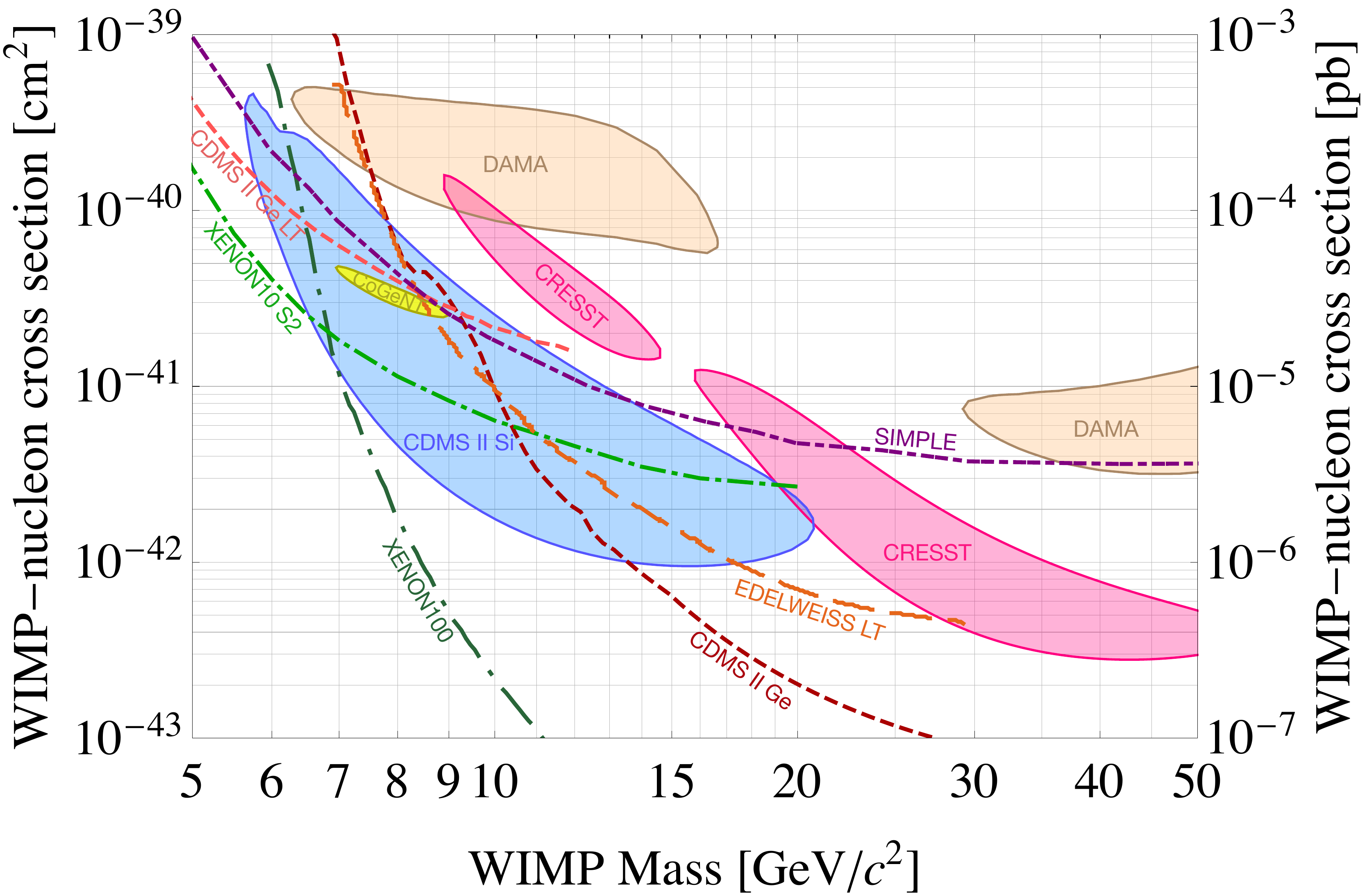}
\caption{\label{fig:SI-low}Expanded plot showing spin-independent WIMP-nucleon cross section limits, including closed contours showing hints for low-mass WIMP signals.}
\end{center}
\end{figure}

The basic methodology for direct detection experiments is to search for rare events that might be the signature of WIMP interactions, namely the ``billiard ball'' elastic scattering of a WIMP from a target nucleus.  The rate of candidate nuclear recoils is converted into a cross section for WIMP-nucleon interactions following a standard prescription that includes the effects of nuclear physics and astrophysical properties~\cite{lewin}.  Experiments can be sensitive to both nuclear spin-independent (SI) interactions and spin-dependent (SD) interactions.  For the range of momentum exchange of interest, the SI interaction is expected to be approximately coherent across the entire nucleus, so for a WIMP with equal coupling to protons and neutrons, the rate scales with the square of the atomic mass of the target nucleus.  Current experiments are therefore more sensitive to SI dark matter than SD dark matter.  Experimental results are usually presented as a plot of WIMP-nucleon cross section versus WIMP mass to allow comparison among experiments.  Fig.~\ref{fig:SI}
shows the current SI landscape, where strict upper limits exist for higher mass WIMPs.  Fig.~\ref{fig:SI-low}
zooms in on the low mass region, where several ``hints'' for dark matter have been observed.

\begin{figure}[h!]%Figure 9
\begin{center}
\includegraphics[width=0.49\textwidth]{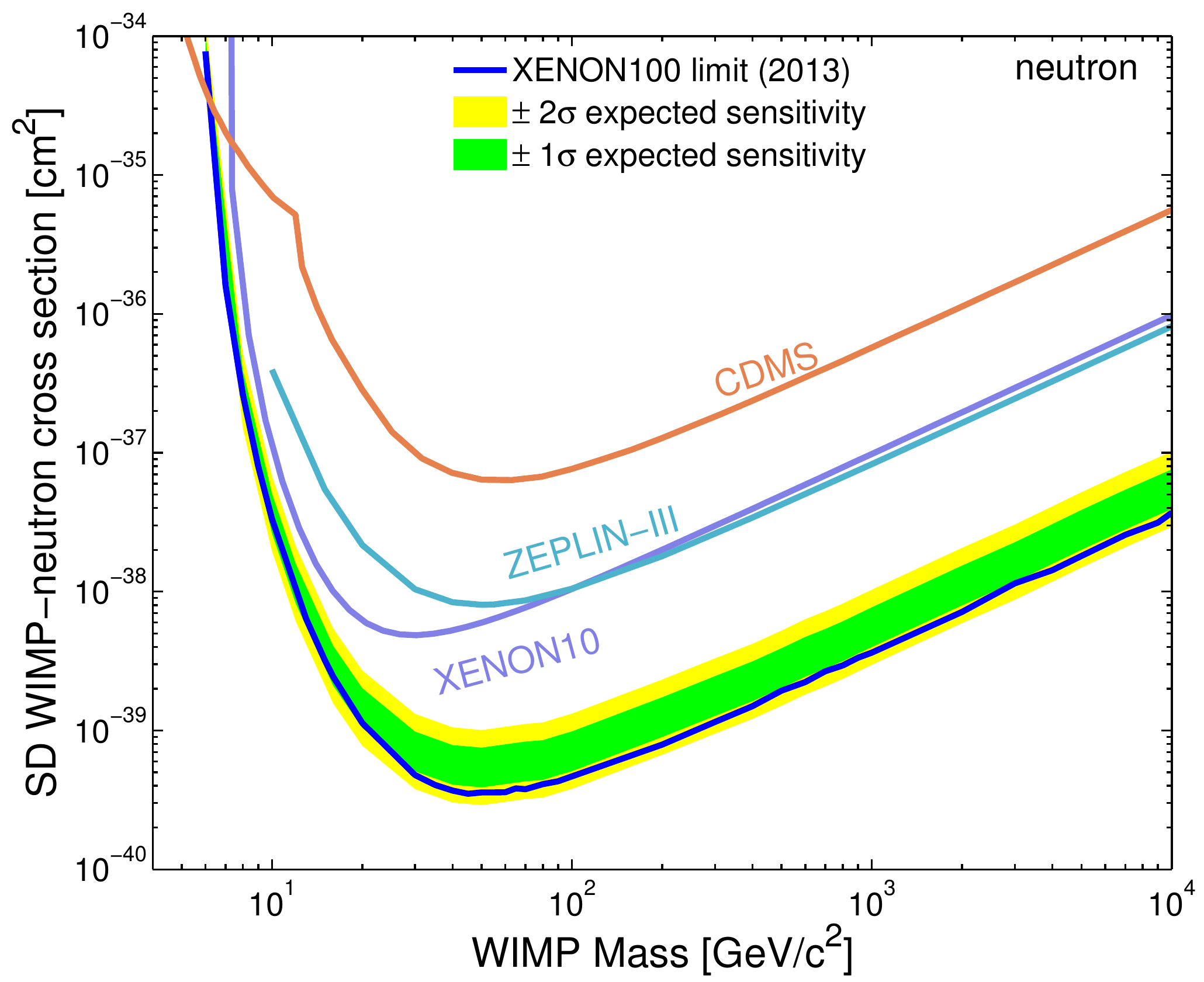}
\includegraphics[width=0.49\textwidth]{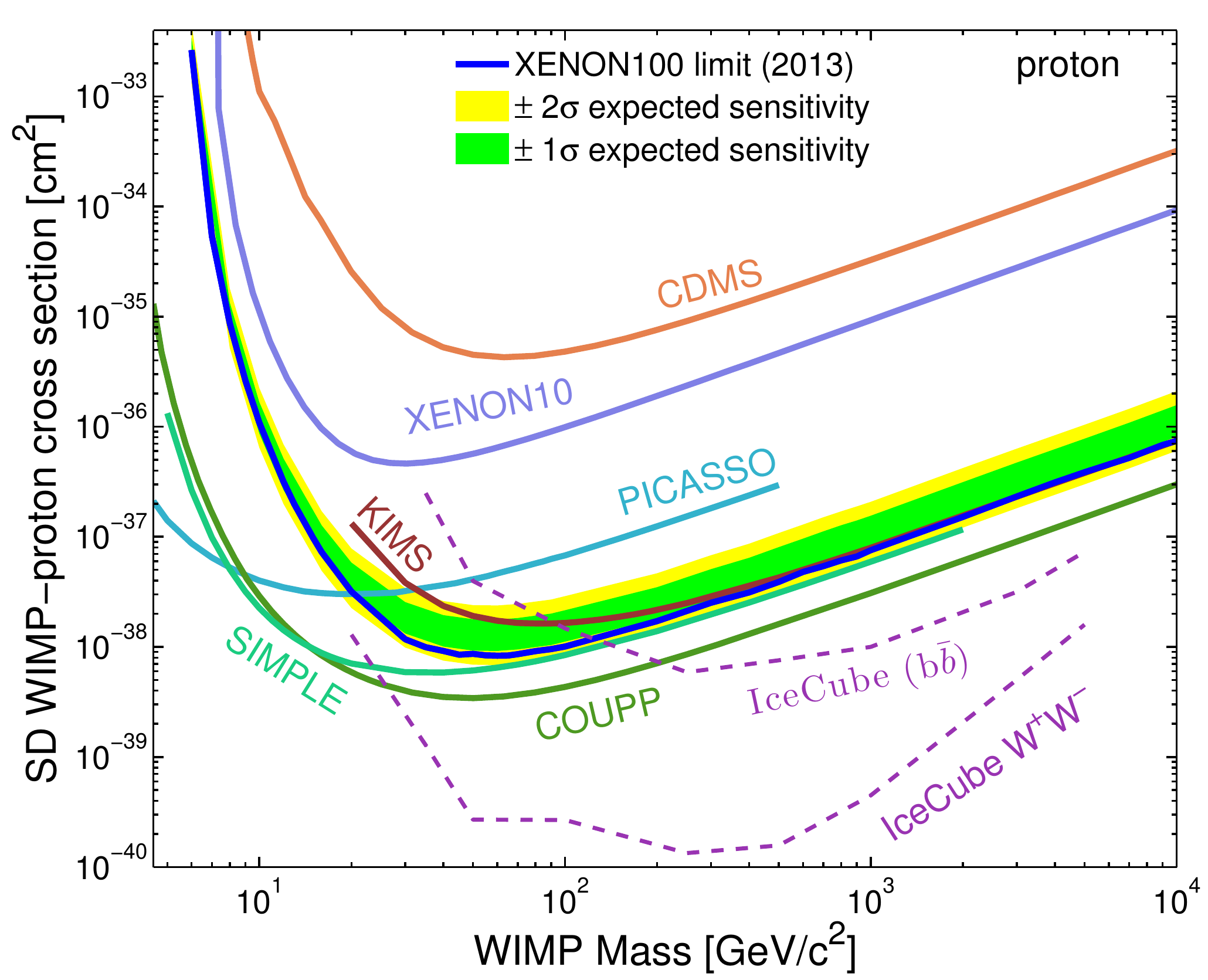}
\caption{Spin-dependent WIMP-neutron (left) and WIMP-proton (right) cross section limits versus WIMP mass for direct detection experiments\cite{Girard:2012zz, Behnke:2012ys, Akimov:2011tj, Aprile:2013doa, Akerib:2006cdms, Archambault:2012pm, Kim:2012rza}, compared with the model-dependent Ice Cube results (model-dependent) as of summer 2013~\cite{Aartsen:2012kia}.}
\label{fig:SD}
\end{center}
\end{figure}

The SD interaction is generally divided into proton and neutron couplings; the current situation is summarized in Fig.~\ref{fig:SD}.
Only direct detection can provide limits on neutron couplings, but solar neutrinos from WIMP annihilation in the sun are stronger for proton coupling.  Other types of interactions are possible, and it is important to have multiple experiments with different targets both in order to cover the parameter space for discovery, as well as to study the interaction type when signals are found.

Nuclear recoils from WIMP scattering result in a featureless energy spectrum, rising exponentially as the energy decreases.  Fig.~\ref{fig:SD} shows the expected interaction rates for some of the typical targets used, and several different WIMP masses, as a function of threshold energy.  Experiments typically do not directly measure the nuclear recoil energy. Instead the energy deposited by a particle interaction must be reconstructed from the experimental measurements as either nuclear-recoil keV$_{\rm r}$ or electron-recoil keV$_e$.  Conversion between the two energies is dependent on the target and experimental technique, and must be calibrated by each experiment.  Typically, radioactive gamma sources are used to provide calibration for electron recoils and neutron sources provide a source of nuclear recoil events.

\begin{figure}[h!]%Figure 10
\begin{center}
\includegraphics[width=0.6\textwidth]{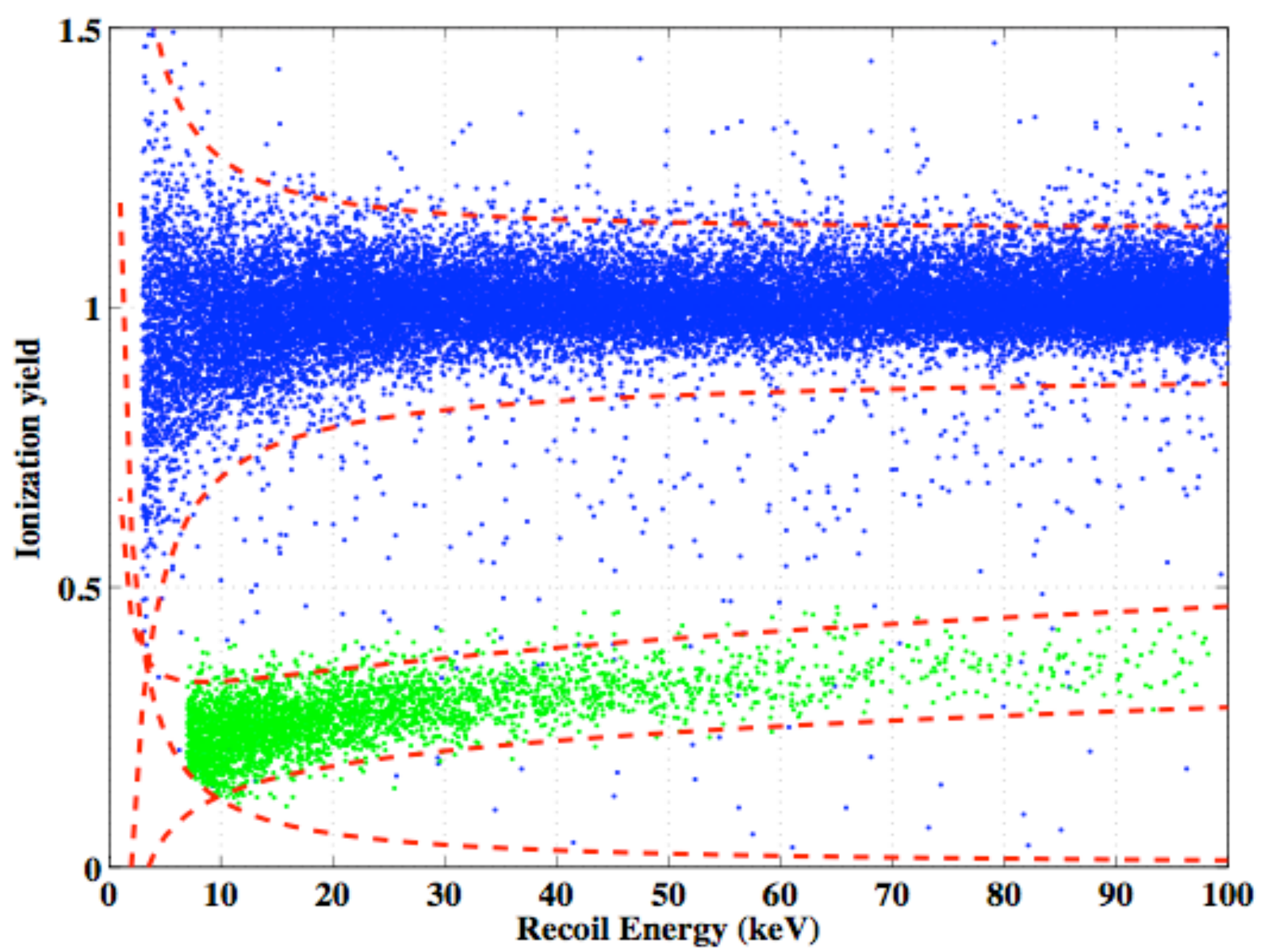}
\caption{\label{fig:cdms-discrimination}Example from the CDMS II experiment of the response to gamma rays (blue) and neutrons (green), showing the difference in ionization versus phonon energy deposition for electron and nuclear recoils.}
\end{center}
\end{figure}

One of the key features of modern experiments is the use of discrimination to select nuclear recoils and reject backgrounds.  The various technologies achieve this discrimination in many ways, but often the signal is split in two components which respond differently to nuclear recoils and backgrounds.  An example showing the difference in ionization versus phonon energy deposition for electron and nuclear recoils in the CDMS-II experiment is shown in Fig.~\ref{fig:cdms-discrimination}.
Other examples include ionization versus scintillation (used in two-phase noble liquid experiments), fast versus slow scintillation (used in liquid argon and liquid neon experiments), and scintillation versus phonons (used by CRESST).

\begin{figure}[h!]%Figure 11
\begin{center}
\includegraphics[width=0.6\hsize]{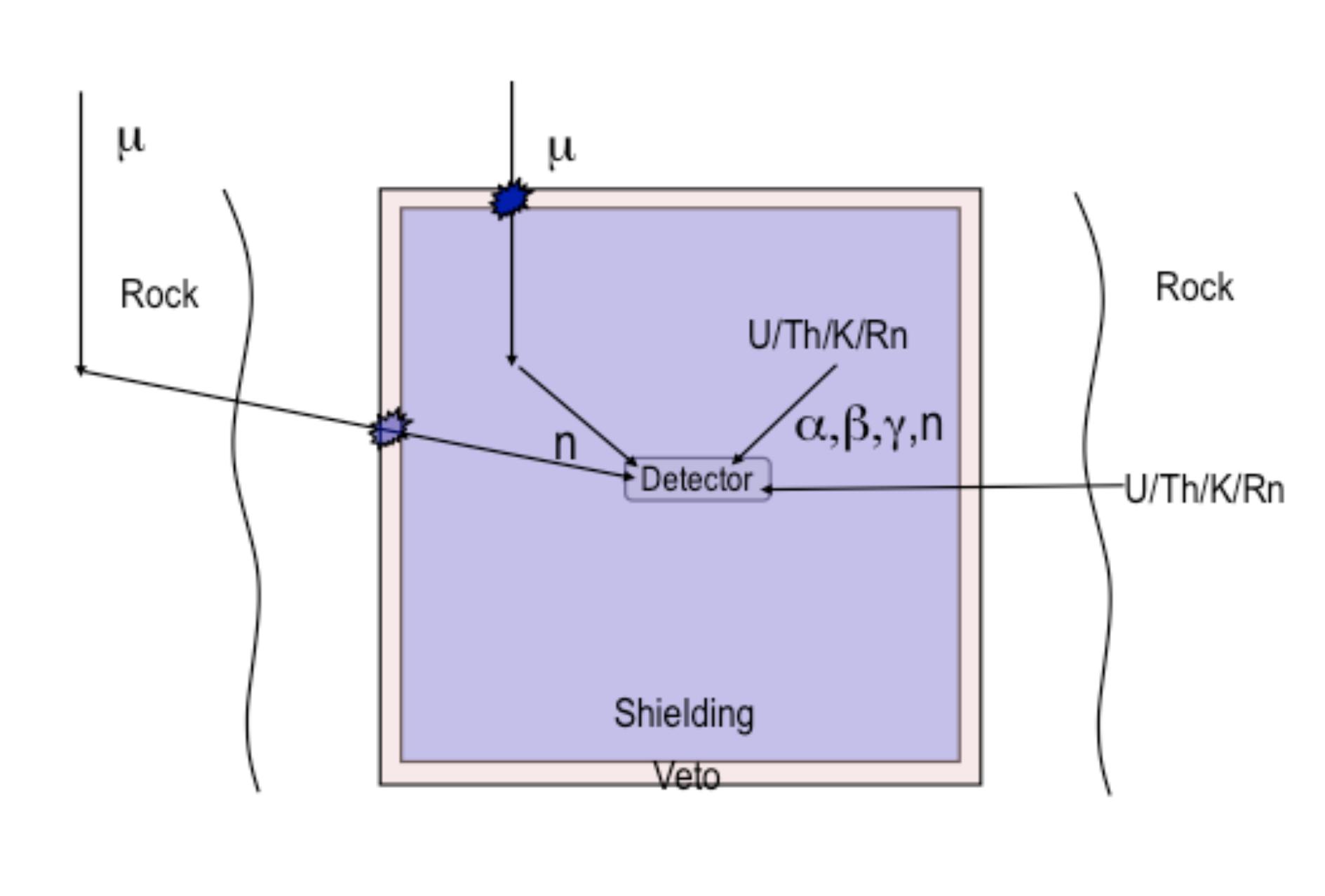}
\caption{\label{fig:background}Background sources and shielding in a typical direct detection experiment}
\end{center}
\end{figure}

Direct detection experiments must be located in deep underground laboratories, to avoid cosmic ray interactions that produce energetic neutrons that could mimic WIMPs.  The experiments must also shield the detectors from the decay products of radioactivity in the environment and in the materials of the experiment itself.  This is especially important for neutrons resulting from fissions or ($\alpha$,$n$) reactions, since a single scatter from a neutron produces a nuclear recoil that is indistinguishable from that produced by a WIMP.  In most cases, the experiments define an ultra-pure active ``fiducial'' volume that is shielded from radioactive decay products produced by impurities in the materials surrounding the detection material.  Fig.~\ref{fig:background}
shows a schematic of the background sources and mitigation strategies.

As dark matter experiments increase their exposure, they become sensitive to neutrinos from various sources. The flux from solar and atmospheric neutrinos forms a fundamental background to dark matter direct detection. For dark matter below 10~GeV/$c^{2}$, the solar neutrino spectrum is the largest background. The recoil energy spectrum from Coherent Neutrino Scattering on a dark matter experiment due to solar neutrinos from the $^{8}$B reaction is almost indistinguishable from a canonical WIMP with a mass of around 6~GeV/$c^{2}$ and a WIMP-nucleon cross section of $5 \times 10^{-45}~\mathrm{cm}^{2}$. Above 20~GeV/$c^{2}$, atmospheric neutrinos become the limiting background, with recoil spectra very similar to WIMPs with a mass of 100~GeV/$c^{2}$ and a WIMP-nucleon cross section of $2 \times 10^{-49}~\mathrm{cm}^{2}$. As can be seen in Figure~\ref{fig:SI-overview}, CNS from solar and atmospheric neutrinos form a fundamental lower bound on the cross section for background-free WIMP detection \cite{2013arXiv1307.5458B}. Next generation experiments will have sensitivity within an order of magnitude of the neutrino signal for most of the mass range, and will actually detect the $^{8}$B solar neutrino signal.

\begin{figure}[h!]%Figure 12
\begin{center}
\includegraphics[width=0.6\hsize]{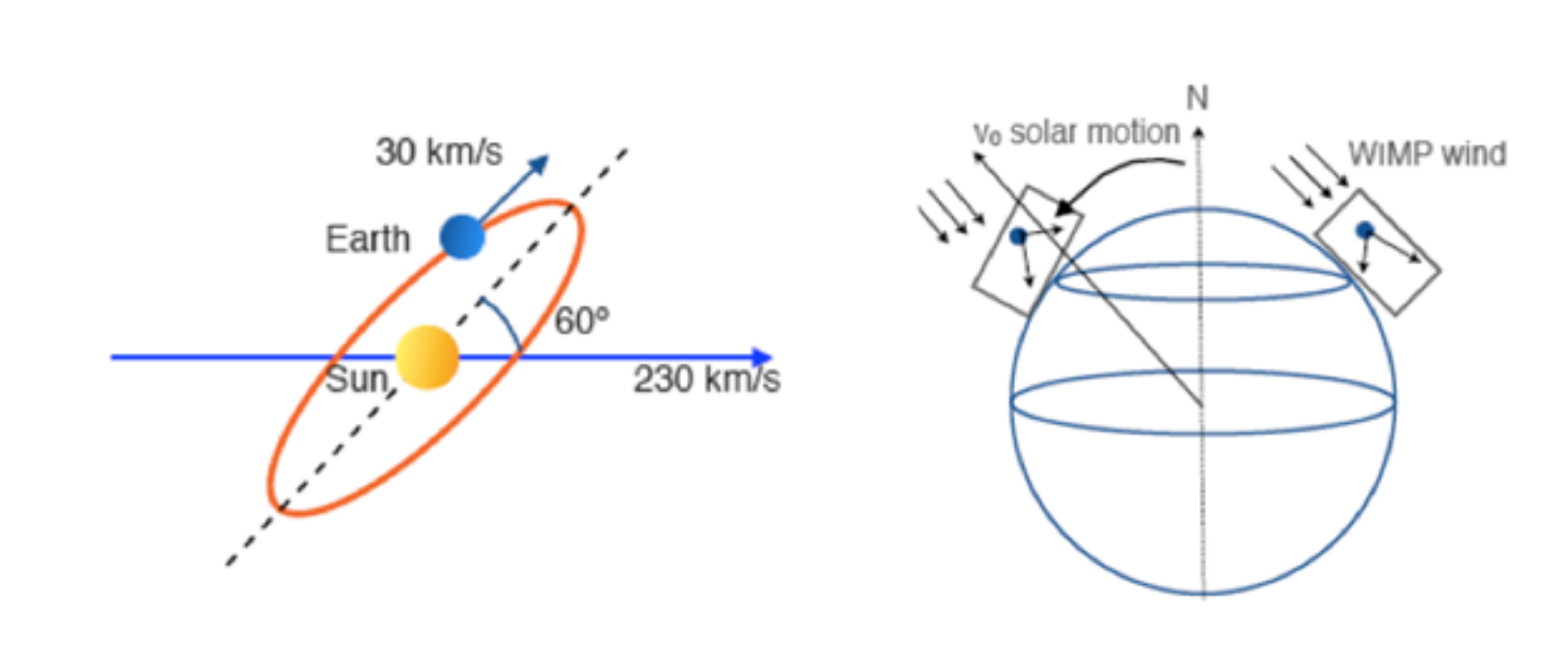}
\caption{\label{fig:direction}Schematic of the possible sources of annual modulation (left) and daily modulation (right) effects if WIMPs are detected in direct detection experiments}
\end{center}
\end{figure}

Finally, another method to deal with backgrounds is to exploit the fact that the Earth is moving through the dark matter that surrounds our galaxy, yielding a ``WIMP wind'' that appears to come from the constellation Cygnus.  This should, in principle, create a small ``annual modulation'' in the detected WIMP rates, as well as a somewhat larger daily modulation, as shown in Fig.~\ref{fig:direction}.
However, if such effects were detected in an experiment, there would still have to be a convincing demonstration that there are no such modulations in background sources.

%%%%%%%%%%%%%%%%%%%%%%%%%%%%%%%%%%%%%%%%%%%%%%%%%%%%%%%%%%%
\section{ Direction Detection Technologies}\label{sec:cf1:ddt}

The first direct detection experiments date back to the 1980s, but the modern era really began with the first experiments to achieve event-by-event active discrimination against backgrounds in the late 1990s.  Since that time, there has been an explosion in the number of technologies used for direct detection searches, and an exponential improvement in experimental sensitivity, with a doubling time of approximately 18~months.  In the remainder of this section, we survey the direct detection technologies, as well as current and future experiments that use them.  The information used to produce this summary was solicited from all active direct detection experiments as part of the Snowmass process.  Further details not contained in this document can be found in the set of white papers on WIMP Direct Dark Matter Detectors prepared for the Snowmass process.

%%%%%%%%%%%%%%%%%%%%%%%%%%%%%%%%%%%%%%%%%%%%%%%%%%%%%%%%%%%
\subsection{Cryogenic Solid State}

The category of cryogenic solid-state detectors for dark matter research includes ionization radiation spectrometers and bolometric detectors using dual-signal readout techniques.

The first published limits on searches for direct detection of cosmologically aged particles gravitationally bound in the Milk Way's dark matter halo came from high
purity germanium (HPGe) spectrometers operated near liquid nitrogen temperature ($\sim$90\,K).  Current research experiments (e.g. CoGeNT, TEXONO and MALBEK \cite{MALBEK}) still employ these techniques that benefit from HPGe detector's advantages of excellent energy resolution ({\it e.g.}, FWHM is $\sim$3\% at 10\,keV), low energy thresholds ($\sim$0.5\,keV$_e$, $\sim$2.0\,keV$_{\rm r}$), and commercial availability \cite{Aalseth:2012if, TEXONO}.  The dominant backgrounds for these types of dark matter detectors are from external $\gamma$-ray radiation due to naturally occurring radioactivity in the materials used in the experimental apparatus.  Careful choice of materials and control of fabrication process have led to reduced background levels in newer detectors.  In addition, point contact detectors with reduced noise make operation possible with lower energy thresholds.  As HPGe ionization spectrometers only measure ionization, electron recoils from gammas and betas are indistinguishable from the nuclear recoils that can be produced by either neutrons or WIMPs. Proposed future experiments using this technique include C-4 and CDEX \cite{C-4,CDEX}.

\begin{figure}[h!]%Figure 13
\begin{center}
\includegraphics[width=0.6\hsize]{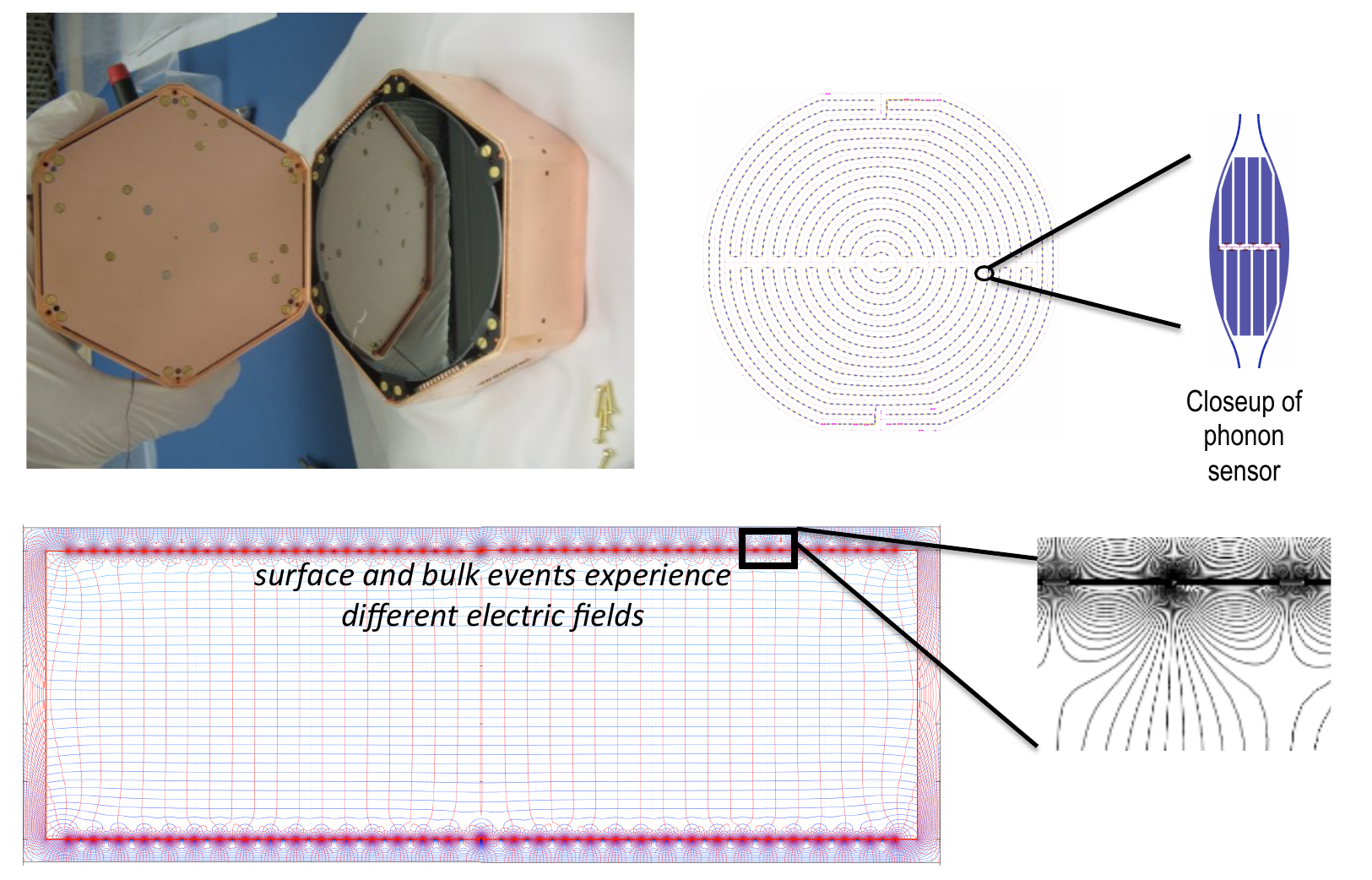}
\caption{\label{fig:cams}Schematic of the CDMS cryogenic solid state detectors.}
\end{center}
\end{figure}

Cryogenic solid-state detectors (e.g. Fig.~\ref{fig:cams})
operated at temperatures $<$100\,mK were developed to distinguish the dominant electron recoil backgrounds from nuclear recoils on an event-by-event basis.  This active background discrimination opened up a whole new regime of WIMP sensitivity by promising ``background free'' exposures.  These detectors were pioneered by the CDMS collaboration, operating germanium and silicon detectors initially at a shallow site at Stanford University and then at the Soudan Underground Laboratory.  European groups further developed the technology with germanium (EDELWEISS) and calcium tungstate (CRESST) \cite{Angloher:2011uu,Armengaud:2011cy,armengaud}.  Generally speaking, the current bolometric detectors use the ratio of energy measured in two different detector readout channels to provide the electron-recoil to nuclear-recoil discrimination.  These two signal readout techniques are phonons (heat) and ionization for SuperCDMS and EDELWEISS, and phonons and scintillation light for CRESST.  These dual-signal readout techniques have made it possible to reject electron-recoil backgrounds by many orders of magnitude.  Neutron backgrounds are mitigated by choice of underground location and shielding.  Surface beta decays, which were the dominant background in earlier versions of these experiments, are strongly suppressed in SuperCDMS iZIP and EDELWEISS-II germanium detectors by interleaving ionization and phonon sensors.  Radiogenic neutrons are expected to be the dominant background in the next generation of these experiments, as with most other technologies.

The cryogenic solid-state experiments have recently demonstrated sensitivity to low-mass WIMPs comparable to that achieved by HPGe experiments, using a combination of reduced backgrounds, improved low energy triggering and better analysis algorithms.  The recent interpretation of excess events in a CDMS silicon detector exposure as a low-mass WIMP signal shows the value of having a second target material (Silicon) with better sensitivity to low-mass WIMPs.  CRESST achieves similar low-mass performance with CaWO$_4$ crystals.

The SuperCDMS collaboration has proposed a new experiment with a payload of 200--400\,kg of Germanium and Silicon as part of the G2 process in the US, to be operational in 2017.  The sensitivity for low mass WIMPS is projected to be competitive with experiments having larger target masses, primarily because of the background discrimination and low energy threshold possible with these detectors. EDELWEISS and CRESST have proposed EURCA, a ton-scale Germanium experiment on a slightly later timescale.

%%%%%%%%%%%%%%%%%%%%%%%%%%%%%%%%%%%%%%%%%%%%%%%%%%%%%%%%%%%
\subsection{Liquid Xenon}

Liquid Xenon (LXe) is an excellent target for WIMP direct detection.  It can be highly purified, is chemically inert, and is radiopure.  The large atomic mass leads to a sizable spin-independent coherent WIMP-nuclear cross section, and the high density and atomic number mean that gamma rays have a very short interaction length.  Thus a large LXe detector is ``self shielding'', consisting of an inner fiducial region with a very low background rate, surrounded by an outer region that scatters gamma rays to energies outside of the region of interest.  Neutrons can be detected by their tendency to multiply scatter in the dense liquid.  LXe scintillates brightly in the VUV range, and it is transparent at this wavelength, which makes it possible to observe a large volume detector effectively with PMTs at the edge.

When operated with no electric field, the electrons and Xe ions are allowed to recombine, yielding scintillation light.  The primary scintillation signal (S1) is measured with PMTs immersed in the liquid.  The scintillation response to nuclear recoils is quenched with respect to equal energy electron recoils.  The XMASS experiment \cite{XMASS} is operating in this mode with an 835\,kg LXe detector.  They plan to use an inner, low background, fiducial volume consisting of 100\,kg of LXe with an analysis threshold of 5\,keV$_e$ for WIMP searches.

\begin{figure}[h!]%Figure 14
\begin{center}
\includegraphics[width=0.6\textwidth]{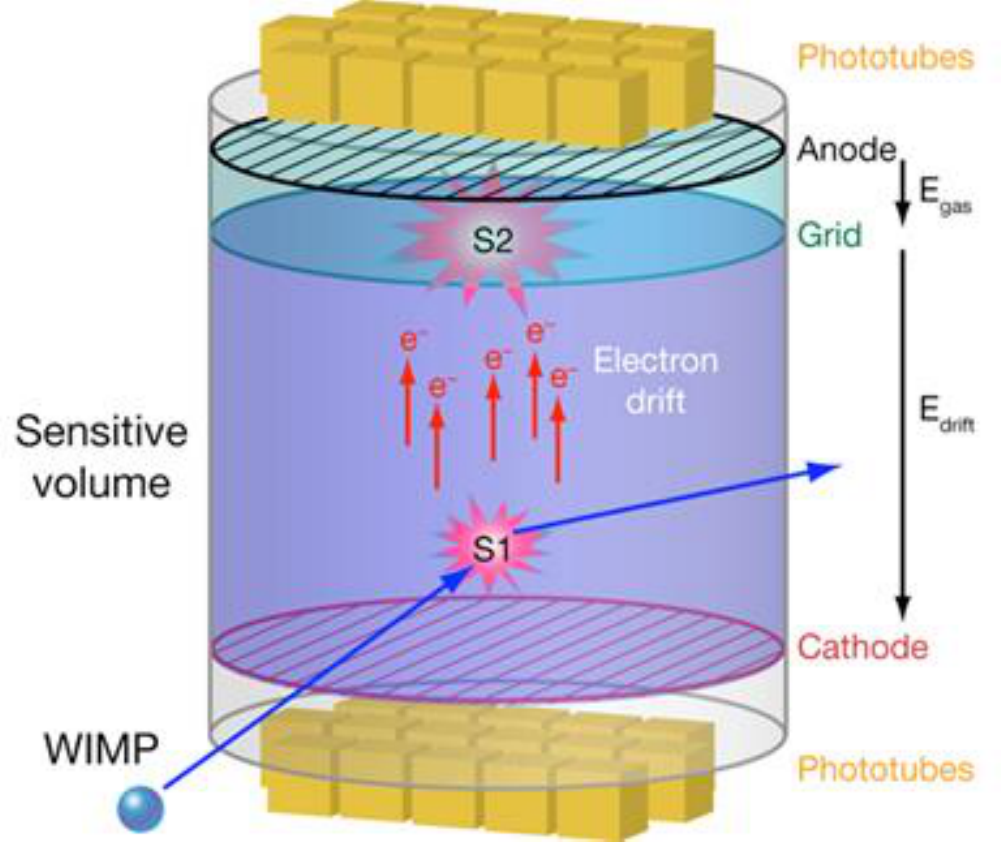}
\caption{\label{fig:xe-tpc}Schematic of a dual-phase LXe TPC.}
\end{center}
\end{figure}

An applied electric field (0.05-4\,kV/cm) causes the ionized electrons in LXe to drift to the surface of the liquid where they are then extracted (using a higher electric field of $\sim$10\,kV/cm) and measured via the scintillation light (S2) they produce in the Xe gas.  In noble liquids, xenon as well as argon, the greater the electric field the more charge is extracted at the expense of light production, because fewer ionized electrons are able to recombine, and therefore the presence of an electric field may be responsible for a fractional loss of light from nuclear recoils.  On the other hand, such a configuration, referred to as ``dual-phase TPC'' (see Fig.~\ref{fig:xe-tpc}),
allows the determination of the interaction position in three dimensions, making it possible to take advantage of the lower background region shielded from external backgrounds.  It also provides the ability to discriminate between electron and nuclear recoils based on the ratio of the prompt (S1) and delayed (S2) scintillation signals.  Event by event discrimination of better than 99.5\% has been demonstrated down to $\sim$5\,keV$_{\rm r}$.  Thresholds of 2\,keV$_{\rm r}$ may be possible using only the S2 signal, and giving up on discrimination of $\beta/\gamma$ events in favor of nuclear recoils \cite{Angle:2011th, Angle:2011th-errata}.

The two-phase LXe approach was pioneered by the Zeplin-II, Zeplin-III and XENON10 experiments \cite{Angle:2011th, ZEPLIN-II, ZEPLIN-III}.  The XENON100 experiment, which is operated by a European-US collaboration in the Gran Sasso Laboratory (LNGS), leads the field in spin-independent WIMP scattering sensitivity.  XENON1T \cite{XENON1T}, designed to operate with a 1-ton fiducial mass, has a target sensitivity of 3$\times$10$^{-47}$\,cm$^2$ after a three-year run starting in 2015.  The LUX experiment is operated by a US-European collaboration in the Sanford Underground Research Facility (SURF).  LUX \cite{LUX} is currently running with 300\,kg of active target at Sanford Lab with the expectation for first results by the end of~2013.  The collaboration is preparing a conceptual design of LZ (6,000\,kg fiducial) as part of the G2 process run by the DOE.  The corresponding sensitivity for LZ is 3$\times$10$^{-48}$\,cm$^2$ in a 3-year run, with a projected start date of 2017.  PandaX-II, planned by a Chinese-US collaboration, will have a fiducial mass of 1,400\,kg installed in the China Jin-Ping Underground Laboratory (CJPL).  The experiment could start operation around 2016 and reach a sensitivity of 2$\times$10$^{-46}$\,cm in a 2-year run \cite{PandaX}.

%%%%%%%%%%%%%%%%%%%%%%%%%%%%%%%%%%%%%%%%%%%%%%%%%%%%%%%%%%%
\subsection{Liquid Argon}

Liquid argon (LAr) experiments employ the same basic design, but replace the LXe with LAr.  As with LXe, single-phase LAr detectors rely upon the collection of the scintillation light induced in argon by energy depositions.  Scintillation light is in the hard UV, requiring a wavelength shifter for use with standard PMTs.  The total amount of scintillation light is essentially proportional to the energy deposited, but the timing structure of the scintillation pulse depends upon the nature of the event.  The ratio of the amount of light stemming from long-lived triplet states to the amount coming from short-lived singlet states is different for nuclear versus electron recoils, allowing for excellent pulse shape discrimination (PSD) for sufficiently large energy depositions.  The relatively large energy threshold for discrimination in argon limits sensitivity to very low WIMP masses.  Argon does have the drawback of an intrinsic radioactive background from $^{39}$Ar, a naturally-occurring radioisotope in atmospheric argon.  The excellent pulse shape discrimination makes this a manageable problem for current detectors, and the $^{39}$Ar actually provides a useful calibration source.  Next-generation experiments may use argon recovered from underground sources \cite{depleted-Ar}, pioneered by DarkSide-50, that has been shown to be substantially depleted in $^{39}$Ar.  This would reduce the otherwise high event rate and possible misidentification of electron recoils as nuclear recoils that such experiments may suffer with atmospheric argon.

Single-phase LAr detectors are typically spherical chambers, for maximal light collection efficiency, although larger detectors may be cylindrical for cost reasons.  In addition to depending on PSD, the energy threshold of the detector is partially determined by how many photons are necessary to reliably reconstruct the position of an event, for the purpose of distinguishing between events in the bulk liquid and those near the vessel walls.  Backgrounds that depend on the position reconstruction (like $\alpha$ activity on inner detector surfaces) should become relatively less important for larger detectors.

Dual-phase argon detectors are very similar in design to their LXe counterparts, with an electric field used to extract the ionization released in a particle interaction.  Dual-phase operation typically requires the use of underground argon depleted in $^{39}$Ar even at the first generation stage.  The energy threshold is not necessarily lower in dual-phase detectors, which can use the drifted charge for the position reconstruction instead of the light: two-phase argon TPCs have two handles on discriminating against electron recoil backgrounds: PSD as in single-phase detectors, and the ratio of charge to light, which is different for nuclear recoils and electron recoils.

There are two experiments currently deploying the single-phase argon technology, MiniCLEAN and DEAP.  The MiniCLEAN detector \cite{MiniCLEAN-a, MiniCLEAN-b}at SNOLAB will feature 500\,kg of argon target with 150\,kg of that intended for the fiducial volume.  The collaboration hopes to fill the detector by the end of 2013 and run with a threshold of 50\,keV$_{\rm r}$, although this will depend on the light yield and the observed discrimination.  It is likely that MiniCLEAN will be background-limited by neutrons from the photomultiplier tubes.  One feature of MiniCLEAN is the intention to use an $^{39}$Ar source to demonstrate the power of the PSD for electron recoil rejection.  A second is the potential use of liquid neon as a target.

Based in Canada, and with UK participation, the DEAP collaboration is currently completing installation of DEAP-3600 at SNOLAB, which will contain a target mass of 3,600\,kg (1,000\,kg fiducial) and has been designed for 0.2~background events per ton-year \cite{DEAP}.  The target date for first cooldown is January~2014.  With excellent PSD, the main background that concerns DEAP is neutrons from their 255~PMTs.  To mitigate these neutrons, they have included 50\,cm acrylic light guides between the target volume and the PMTs as a shield.  They hope to run at a threshold of 60\,keV$_{\rm r}$, reaching an ultimate sensitivity of 10$^{-46}$\,cm$^2$ after a 3-years run.  Use of underground argon depleted in $^{39}$Ar may allow for reduction in threshold.

\begin{figure}[h!]%Figure 15
\begin{center}
\includegraphics[width=0.6\textwidth]{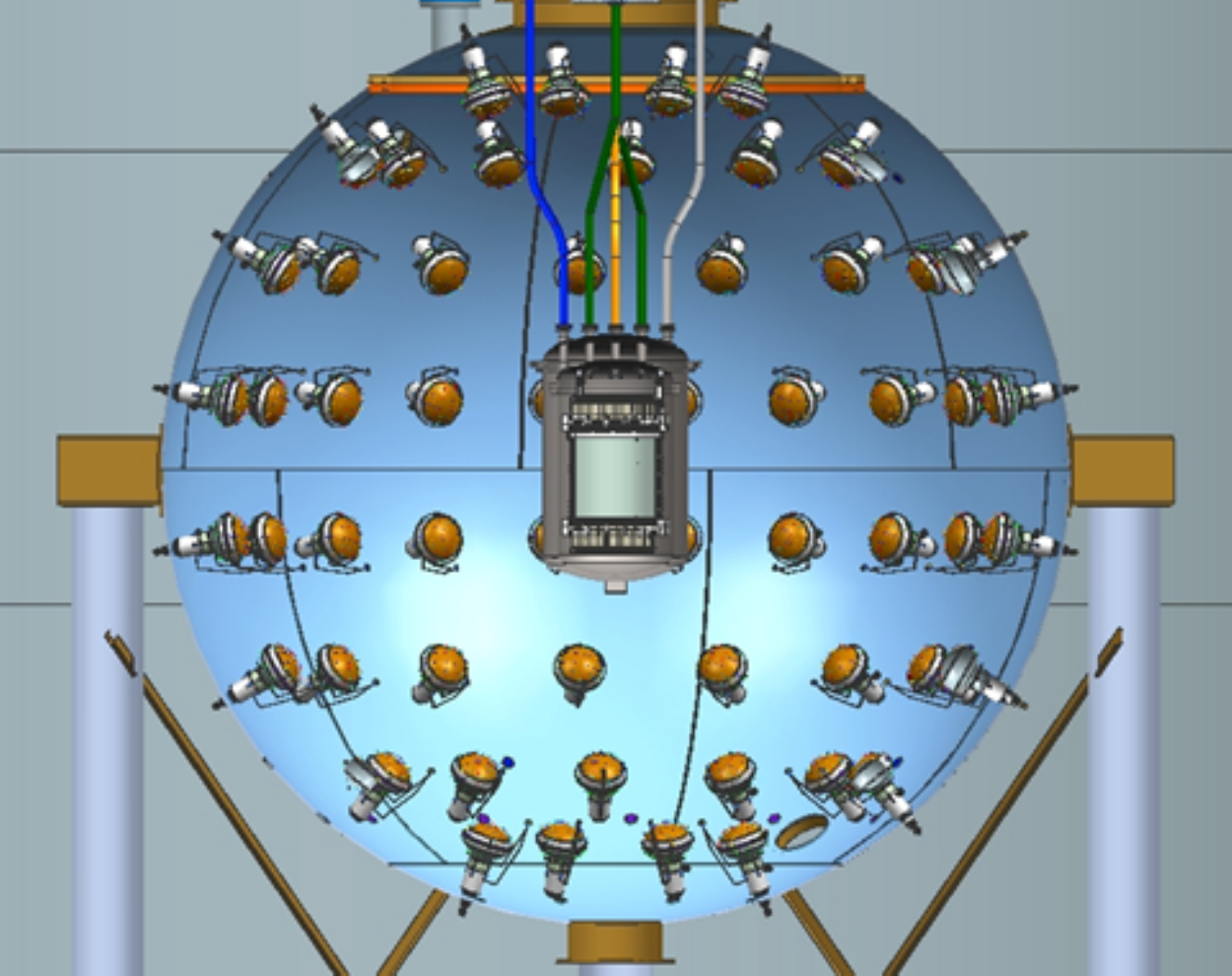}
\caption{\label{fig:darkside}Schematic of the Darkside-50 liquid argon dual-phase TPC, nestled within a 4$\pi$, 30-tons liquid scintillator neutron veto counter.  The liquid scintillator neutron veto counter is hosted inside a 1,000-tons active water \v{C}erenkov muon veto.}
\end{center}
\end{figure}

The major US-led collaboration deploying dual-phase liquid argon is DarkSide \cite{DarkSide}.  They are currently commissioning DarkSide-50 (Fig.~\ref{fig:darkside})
at LNGS, featuring a 50,kg fiducial mass and an expected sensitivity of 2$\times$10$^{-45}$\,cm$^2$ in a 3-years run, with an energy threshold of 25\,keV$_{\rm r}$.  The Collaboration has identified sources of underground argon with less than 0.65\% of the atmospheric concentration of $^{39}$Ar, which will be exploited as the target for the DarkSide-50 dark matter search.  They are also taking advantage of advances in PMT photocathode technology to use higher quantum efficiency PMTs and developing lower-background PMTs for a second generation experiment.  DarkSide-50 will be the first dark matter experiment with a 4$\pi$ active neutron veto using boron-loaded liquid scintillator, housed inside a 1,000 tons water \v{C}erenkov muon veto, both sized for a second generation detector: this vetoing technique is estimated to reject neutron-induced events in the active argon from trace radioactivity in detector components by a factor of about 100, and allows detailed studies of residual neutron backgrounds.  The DarkSide collaboration is also preparing for a second generation experiment, DarkSide-G2, featuring 5,000\,kg (2,800\,kg fiducial) of underground argon depleted in $^{39}$Ar, with a reach of 2$\times$10$^{-47}$\,cm$^2$ in a 5-years run.

ArDM is a European collaboration deploying a 1,000\,kg dual-phase argon TPC in the Canfranc laboratory. The detector was commissioned at CERN and is now being moved underground \cite{ArDM}.

%%%%%%%%%%%%%%%%%%%%%%%%%%%%%%%%%%%%%%%%%%%%%%%%%%%%%%%%%%%
\subsection{Superheated Liquid Detectors}

Threshold detectors are a class of experiments currently encompassing bubble chambers and superheated droplet detectors.  These detectors are weakly superheated in the sense that the local energy deposition from a nuclear recoil will induce bubble nucleation.  However, $\beta/\gamma$ events, which lose their energy over a comparatively longer range, will not induce bubbles.  Additional advantages of these detectors include excellent spin-dependent sensitivity, and the ability to operate with interchangeable target liquids.  Discrimination against $\alpha$'s, which can deposit sufficient energy density to produce bubbles, has been established through the use of acoustic signals.  Threshold experiments are continuing to refine target material choices with regard to the chemical stability and compatibility of the target fluids.  This technology is still under development for scaling single modules up to moderate target mass (100-1000\,kg).  One of the drawbacks of this technology is the lack of an energy measurement on a per event basis.  These detectors must instead perform threshold scans by varying the operational temperature and pressure, in order to produce an energy spectrum.  This arguably makes the investigation of unexpected backgrounds a more lengthy process.

Two collaborations, PICASSO and SIMPLE, utilize superheated droplet detectors \cite{Archambault:2012pm,Girard:2012zz}.  SIMPLE operates at LSBB in Southern France with 0.2\,kg active volume of superheated C$_2$ClF$_5$ (freon) droplets in a gel matrix.n  SIMPLE has achieved a threshold of 8\,keV$_{\rm r}$, and published a spin-dependent limit at a WIMP mass of 35\,GeV of 5.3$\times$10$^{-39}$\,cm$^2$ and a spin-independent limit of 4.8$\times$10$^{-42}$\,cm$^2$.  The dominant backgrounds are alphas from environmental radon, followed by neutrons created via U and Th initiated ($\alpha$,$n$) reactions in the detector materials.  SIMPLE is planning to operate a second experiment soon, which will replace their superheated droplet detectors with a more conventional bubble chamber.  They will have a target volume of 1-2\,kg and will use C$_3$F$_8$ CF$_3$Br, in addition to C$_2$ClF$_5$.

PICASSO has 32 detector modules, which represent 2.7\,kg of the target material C$_4$F$_{10}$, operating at SNOLAB.  PICASSO pioneered the use of acoustic sensors to reject $\alpha$'s.  This technique has subsequently been adopted by other threshold experiments.  The acoustic discrimination has enabled them to achieve good sensitivity to spin-dependent WIMPs.  The collaboration reports that the most recent runs of PICASSO are becoming limited by $\alpha$'s in the gel, where the acoustic discrimination does not work as effectively.  The collaboration believes that scaling up from the present configuration requires moving to the bubble chamber technology.  They are actively developing a prototype chamber that is self-regulating (requires no external pressurization).

\begin{figure}[h!]%Figure 16
\begin{center}
\includegraphics[width=\textwidth]{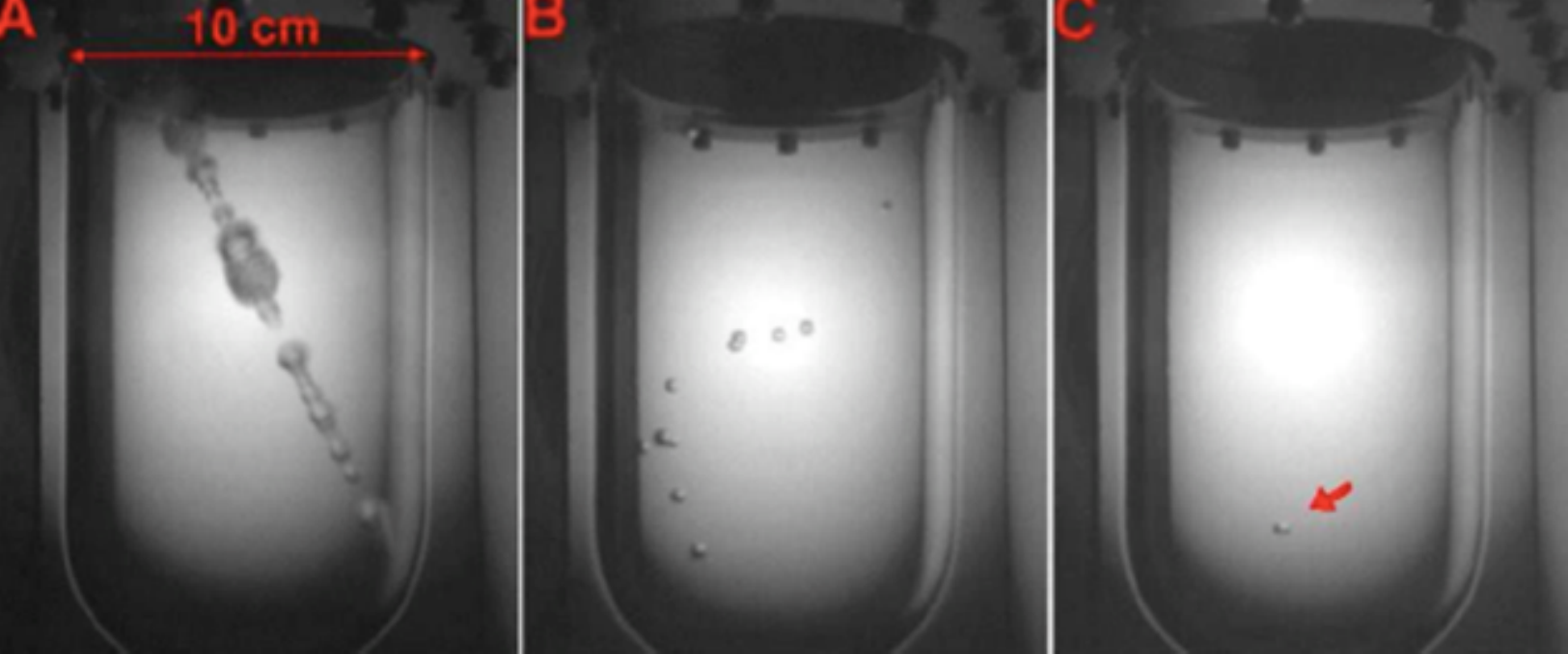}
\caption{\label{fig:coupp}Photograph of bubbles from a muon (left), neutron multiple scatter (center) and neutron single scatter (right) in a COUPP bubble chamber.}
\end{center}
\end{figure}

In contrast to the droplet detectors, the COUPP experiment \cite{COUPP} uconsists of a monolithic body of superheated fluid contained within a single pressure-controlled vessel (a bubble chamber).  Bubbles are recorded by cameras, triggered by a change in the appearance of the chamber.  Like the droplet detectors, COUPP also records the acoustic signal of the nucleated bubbles and has successfully utilized this information to reject alphas to better than 99.3\%.  Because of the images, COUPP also has excellent position reconstruction, and can easily identify multiple scattering events characteristic of neutrons (see Fig.~\ref{fig:coupp}).
To date, published WIMP searches have been performed with a single chamber filled with up to 4\,kg of CF$_3$I.  In these searches, COUPP has been most competitive in spin-dependent searches, like its droplet detector counterparts.  The most recent results were limited by neutron backgrounds, thus the collaboration is focusing on improving the radiopurity of the experimental setup.  Currently they are also commissioning a 60\,kg chamber, which is filled with CF$_3$I and is installed at SNOLAB.

The COUPP and PICASSO groups have agreed to merge their efforts towards a single ton-scale experiment, known as PICO.  They are now working jointly on construction and operation of a 4\,kg bubble chamber, which will be filled with the target material C$_3$F$_8$.  This chamber will combine the bubble-chamber operational experiences of the COUPP experiment and fluid handling expertise from PICASSO.  The designated fluid possesses better spin-dependent sensitivity than CF$_3$I, due to the high concentration of fluorine.  The collaborations have demonstrated thresholds down to a few keV with C$_3$F$_8$.  Thus, this joint effort is expected to have significant sensitivity to WIMPs over a broad mass range for both spin-dependent and spin-independent scattering.

%%%%%%%%%%%%%%%%%%%%%%%%%%%%%%%%%%%%%%%%%%%%%%%%%%%%%%%%%%%
\subsection{Scintillating Crystal Detectors}

Scintillating inorganic crystals, such as NaI and CsI, are the basis of one important class of dark matter experiments.  These crystals are high efficiency scintillators producing many optical photons even for quenched WIMP-induced nuclear recoils.  They benefit from an attractive combination of low target costs and existing well-understood technologies for light collection and detection over wide areas.  Scintillation pulse shape discrimination provides modest levels of EM background discrimination (90--99\%) in the currently deployed NaI(Tl) and CsI(Tl) experiments.  Additional background rejection power comes from detection of coincidences between elements of multi-crystal arrays and from the exploitation of the predicted annual modulation properties of the WIMP signal.  The main experimental challenges for these types of detectors are obtaining a low overall background rate and, more importantly, maintaining detector stability over a period of years.

The DAMA/LIBRA \cite{Bernabei:2010mq} collaboration began a search for dark matter with an array of NaI crystals in 1995 and has been operating the current DAMA/LIBRA setup of 250\,kg of NaI since 2003.  They have observed an annual modulation in their data at greater than 9$\sigma$ significance, with a phase consistent with that expected from galactic dark matter interactions.  If the signal is interpreted as evidence of spin independent (or spin dependent) scattering of WIMP dark matter, it is in strong tension with results from many other searches.  The collaboration maintains that this signal represents a model-independent observation of dark matter interactions.  To date, no successful experimental or theoretical explanation for the annual modulation signal has achieved consensus in the community.

Several other collaborations are now attempting to test the DAMA/LIBRA signal using crystal detectors.  The main challenge facing these experiments is to reproduce, or improve on, the ultra-low background levels achieved in the DAMA/LIBRA NaI.  The ANAiS collaboration aims to build a 250 kg ultrapure NaI (T1) array at the Canfranc Underground Laboratory \cite{ANAiS}. The KIMS experiment \cite{Kim:2012rza} is currently operating 100\,kg of CsI, with future plans to deploy two detectors, a lower-background array of NaI (KIMS-NaI) and cryogenic CaMoO$_4$ bolometers (AMORE-DARK).  The CINDMS collaboration plans to deploy 100\,kg of CsI(Na).  They expect to achieve a factor of 10$^9$ reduction for $\beta/\gamma$ background reduction by exploiting the dramatically better pulse shape discrimination properties of CsI(Na) compared to CsI(Tl) and NaI(Tl). The PICO-LON collaboration is developing a 100 kg NaI (T1) detector with thin layers, allowing gamma ray background to be reduced by anti-coincidence \cite{PICO-LON}.  A Princeton-based group is developing NaI(Tl) with better radiopurity than the DAMA/LIBRA material.  Finally, DM-Ice \cite{DM-Ice} is currently running 17\,kg of NaI at the bottom of two phototube strings in the IceCube array at the South Pole, with plans to upgrade to 250\,kg by the end of 2015.  DM-Ice's location in the southern hemisphere will directly test whether the modulation observed in DAMA/LIBRA correlates with seasonal effects (out of phase in southern versus northern hemispheres) or with the earth's velocity relative to the galaxy (in phase in both southern and northern hemispheres).

\begin{figure}[h!]%Figure 17
\begin{center}
\includegraphics[width=0.6\textwidth]{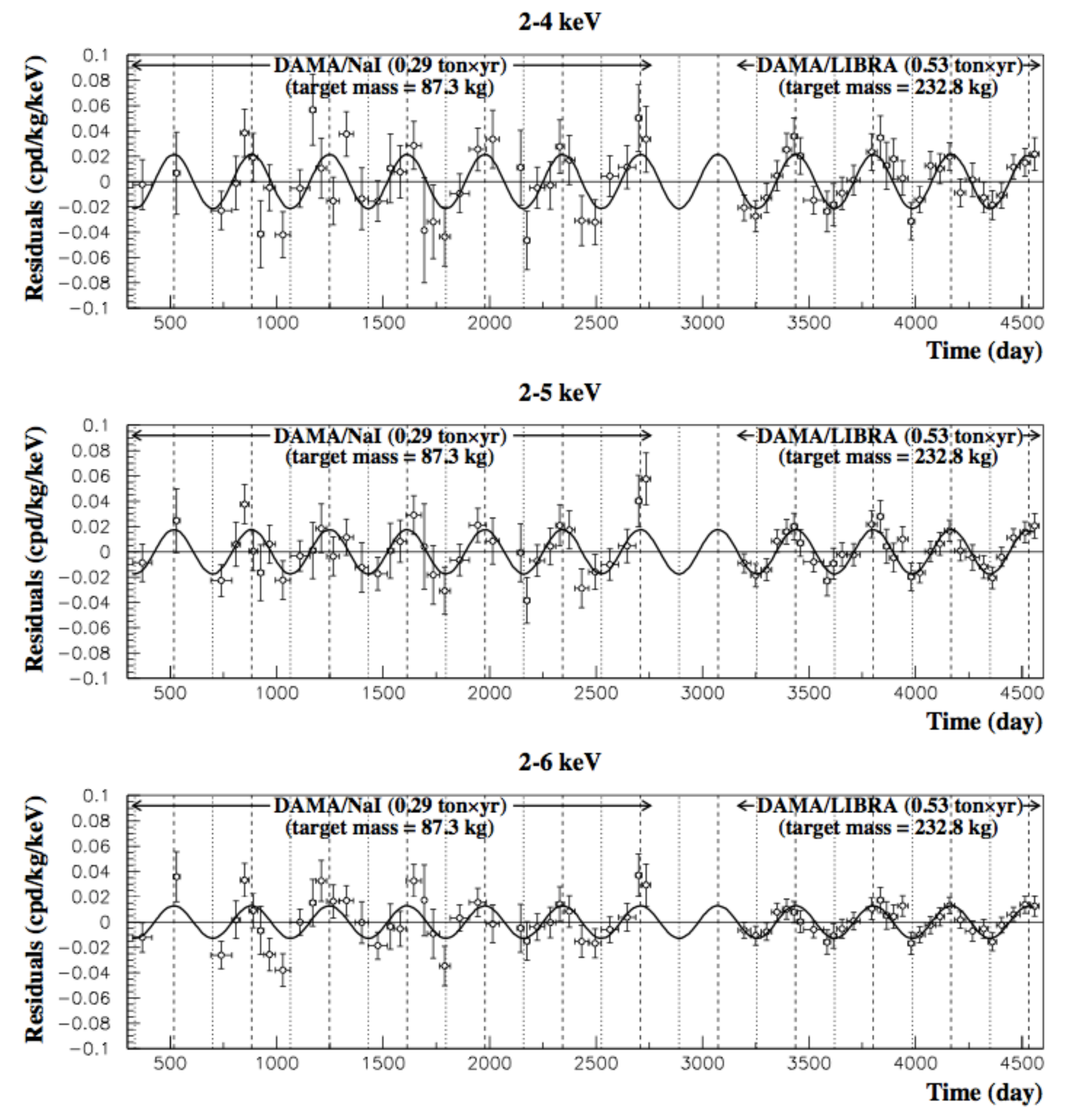}
\caption{\label{fig:dama}Residual rate in counts/kg/keV/d from the DAMA/LIBRA experiment as a function of time, showing the annual modulation signal that the collaboration interprets as evidence for detection of dark matter.}
\end{center}
\end{figure}

It is very important for more progress to be made in resolving the DAMA/LIBRA claims of a dark matter signal, and several independent measurements in different locations will provide very important observations that will significantly strengthen or weaken these claims.  Fig.~\ref{fig:dama}
shows the annual modulation signal from DAMA/LIBRA.

%%%%%%%%%%%%%%%%%%%%%%%%%%%%%%%%%%%%%%%%%%%%%%%%%%%%%%%%%%%
\subsection{Directional Detectors}

The motion of earth through the galaxy creates a ``WIMP wind'' emanating from the direction of the constellation Cygnus \cite{Drukier:1986tm}.  Since backgrounds other than those from neutrino-induced events are expected to be isotropic, detection of a signal with a preferred direction could provide a powerful additional discriminant against backgrounds.  Directional detectors attempt to exploit this effect by sensing the vector direction of nuclear recoils, and thereby inferring the direction of the WIMP that created them.

Thus far, most attempts at directional detection have focused on low-pressure gas time projection chambers (TPCs), in order to provide the 3-D track reconstruction and energy resolution needed to identify low energy nuclear recoils.  The primary challenge facing such detectors is deploying sufficient gas target mass to be sensitive to the expected low rate of WIMP interactions. With the typical gases used ({\it e.g.}. CF$_4$), at pressures of $\sim$50\,mbar, even a large (1\,m$^3$) TPC module provides only $\sim$100\,g of target mass, compared with the 100's of kg solid and liquid targets deployed by non-directional detectors.  Another significant challenge is pushing the detection technique down to low energy threshold.  Current thresholds range from are high ($\sim$50\,keV$_{\rm r}$), which limits the low-mass reach of these detectors.  Directional detectors also face the same low radioactivity materials requirements as with other techniques, particularly in the actual detection material where fiducialization of events is more difficult.  Finally, achieving sufficient angular resolution and, especially, distinguishing the vector direction of the recoiling nucleus are crucial to the background discrimination for directional detectors.

The field was pioneered by the DRIFT collaboration, with a series of detectors operating in the Boulby mine that have used negative ion drifting in CS$_2$ and CF$_4$ to minimize track diffusion and thus improve angular resolution \cite{DRIFT}.  The detection of drifting charges in DRIFT is accomplished with a low-background multi-wire proportional chamber (MWPC).  The currently-operating DRIFT-IId detector contains 139\,g of target gas and is limited by backgrounds from the MWPC wire grid.  A similarly-sized DRIFT-IIe detector will be deployed soon and is expected to have a substantial reduction in backgrounds.  NEWAGE \cite{NEWAGE} and MIMAC \cite{MIMAC} are low pressure TPC detectors with comparable dimensions, but using micro-PIC and Micromegas detectors, respectively, in place of the MWPC readout.  DM-TPC \cite{DMTPC} is currently running a TPC that uses CCDs to image the scintillation light from drifting charge that reaches an amplification region, thus reconstructing the vector direction of the nuclear recoil track. Finally, the D$^3$ collaboration \cite{D3} has demonstrated prototypes of a novel TPC using gas electron multiplier (GEM) readout that may allow full nuclear recoil track detection at substantially lower energies.

\begin{figure}[h!]%Figure 18
\begin{center}
\includegraphics[width=0.4\textwidth]{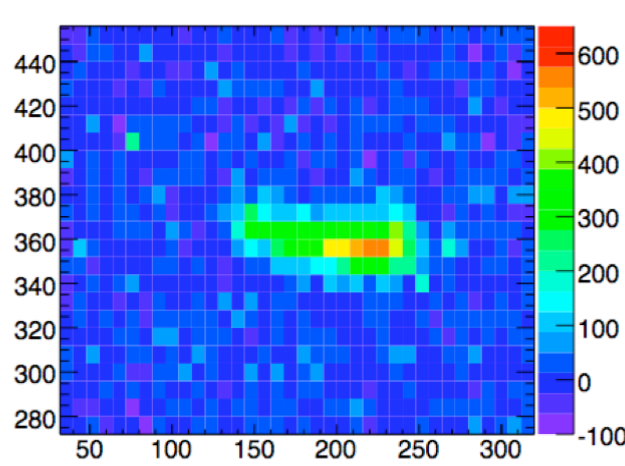}
\caption{\label{fig:dmtpc}Nuclear recoil track from DMTPC.}
\end{center}
\end{figure}

Fig.~\ref{fig:dmtpc}
shows a plot of directional detection of a nuclear recoil from the DMTPC experiment.

%%%%%%%%%%%%%%%%%%%%%%%%%%%%%%%%%%%%%%%%%%%%%%%%%%%%%%%%%%%
\subsection{New Ideas}

While ``traditional'' detectors are getting increasingly larger in size and lower in background, there has been no confirmed WIMP signal at the canonical supersymmetry mass scale ($\sim$100\,GeV), although there are several hints of low-mass WIMP signals.  There are new ideas to extend the search parameters for direct dark matter detection, both for high mass and low mass, as well as new ways to add direction information to the events.

Both DAMIC and Liquid $^4$He are promising ways to achieve better sensitivity to light WIMPs (1--10\,GeV).  DAMIC \cite{DAMIC} uses CCD chips as the target mass and records the ionization produced in Si.  DAMIC test runs at MINOS Near Hall set a limit of 2$\times$10$^{-38}$\,cm$^2$ at 1\,GeV and 2$\times$10$^{-39}$\,cm$^2$ at 10\,GeV.  An array with 10\,g was installed at SNOLAB in December 2012, and is currently operating with 40\,eV$_e$ threshold and has been in physics run since February~2013.  Technology exists that should lower the threshold to 4\,eV$_e$.  The collaboration submitted to SNOLab a proposal to increase the mass to 100\,g.

Since liquid $^4$He is a low-mass target, it has good kinematics for light WIMPs \cite{guo}.  Additionally, if signal is observed by other experiments in the community, a low-mass target provides one end of the target mass lever arm for determining WIMP mass and mean velocity.  Other advantages include multiple signal channels for characterizing and reducing background through discrimination, even at low WIMP masses where most technologies maintain only one signal channel.  In addition to the usual light (S1) and charge (S2) used in other noble gas detectors, the proponents are exploring the use of triplet excimers (S3), and possibly rotons and phonons.  The expected sensitivity with S1 and S2 readout, 0.1--100\,kg target mass (fiducial mass order 0.05--50\,kg), and analysis threshold $\sim$2--4\,keV$_{\rm r}$ is 10$^{-43}$\,cm$^2$ at 2\,GeV, 10$^{-44}$\,cm$^2$ at 5\,GeV, 10$^{-44}$\,cm$^2$ at 10\,GeV, and possibly better with rotons.

The NEXT/Osprey \cite{NEXT-Osprey} proponents are exploring the feasibility of using a xenon TPC to exploit atomic/molecular processes in xenon-TMA (trimethylamine) or TEA (triethylamine) to obtain directionality through columnar recombination.  The directionality signal might be contained in the ratio of Recombination/Ionization (R/I), with no track visualization required.  Although the prototype target mass is currently quite small, it appears possible to scale this technique to large gas target masses.

The use of nuclear emulsion \cite{nuclear-emulsion} as a WIMP target is being developed by groups in Nagoya, Japan, and Naples and LNGS, Italy.  It is currently in R\&D stage, studying backgrounds with an eye to minimizing detector sensitivity to gamma and beta backgrounds.  The goal is to operate a detector of size few tens of kg in 2~years.  They do not anticipate any difficulty in producing a kg-scale detector. Directionality would be achieved by looking for tracks produced by silver grains in emulsion and placing the detector on an equatorial telescope.

Another new idea is a DNA-based detector \cite{DNA}.  The approach is to have a field of single RNA strands with target material (currently gold) nearby.  When WIMPs knock off a nucleus in the target material, it recoils, cutting strands of DNA.  These are then sequenced with industrial techniques so the recoil track can be mapped.

%%%%%%%%%%%%%%%%%%%%%%%%%%%%%%%%%%%%%%%%%%%%%%%%%%%%%%%%%%%
\subsection{ Detector R\&D}

A continual program of detector R\&D has been a vital source of new ideas in direct detection and we strongly believe this must continue.  It is especially important to pursue directional detection techniques that show promise of lower energy thresholds, since these will be much easier to deploy in case a signal is seen in counting experiments.  Some of the ``new ideas'' discussed in the previous section may allow searches for WIMPS with mass $<$1\,GeV, which would be territory that is not currently accessible to conventional experiments.  Finally, areas of detector R\&D that are of interest for other areas of physics ({\it e.g.}, TES sensors for CMB experiments, as well as dark matter experiments) should also be supported.

%%%%%%%%%%%%%%%%%%%%%%%%%%%%%%%%%%%%%%%%%%%%%%%%%%%%%%%%%%%
\section{Energy Scale}

Direct detection experiments seek to identify dark matter via interactions with target nuclei contained within experimental instruments. The dark matter particle scatters off a target nucleus, imparting energy in the form of a nuclear recoil. In nearly all cases the energy of the nuclear recoil is converted to another form in the process of measurement, such as into thermal motion (heat), electronic ionization, scintillation light emission, and even sound. In all cases, knowing the energy partitioning from the nuclear recoil into these secondary modes is crucial for understanding the response of the detection instrument. The detector response is used in inferring the putative dark matter particle's mass and interaction cross section which are the principal measurement objectives for all direct detection dark matter experiments.

The energy partitioning from nuclear recoil energy to secondary modes is thus a necessary experimental determination. The canonical method is to use a mono-energetic neutron beam to create nuclear recoils in a detector calibration set-up. The recoil energy is selected by detecting the neutron scattering angle which a moveable secondary neutron-tagging detector. Other novel methods for creating and selecting nuclear recoils of known energy are feasible. For example, these methods are used to study the ionization "quenching factor" in solid state materials such as germanium and silicon, the scintillation efficiency $L_{eff}$ and charge yield $Q_{y}$ in liquid noble gases, and scintillation quenching in NaI and CsI detectors.

On-going R\&D to improve the understanding of the energy partitioning from nuclear recoil energy to secondary modes is necessary for improving the reconstruction of the inferred nuclear recoil energy spectrum and is particularly important for expanding sensitivity reach to lower energy nuclear recoils.  A firm understanding of how the nuclear recoil energy is transferred to the measured mode is necessary to ensure robust experimental science results.

To highlight the potential perils, it is worth to note, for example, that all limits obtained in xenon and argon TPC dark matter searches in the last decade were based on the expectation that any decrease in light yield due to the application of a drift field would be negligible ($<$5\%), based on a single precise measurement well above the energy range of interest for dark matter detection in xenon~\cite{aprile} and on a set of measurements of lower energy but also lower precision~\cite{manzur}.  The SCENE Collaboration, making use of monochromatic, low-energy, pulsed neutron beam, recently reported instead the observation in argon of a strong field-dependent quenching of scintillation light~\cite{scene} in the energy range ($\sim$10\,keV) crucial for xenon results.  If confirmed for xenon, this will require to revisit the set of exclusion plots produced insofar, with expected impact in the region of exclusion at low WIMP masses.

%%%%%%%%%%%%%%%%%%%%%%%%%%%%%%%%%%%%%%%%%%%%%%%%%%%%%%%%%%%
\section{Other Physics With Dark Matter Experiments}\label{sec:cf1:optcbdwdme}

%%%%%%%%%%%%%%%%%%%%%%%%%%%%%%%%%%%%%%%%%%%%%%%%%%%%%%%%%%%
\subsection{Axion Searches}

The Peccei-Quinn theory originally proposed the axion as part of the solution to the strong CP problem in QCD.  However, it has since been generalized to a class of particles that may also explain dark matter.  Several experiments designed for WIMP detection have set compelling limits on axions.  Typically, these experiments search for astrophysical sources of axions through the Primakov effect, whereby an axion is converted into a photon within the detector via coupling to the Coloumb field of an atom.  In the case of solar axion searches, the probability for scattering is enhanced by Bragg diffraction.  Knowledge of the crystal orientation allows one to look for this enhancement over backgrounds.

The most recent competitive axio-electric coupling limits from direct-detector experiments have been set by EDELWEISS, CDMS, CoGeNT and DAMA.  These limits are not as competitive as axion helioscopes at masses $<$1\,eV, but do provide a complementary technique to other axion searches.  Axion searches with next generation direct detection experiments will benefit from the increased mass and reduced backgrounds.

%%%%%%%%%%%%%%%%%%%%%%%%%%%%%%%%%%%%%%%%%%%%%%%%%%%%%%%%%%%
\subsection{Other non-WIMP dark matter searches}

In addition to axions, there are many classes of non-WIMP dark matter candidates that may be detected by experiments designed to search for WIMPs.  These include searches for dark photons, inelastic dark matter, lightly ionizing fractionally charged particles (LIPs) and may more.  Searches for these particles are distinct from the traditional WIMP search in that the WIMP interacts with the detector through a means other than elastic scattering off nuclei.  This may include, but is not limited to, ionization, elastic scattering off electrons, or excitation of the molecular system.  Often the resulting interaction is detected as an electron recoil.  One set of models, lumped under the description of ``sub-GeV dark matter''~\cite{essig} includes the ``WIMP-less'' scenario, axinos, and gravitinos.  XENON10 has published limits based on a search for particles that induce ionization of individual electrons and was able to set limits on particle interactions that produce one two or three electrons \cite{sorensen-etal}.  Ultra-low threshold detectors such as CDMS-lite are also expected to have excellent sensitivity to such models.  Searches for LIPs require identification of minimum ionizing particles that produce a track within the detector or detecting array.  In many cases these searches are complementary to searches at colliders, fixed target experiments or underground neutrino experiments.

%%%%%%%%%%%%%%%%%%%%%%%%%%%%%%%%%%%%%%%%%%%%%%%%%%%%%%%%%%%
\subsection{Searches for neutrinoless double beta decay}

In principle, the detectors utilized for direct dark matter searches may also be deployed to search for neutrinoless double beta $0\nu2\beta$ decay.  In both cases, the experiments require long, stable operation of detectors that are built from extremely radiopure materials, located deep underground to mitigate cosmogenic sources of background and have low noise installations.  At present, searches for dark matter and $0\nu2\beta$ are typically executed by separate groups.  While this may be partly historical, the needs of these two groups diverge in crucial ways on a few issues.  $0\nu2\beta$ searches often require detectors to be constructed with isotopically enriched materials.  This significantly increases the cost for the detecting material.  The excellent energy resolutions required to resolve peaks near the endpoint in $0\nu2\beta$ experiments also puts unnecessary and costly constraints on dark matter experiments.  Direct detection experiments must be performed at lower energies (typically a few to tens of keV), while the signal region for $0\nu2\beta$ is typically in the MeV range.  This results in different requirements on data acquisition and electronics.  Despite these differences, the communities have succeeded in working closely together in that much of the transferrable knowledge is actively shared.  For example, the communities share the same material screening facilities and jointly maintain databases that record the radiopurity of common materials utilized in the construction of these experiments.

The CoGeNT experiment, which utilizes a prototype design for the Majorana experiment, is an example of how both groups can benefit from improvements to the chosen detector technology.  In this case, the point contact design reduces noise and allows for a very low threshold for the dark matter experiment.  The CUORE experiment, now under construction, will be capable of conducting both $0\nu2\beta$ decay and dark matter searches using TeO$_2$ crystals.  Previous generations of this technology have demonstrated capability in both these areas.  While CUORE's dark matter searches have not been as competitive as the $0\nu2\beta$ decay results to date, CUORE continues to refine its ability to look for dark matter, by pushing its energy threshold lower.  Large Xe-based dark matter experiments may also be used to search for $0\nu2\beta$ decay.  Sensitivity is projected to be significant, even for a natural Xe target with $^{136}$Xe isotopic abundance of 8.9\%.  Self-shielding significantly reduces gamma-ray backgrounds, which are primarily due to photomultipliers and other detector materials.  Compared to current limits, the LZ experiment is projected to give an order of magnitude improved sensitivity to the half-life of $^{136}$Xe $0\nu2\beta$ decay.

%%%%%%%%%%%%%%%%%%%%%%%%%%%%%%%%%%%%%%%%%%%%%%%%%%%%%%%%%%%
\subsection{Coherent Neutrino Scattering Science}

Although not yet detected, Coherent Neutrino Scattering (CNS) off target nuclei is a well-predicted standard model process which shares much of the same features of canonical WIMP-nucleus scattering.  Both types of searches benefit from detectors with low thresholds, low backgrounds, and large target masses. CNS can be studied using neutrinos from fabricated sources (neutrino beams, nuclear reactors, or radioactive sources) or from neutrinos of cosmic origin (solar and atmospheric neutrinos). Detectors must be optimized for each source given its neutrino energy spectrum and the flux at the detector.

For solar and atmospheric neutrinos, CNS detection requires combinations of low-background, low-threshold, and large target mass which are similar to those needed for G2 WIMP sensitivities. The G2 exposures and thresholds will lead to hundreds of detected $^{8}$B solar neutrino events. This signal will form an excellent calibration source for the experiments, and will allow an initial measurement of the CNS cross section which can be cross-calibrated the neutral current measurement from SNO. A measurement of the CNS cross section with $\sim$~10\% uncertainty results in a $\sim$~5\% measurement of $\sin^{2}\theta_{W}$ at $Q \sim 0.04$~GeV/$c$, momentum transfers which are much lower than all previous neutrino scattering measurements, including NuTeV's at $Q \sim 4$~GeV/$c$ \cite{2006PhRvD..73c3005S,2003PhRvL..90w9902Z}. New physics can also be probed through a search for Non-Standard Interactions (NSI). NSI extra terms can be added to the standard model Lagrangian by various sources, including incorporating neutrino mass into the standard model and supersymmetry. The $^{8}$B CNS measurement would be more than an order of magnitude more sensitive to several NSI terms than existing limits \cite{2007PhRvD..76g3008B}.  Finally, knowledge of the coherent neutrino cross section is essential in supernova explosion modeling, as $\sim$~99\% of the energy is carried away by neutrinos, and the opacity of the electron/nucleus plasma to neutrinos in the core during collapse is dominated by the CNS cross section.

Many of the dark matter technologies in this report are being adapted to dedicated CNS measurement proposals at beams, reactors, and with electron-capture sources. Scintillating crystals, Point-contact high-purity germanium detectors, cryogenic phonon-mediated detectors, two-phase xenon detectors, and two-phase argon detectors are all being pursued. These proposals aim to do all the science described above, but with more control over flux and neutrino energies and potentially higher significance measurements. In addition, these terrestrial sources enable searches for sterile neutrinos through both energy-dependent and position-dependent disappearance searches \cite{2012PhRvD..86a3004A}. Since CNS is a flavor-blind process, the detection of a disappearance oscillation signal would be a smoking-gun measurement of a sterile process. Finally, high precision measurements of the CNS cross section allow for measurement of neutron form factors \cite{2013IJMPE..2230013P}.

%%%%%%%%%%%%%%%%%%%%%%%%%%%%%%%%%%%%%%%%%%%%%%%%%%%%%%%%%%%
\section{ Common Infrastructure}\label{sec:cf1:ci}

Direct detection experiments require an infrastructure that has much in common with other areas of physics, especially neutrinoless double beta decay experiments and, to a lesser extent, proton decay searches.  These include the need for large underground spaces, to avoid cosmic ray interactions, exclusion of radioactive contaminants at an extreme level, and detailed simulation of the effects of backgrounds.  These experiments all share with other particle physics experiments needs for high-throughput data acquisition and analysis systems. In this section, we briefly discuss these common infrastructure needs.

%%%%%%%%%%%%%%%%%%%%%%%%%%%%%%%%%%%%%%%%%%%%%%%%%%%%%%%%%%%
\subsection{Underground Facilities}

Dark matter needs for underground space and related infrastructure will evolve with time as the scale of experiments grows and as R\&D leads to new avenues of experimental exploration.  The space and infrastructure budget includes not only the expected major large-scale detectors, but also required auxiliary efforts, such as low-background screening and storage, as well as for R\&D on new or alternative techniques.  One needs to account for several similarly large neutrinoless double beta decay experiments worldwide that compete for the same space, but might share infrastructure such as radiopurity assay.

While it appears that there is enough space worldwide for the G2 experiments, G3 projects will a major driver of underground space and related infrastructure.  These experiments will be physically large due to the substantial shield/veto systems required to mitigate radiogenic and cosmogenic backgrounds.  The DUSEL S4 engineering studies provide a rough envelope for the size of these experiments, which we can take to
be 25\,m$\times$15\,m$\times$15\,m for this discussion.

Currently, there are no U.S. sites capable of hosting experiments of this size.  The available caverns at the Soudan Lab (2,100\,meters of water equivalent, m.w.e.) and the Sanford Underground Research Facility at Homestake (4,200\,m.w.e.) are too small and would require new excavation.  However, there are a number of internationally available options.  The Laboratori Nazionali del Gran Sasso (LNGS) has extensive underground space capable of hosting G3 experiments at a 3,000\,m.w.e. effective depth.  The Laboratoire Souterrain de Modane (LSM) has a depth of 4,800\,m.w.e.: a major expansion with a new volume of 17,500\,m$^3$ is expected to become available in 2017 which could house one G3 experiment.  The Cryopit and Cube Hall caverns at SNOLAB (6,100\,m.w.e.) are large enough, but their availability on the timescale of such experiments is not clear.  Jin Ping Laboratory in China (CJPL) will be a site of very substantial depth (7,000\,m.w.e.) and is expected to have 100,000\,m$^3$ of space in a format large enough for G3 experiments on the 2016 timescale.  The India Neutrino Observatory (INO) may have an available volume for non-neutrino experiments of about 26000\,m$^3$, but there is some uncertainty about the 2018 timescale.  Another site in the Andes is being considered as well.

\begin{figure}[h!]%Figure 19
\begin{center}
\includegraphics[width=0.49\textwidth]{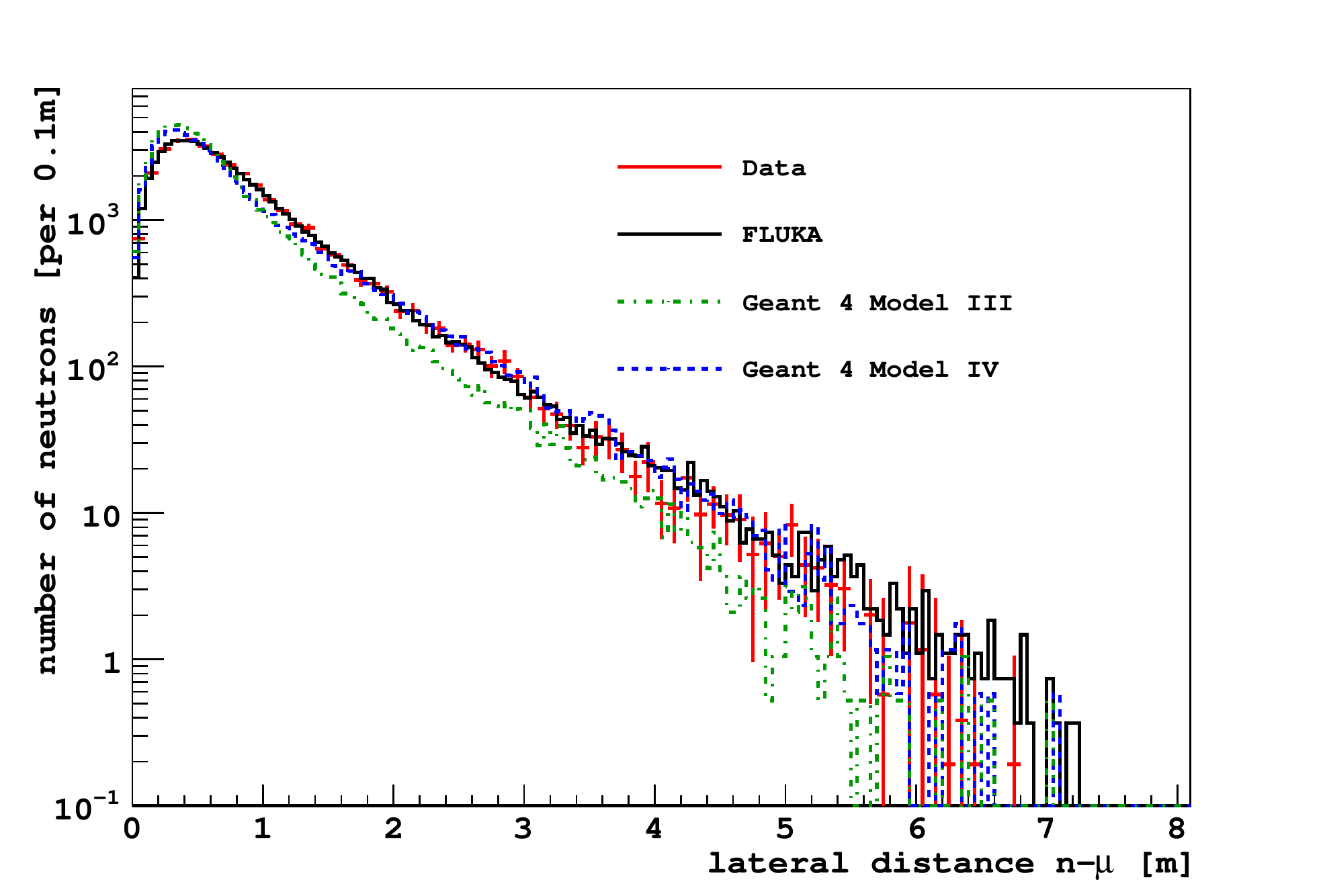}
\includegraphics[width=0.49\textwidth]{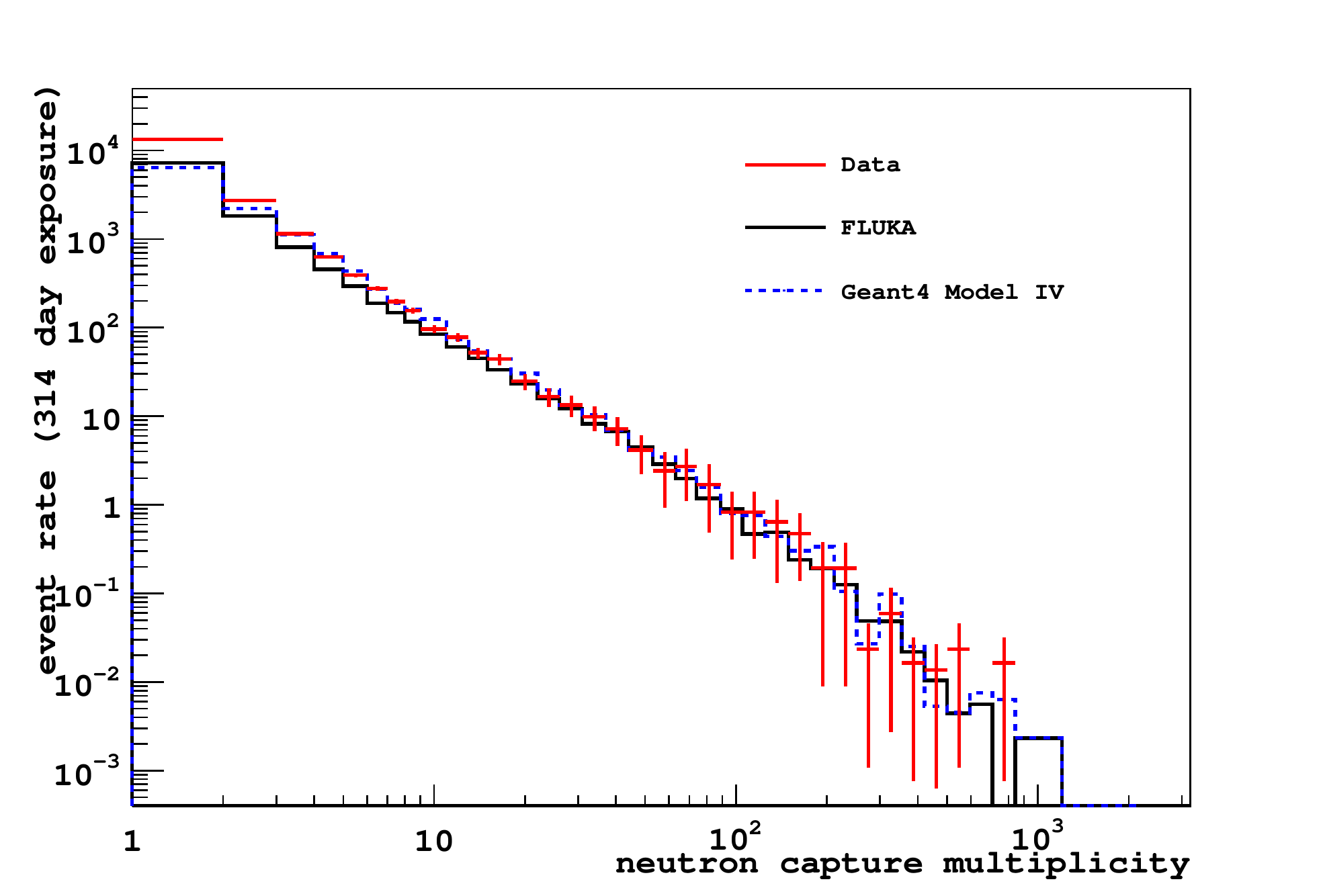}
\caption{\label{fig:borexino}Left: lateral distance between neutron capture points and the parent muon track as measured in Borexino, and compared with predictions obtained with GEANT-4 and FLUKA.  Right: neutron capture multiplicity as measured in Borexino, and compared with predictions obtained with GEANT-4 and FLUKA.  From Ref.~\cite{borexino}.}
\end{center}
\end{figure}

The study solar neutrino detectors with large volumes of scintillator and water has permitted the precision measurement of cosmogenic neutrons, as well as the fine-tuning of GEANT-4 and FLUKA models used to study cosmogenic backgrounds: see Fig.~\ref{fig:borexino}
and Ref.~\cite{borexino}.  Studies with Monte Carlo codes validated through this effort demonstrate that cosmogenic backgrounds for G2 detectors at moderate depths can be fully abated by the use of large ($\sim$10 m diameter) active vetos~\cite{empl}.  The same scheme can be used profitably for G3 experiments with larger shields, whose ultimate size will be determined with the additional information coming from operation of the G2 detectors.  It thus appears that depth by itself is not the determining factor in the siting of a large G3 experiment, but risk of unexpected cosmogenically-induced backgrounds in these ultra-sensitive experiments is greater at shallower depths.  In addition, sites at deeper locations may require a smaller footprint for G3, since their active shielding could be more compact.

It is expected that there will be between two and four G3 dark matter experiments worldwide, depending on the U.S., European, and Japanese level of support, and whether new international participants (e.g., China, India) can provide additional funding.  As advocated by IUPAP, open-access policies at major international laboratories are essential for an optimal scientific program worldwide.  In the field of dark matter in particular, open-access has made it possible for U.S.-led G1 and G2 experiments to use many sites outside the US.  Moreover, these international underground laboratories recognize the high scientific priority of dark matter experiments and are interested in housing international G3 dark matter experiments.

We believe that it is important for the health of worldwide and US underground science, and our own long term access to international facilities, that the US contribute its share for support of a frontier US, or cooperative US-international, underground facility, capable of hosting G3 dark matter experiments.

%%%%%%%%%%%%%%%%%%%%%%%%%%%%%%%%%%%%%%%%%%%%%%%%%%%%%%%%%%%
\subsection{Low-Background Materials and Assay Requirements}

The next generations of direct detection dark matter experiments will require stringent radiopurity in their target materials, internal instrumentation, and shielding components.  A survey of major material assay facilities world-wide and of anticipated assay needs of G2 experiments indicate current assay capability is marginally adequate in sensitivity and inadequate in throughput (see Ref.~\cite{assay} for more information).  High purity germanium (HPGe) gamma-ray assay has been the workhorse for material screening.  It appears hundreds of samples will require screening at the limits of current achievable HPGe sensitivity.  Mass spectroscopy methods and neutron activation analysis have demonstrated superior assay sensitivity in some specific cases, however, they are also currently throughput limited.  In addition, both alpha/beta screening and radon emanation analysis at or beyond the current achievable sensitivity levels will be needed for dozens of samples.

Experiments that follow G2 will have even more stringent material radiopurity specifications exceeding current assay capabilities in both sensitivity and throughput.  These screening needs must precede the commissioning of experiments by 3-5 years to inform design and quality control of components.  Achieving the radiopurity goals of the next decade's experiments will require investment in new techniques and tools to improve material assay sensitivity and throughput.  Engineered low background materials are a significant R\&D expenditure, requiring developmental lead-time, and thus should benefit many end-user experiments.  Reserving underground real estate for assay methods affected by cosmic rays (e.g. HPGe) and engineered radiopure materials (e.g. electroformed copper) activated by cosmic rays, is judicious forward planning.  In many cases, these spaces will require radon-suppressed sample preparation and material storage space.

The above-mentioned survey of material assay facilities indicated very little development is taking place to improve sensitivity or throughput capability.  In order to create the necessary infrastructure, it may help to form a Consortium of low background assay centers in the U.S that is managed by a scientific board with representatives from the community.  This would provide for common use of existing screening facilities and a unified plan for new infrastructure and site development (see the Snowmass Facilities White Paper).  Such a loose organization already exists as AARM (Assay and Acquisition of Radiopure Materials), which is currently populating a community-wide materials database with published assay information.  A more formal organization such as the Consortium is being actively investigated with DOE and NSF.  Eventually, all screening done in assay centers controlled by the Consortium could directly input results to the open-access database, which would reduce duplication and provide for more efficient vendor selection.

%%%%%%%%%%%%%%%%%%%%%%%%%%%%%%%%%%%%%%%%%%%%%%%%%%%%%%%%%%%
\subsection{Radon Mitigation and Sensing}

Radon and its radioactive decay daughters require special attention as a potential background source for direct detection dark matter and neutrinoless double-beta decay experiments.  Present in air, radon impacts experiments through three distinct vectors: (1) radon daughter deposition on detector materials during storage, (2) radon and radon daughter infiltration during experimental assembly and operation, and (3) radon contamination of liquids and gases, such as the liquid noble gases commonly used as dark matter targets.

Experimental backgrounds from radon and radon daughters can appear in a variety of ways.  For example, short-lived radon daughters may act as a direct gamma-ray background, radon daughters deposited from the air onto detector surfaces or support materials may generate backgrounds from low-energy beta decays, $^{210}$Po $\alpha$-decays may create backgrounds from recoiling $^{206}$Pb nuclei, and radon daughter $\alpha$-decays may generate ($\alpha$,n) neutron backgrounds.

Better mitigation against radon and radon-daughter backgrounds requires improvement of (1) reduced-radon storage capability, (2) reduced-radon laboratory spaces for detector assembly, (3) specialized radon filtration systems for liquids and gases used in detectors, (4) surface screeners sensitive to the non-penetrating radon daughters, and (5) methods of removing implanted radon daughters from surfaces.  See Ref.~\cite{radon} for more details.

%%%%%%%%%%%%%%%%%%%%%%%%%%%%%%%%%%%%%%%%%%%%%%%%%%%%%%%%%%%
\subsection{Computing and DAQ}

The data acquisition (DAQ) and computing requirements of the current generation of direct dark matter search experiments have been addressed on an experiment by experiment basis.  Larger direct detection experiments can be expected to benefit from techniques developed in other areas of particle physics.  In this section, we address two key areas: the evolution of requirements for simulation of the experiments, and the future needs in the area of offline computing and software.

%%%%%%%%%%%%%%%%%%%%%%%%%%%%%%%%%%%%%%%%%%%%%%%%%%%%%%%%%%%
\subsection{Simulation}

Current direct-detection dark matter experiments use some or all of the following codes to simulate their experiments:  FLUGA, GEANT-4, MCNPX, MUSUN, MUSIC and SOURCES.  Over the next 20 years the role of simulations in these experiments will increase and simulation codes must be able to adapt and expand to meet the need.

GEANT-4 is a software toolkit especially suited to such adaptation.  Its flexible physics and geometry make it useful at almost all stages of experiment simulation, and its imminent upgrade to multi-threading will allow more efficient use to be made of computing resources.  GEANT-4 development is driven largely by funding from the HEP community.  Funding from the dark matter community should be made available to ensure that their simulation needs are met.

Simulation codes are one example of resources shared among experiments.  Another might be simulation frameworks.  A framework connects collections of simulation tools and allows access to them through a common interface.  Historically, each dark matter experiment has developed its own framework, but as complexity and cost increase, development of a common, adaptable framework should be discussed.

Development and coordination of software resources common to all direct-detection experiments should be encouraged.  An organization could be imagined which would foster more extensive collaboration between code developers and experimenters, serve as a clearinghouse for available software tools, and fund the development of new, commonly desired code.

%%%%%%%%%%%%%%%%%%%%%%%%%%%%%%%%%%%%%%%%%%%%%%%%%%%%%%%%%%%
\subsection{Offline Computing And Software}

The trend towards larger, and more complex, direct detection experiments means that the requirements on the offline computing and software are changing in a qualitative way.  Already for G2 experiments, data volumes will grow to the PetaByte scale, demanding new requirements in automation of data bookkeeping, processing and analysis.  Simulated and real data will benefit from being accessible through a unified framework to facilitate complex and sophisticated physics analysis.  The offline software will reflect this increased complexity.  At the same time, the collaborations, on an absolute scale, are not very large with few software professionals and usually without any dedicated offline computing funding.  Thus the challenge is scaling up offline computing with limited manpower, resources and funding.

Cooperation among DOE laboratories will be very important.  At the same time, the needs of many of the small Cosmic and Intensity Frontier experiments are quite similar.  It would be worthwhile investigating a more inclusive support model, for example through a consortium.  This approach may make it easier for smaller groups or individuals to contribute and make overall support from DOE more cost effective.

%%%%%%%%%%%%%%%%%%%%%%%%%%%%%%%%%%%%%%%%%%%%%%%%%%%%%%%%%%%
\section{ Sensitivity and discovery potential}\label{sec:cf1:sadp}

A direct detection experiment's expected sensitivity is a combination of the total number of WIMP interactions that can potentially be observed (the integrated WIMP nuclear scattering rate above the energy threshold of the experiment) and the total number of expected background events.  Expected sensitivity is usually defined as the expected 90\% confidence level (C.L.) limit on the WIMP interaction cross section with nucleons as a function of WIMP mass, in the absence of any signal but without background subtraction (i.e. by treating all events in the WIMP region of interest as WIMP candidates).  The sensitivity is maximal if the experiment is projected to be free of background, and degrades roughly proportionally with the expected number of background events.

By contrast, discovery potential can be loosely defined as the expected ratio of signal to background events as a function of WIMP mass for a given WIMP-nucleon cross section.  Following HEP tradition, evidence for WIMPs may be claimed if the estimated number of signal events compared with the expected background rises to the 3$\sigma$ level, and a WIMP discovery may be claimed if a 5$\sigma$ level is reached.  For small numbers of events, these levels are determined from Poisson statistics.  The minimum number of observed events (including both signal and background events) for a discovery rises rapidly with the expected background, from 6\,events for an expected background of 0.1\,events to 10\,events for an expected background of 1\,event.  Overall, the minimum number of observed events increases as the 0.7$^{\rm th}$ power of the expected number of background events.

Both sensitivity and discovery potential depend on several experimental factors, including the exposure, background performance, energy threshold and the understanding of nuclear recoil energy scales.

The exposure for an experiment is defined as the product of the usable target mass and the length of time that target mass is actually searching for WIMPs.  The usable target mass is commonly referred to as ``fiducial'' mass, where typically only a clean ``inner'' volume is used, with an outer volume serving as passive or active shielding against backgrounds. The length of time is determined from the livetime of the experiment and the efficiency of the analysis used to search for WIMPs.  The sensitivity~of an experiment to WIMPs in all mass ranges~depends linearly on the exposure until the experiment becomes limited by irreducible backgrounds.

Background performance is a combination of reduction of known background sources from passive and active shielding, and active discrimination of the detectors against residual backgrounds.  The emphasis depends on the technology used; experiments with modest active discrimination must necessarily employ state-of-the-art background reduction, while those with exceptional active discrimination against backgrounds may have to worry less about reducing those backgrounds.

\begin{figure}[h!]%Figure 20
\begin{center}
\includegraphics[width=\textwidth]{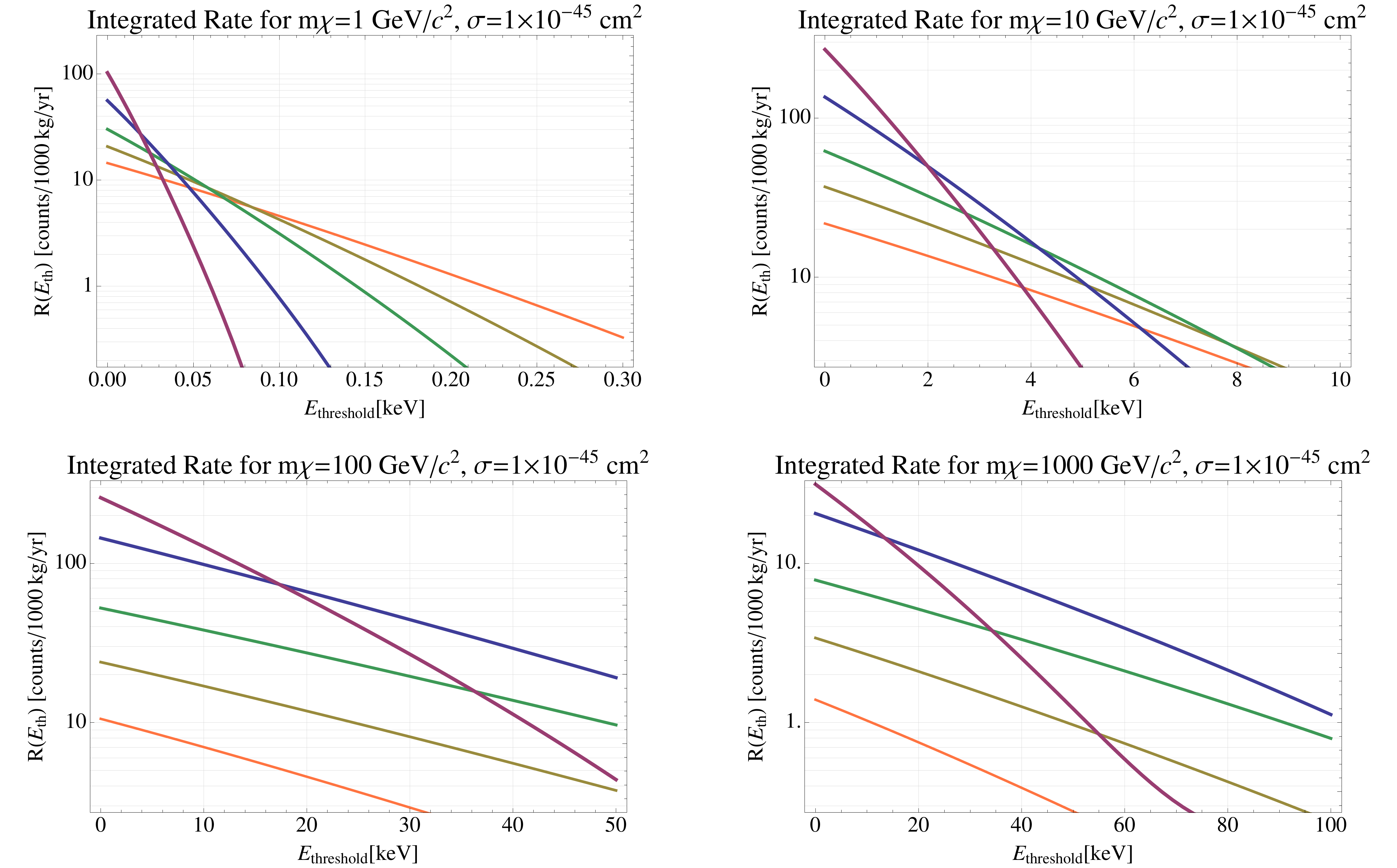}
\caption{\label{fig:integral}Integrated rate versus energy threshold for 1 (upper left), 10 (upper right), 100 (lower left) and 1,000\,GeV (lower right) WIMP masses, assuming a spin-independent WIMP-nucleon cross section of 10$^{-45}$\,cm$^2$, shown for Xenon (blue), Germanium (purple), Argon (green), Silicon (brown) and Neon (orange) target.}
\end{center}
\end{figure}

\begin{figure}[h!]%Figure 21
\begin{center}
\includegraphics[width=\textwidth]{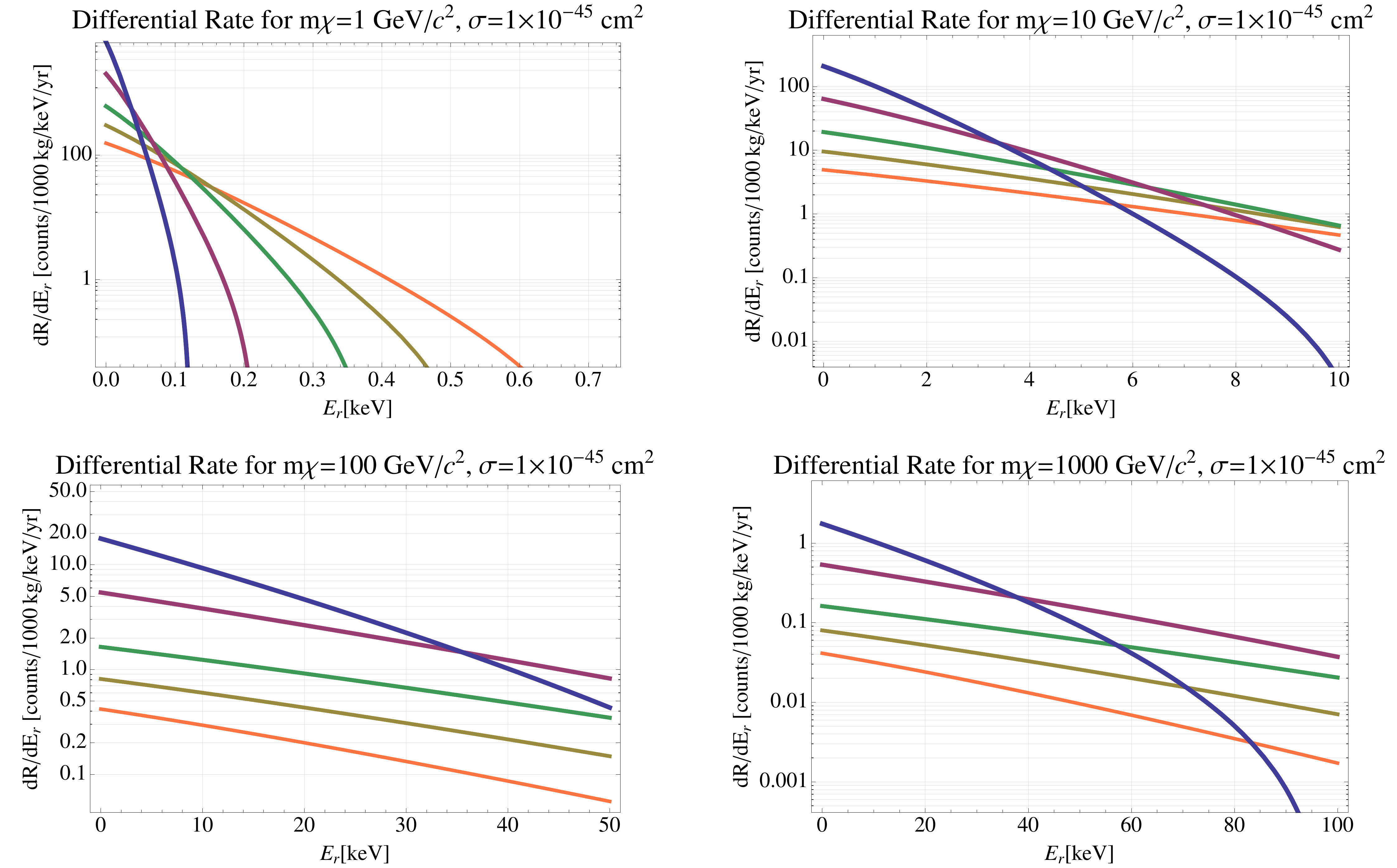}
\caption{\label{fig:differential}Differential rate versus nuclear recoil energy for 1 (upper left), 10 (upper right), 100 (lower left) and 1,000\,GeV (lower right) WIMP masses assuming a spin-independent WIMP-nucleon cross section of 10$^{-45}$\,cm$^2$, shown for Xenon (blue), Germanium (purple), Argon (green), Silicon (brown) and Neon (orange) target.}
\end{center}
\end{figure}

Nuclear recoil energies deposited by WIMP interactions are quite small, in the 0.1--100\,keV$_{\rm r}$ range.  Direct detection experiments do not actually measure nuclear recoil energies directly, but instead calculate them from the observables such as ionization, phonon or scintillation signals.  Systematic errors in conversion between measured energies and nuclear recoil energies can arise due to poor understanding of the physical process of nuclear recoil energy deposition or incorrect measurements.  There are energy thresholds associated with each of these observable signals from which the nuclear recoil energy threshold must be calculated.  The lower an experiment can push its thresholds, the larger its sensitivity to WIMPS, since the WIMP nuclear recoil energy spectrum rises exponentially towards low energy.  This effect is most evident for heavy target isotopes, and becomes increasingly important for low mass WIMPS, as demonstrated in Fig.~\ref{fig:integral}
and \ref{fig:differential}.

Delivering improved performance, by an order of magnitude or more, for the new direct detection pathfinder experiments clearly involves some technological risk.  It is possible that the projected improvement in sensitivity for a given experiment will not be realized in practice, due to an unexpected background, or other systematic effect.  Using different detector technologies in multiple pathfinder experiments helps manage this technological risk, and improves the reliability of the progress of the program over time.  It would be optimal for the US direct detection program to support at least two detector technologies when pursuing improvements in each of the main dark matter particle frontiers.

%%%%%%%%%%%%%%%%%%%%%%%%%%%%%%%%%%%%%%%%%%%%%%%%%%%%%%%%%%%
\section{ Direct Detection Program Roadmap}\label{sec:cf1:ddpr}

%%%%%%%%%%%%%%%%%%%%%%%%%%%%%%%%%%%%%%%%%%%%%%%%%%%%%%%%%%%
\subsection{Projected Progress in WIMP Direct Detection}

\begin{figure}[h!]%Figure 22
\begin{center}
\includegraphics[width=0.6\textwidth]{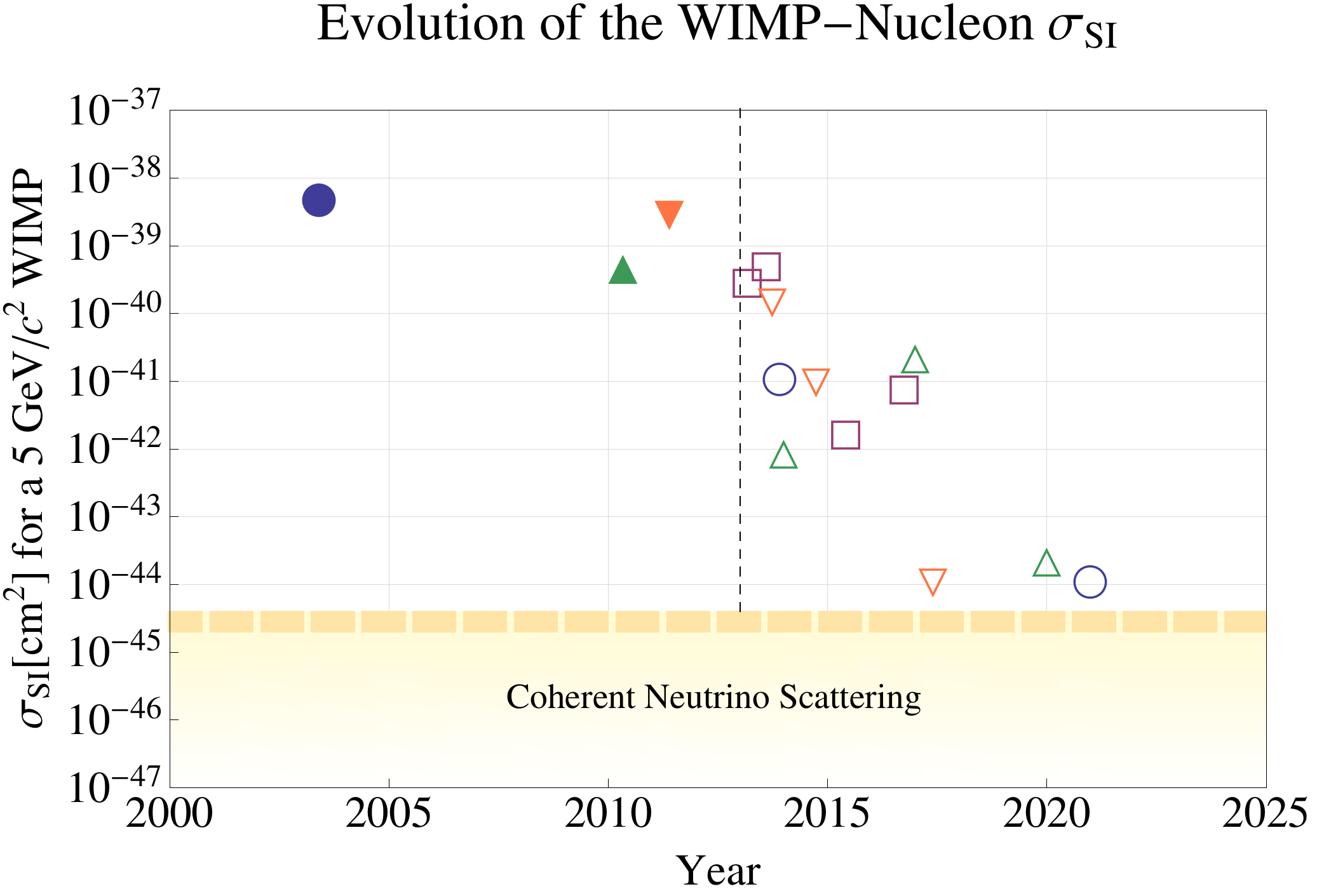}
\caption{\label{fig:5GeV}History and projected evolution with time of spin-independent WIMP-nucleon cross section limits for a 5\,GeV WIMP.  The shapes correspond to technologies: cryogenic solid state (blue circles), crystal detectors (purple squares), liquid argon (brown diamonds), liquid xenon (green triangles), and threshold detectors (orange inverted triangle).  The brown and blue rectangular shaded regions indicate classes of theoretical models which have been, or will be challenged by direct detection results.  Below the yellow dashed line, WIMP sensitivity is limited by coherent neutrino-nucleus scattering.}
\end{center}
\end{figure}

\begin{figure}[h!]%Figure 23
\begin{center}
\includegraphics[width=0.6\textwidth]{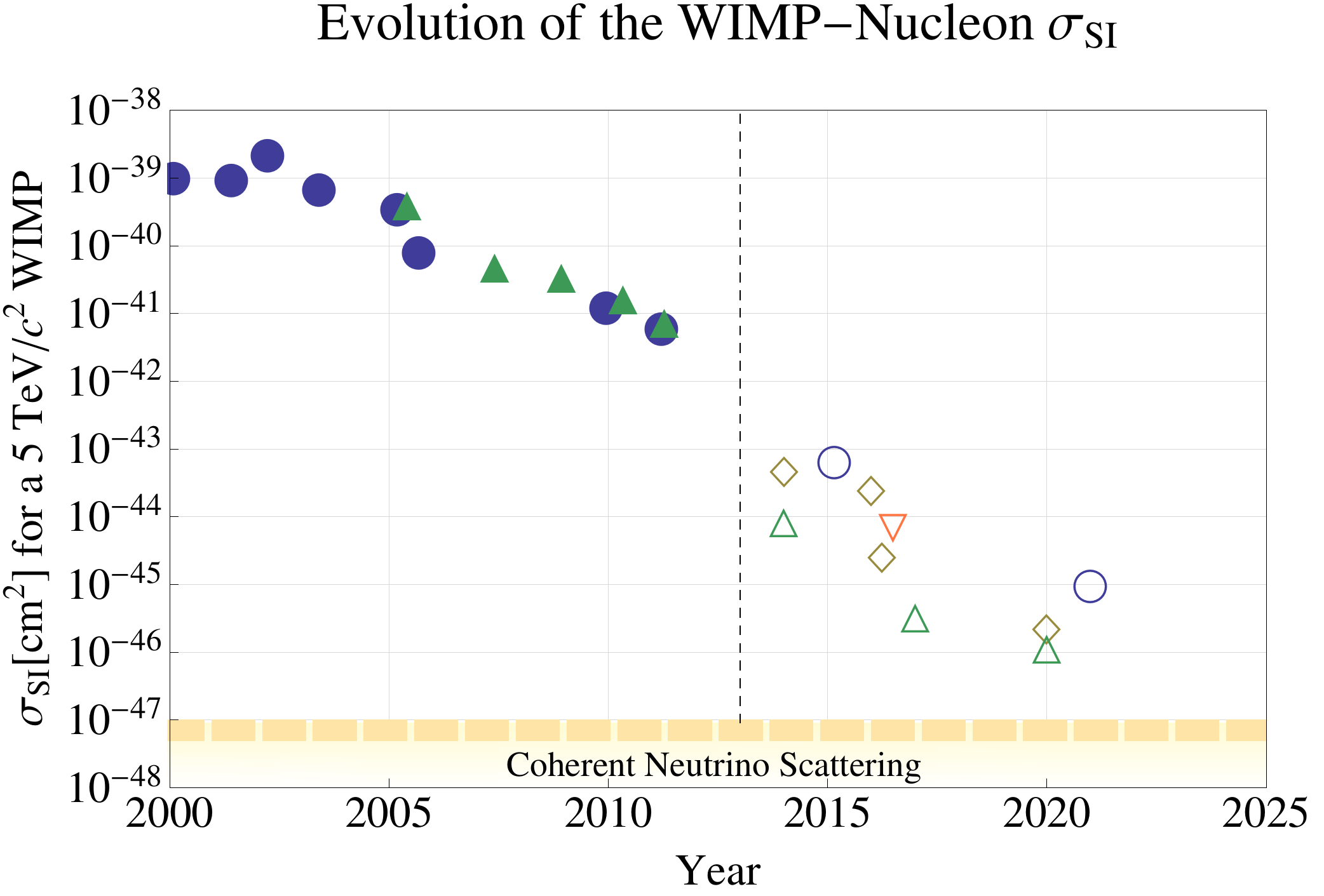}
\caption{\label{fig:5TeV}History and projected evolution with time of spin-independent WIMP-nucleon cross section limits for a 5\,TeV WIMP.  Technology is coded as in Fig.~\ref{fig:5GeV}.}
\end{center}
\end{figure}

The first two decades of dark matter direct detection experiments have yielded a diverse and successful program, although not yet definitive evidence for WIMPs.  Starting with just a few experiments using solid-state targets, the technologies used for these experiments have grown considerably.  There has been a remarkable improvement in WIMP sensitivities, especially in that range where the WIMP mass is comparable to the atomic mass of the target nuclei.  A selection of spin-independent results from the first two decades of these experiments, and projections for the coming decade are shown in Fig.~\ref{fig:5GeV},
\ref{fig:50GeV}, \ref{fig:5TeV},
for three representative WIMP masses.  A ``Moore's law'' type improvement is particularly evident for larger WIMP masses, with a sensitivity doubling time of roughly 18~months.  Note that direct detection experiments have sensitivity to very large WIMP masses, much larger than those accessible to the LHC.  More recently, there has also been rapid progress in sensitivity to low-mass WIMPs.

Sensitivity projections are subject to uncertainties from many factors, including technical issues with the experiments, the appearance of unexpected backgrounds and delays in funding.  Despite these uncertainties, the history of the field gives us confidence that progress will continue unabated through the next decade.  Beyond this point, sensitivity gains will begin to be limited by solar and atmospheric neutrino backgrounds.

%%%%%%%%%%%%%%%%%%%%%%%%%%%%%%%%%%%%%%%%%%%%%%%%%%%%%%%%%%%
\subsection{ Establishing a Discovery}

Direct detection experiments must be able to detect the tiny ($\sim$keV) energy depositions of dark matter while simultaneously excluding the background from standard model interactions at extraordinary levels ($<$1\,event/ton/year).  In order for an observation of signal candidates in a given experiment to be convincing evidence for WIMPs to the experimenters themselves, the results must be statistically significant ($>$3$\sigma$) and the estimate of the known backgrounds must be robust.  In order to convince the community that WIMP dark matter has been discovered, at least two such experiments with different targets and different systematic effects are required to provide evidence at the 5$\sigma$ level that is compatible with a single WIMP model, cross section, and mass.  Making some form of the experimental data public to the will help establish trust in the result.

Robust estimate of experimental backgrounds requires use of in-situ experimental data to estimate known backgrounds and to reduce the probability of the existence of any unknown backgrounds.  Independent means to tag neutron-induced events, reducing the dependence on the accuracy of Monte Carlo simulations, is one example of such robust background estimates.  In general, the more data outside the signal region can be used to constrain backgrounds, the more robust the background estimate.  Ultimately, measurements of all known processes that may mimic a signal are desired since estimations based on extrapolations or simulations may not be accurate, especially for rare processes.

Possible WIMP signatures can vary significantly in the degree to which they are convincing.  Observing an annual or diurnal modulation may be very convincing, especially if done in conjunction with a counting signal that is not dominated by backgrounds.  A modulation signal in the presence of considerable background requires the challenging proof that the background itself is not modulating. In a counting experiment, it is important that any signal candidate events appear uniformly in position throughout a detector and have distributions in energy and discrimination parameters consistent with a dark matter model. Generally speaking, the more measures in which WIMP signals and backgrounds can be separated, the more convincing any signal claim is likely to be.  Conversely, if an unknown background could conceivably show up only in the signal region without being distinguishable in a sideband, the potential signal is not going to be believable.

%%%%%%%%%%%%%%%%%%%%%%%%%%%%%%%%%%%%%%%%%%%%%%%%%%%%%%%%%%%
\subsection{Post-Discovery Dark Matter Particle Physics}

Once a convincing WIMP signal is detected, attention will turn to measuring the particle properties.  With sufficient statistics on multiple targets, information on the WIMP's mass and the nature of its coupling to nucleons can be gleaned from direct detection experiments.

\begin{figure}[h!]%Figure 24
\begin{center}
\includegraphics[width=0.49\textwidth]{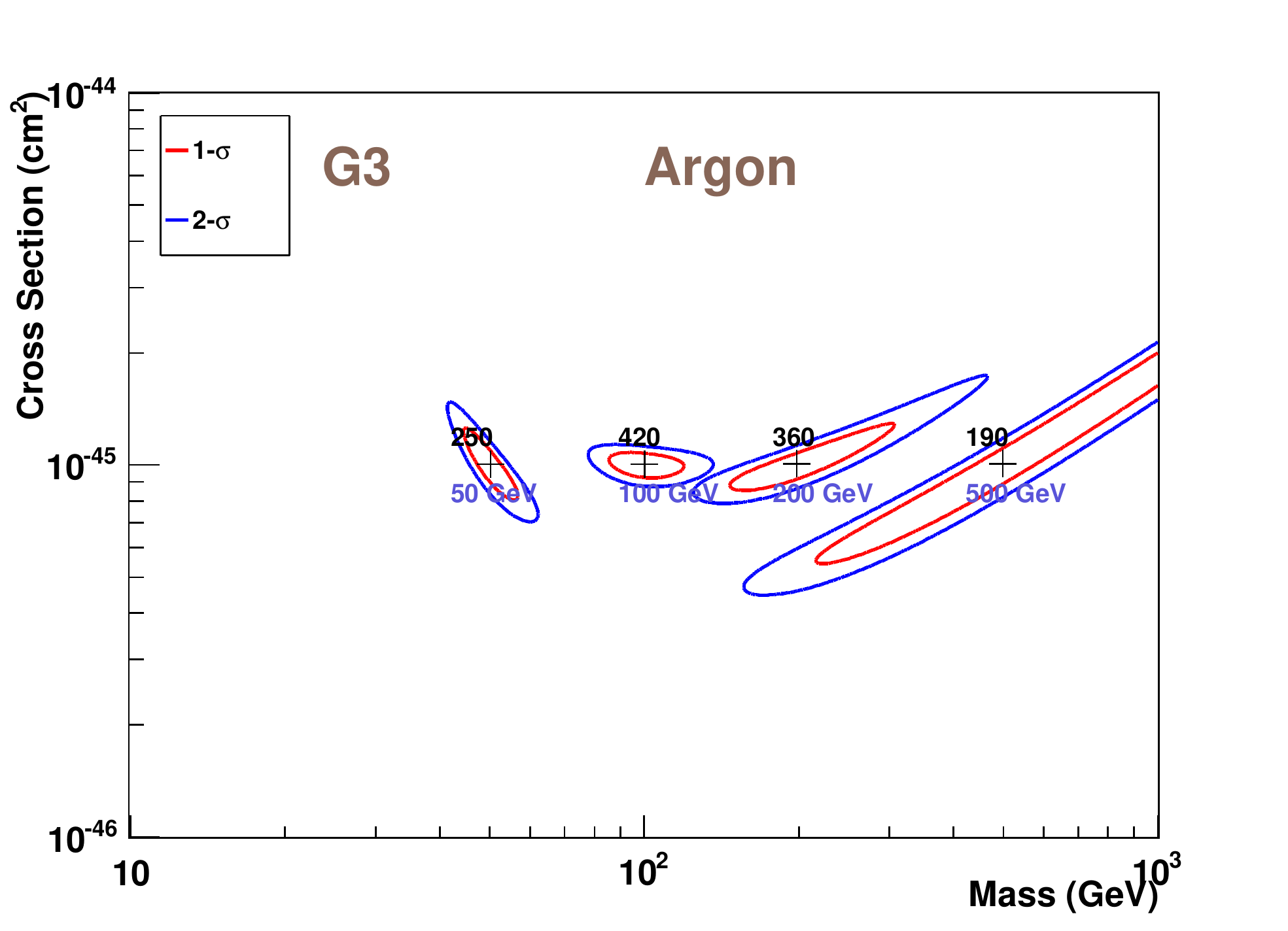}
\includegraphics[width=0.49\textwidth]{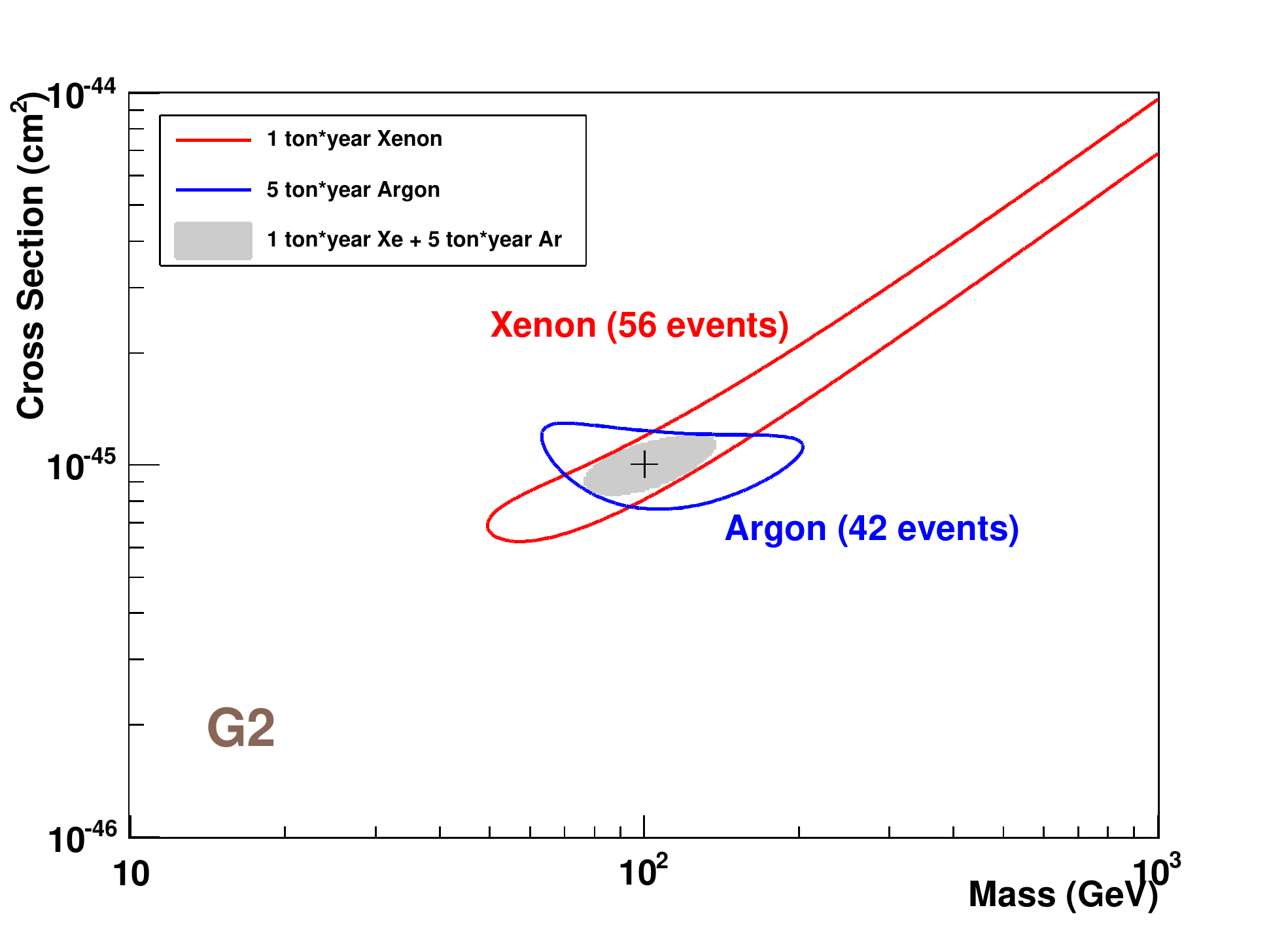}
\caption{\label{fig:xax}Allowed spin-independent and mass parameter space from a joint analysis of simulated xenon and argon experiments~\cite{xax}.  On the left, the WIMP mass varies between 50 and 500\,GeV.  On the right, is it 100\,GeV.  The cross section for all case studies is 10$^{-45}$\,cm$^2$.}
\end{center}
\end{figure}

A WIMP discovery naturally leads to an extensive campaign to collect high-statistics recoil energy spectra on multiple target nuclei.  Generically, all WIMP recoil spectra are approximately exponential in shape, with the slope of the exponential depending on the mass of the WIMP and the target nucleus.  If the WIMP is lighter than the target, significant constraints on WIMP mass and cross section can be derived from a small number of events.  However, the mass and cross section become degenerate if the WIMP mass is larger than the target nucleus mass, due to kinematical and nuclear form factor effects.  For WIMPs with mass about the same as the target, 100~detected events constrains the mass to about 50\%, assuming the WIMP velocity distribution is known~\cite{green}.  The degeneracy between mass and cross section can be partially broken by combining data on multiple targets, with the best leverage coming from combinations of heavy and light target nuclei (see Fig.~\ref{fig:xax}
for an example).  Unfortunately, if the WIMP is heavier than 300\,GeV, the spectral shape retains very little information on the mass no matter what target nucleus is chosen and in this case direct detection is unlikely to ever yield more than a lower limit on the mass.

\begin{figure}[h!]%%Figure 25
\begin{center}
\includegraphics[width=0.5\textwidth]{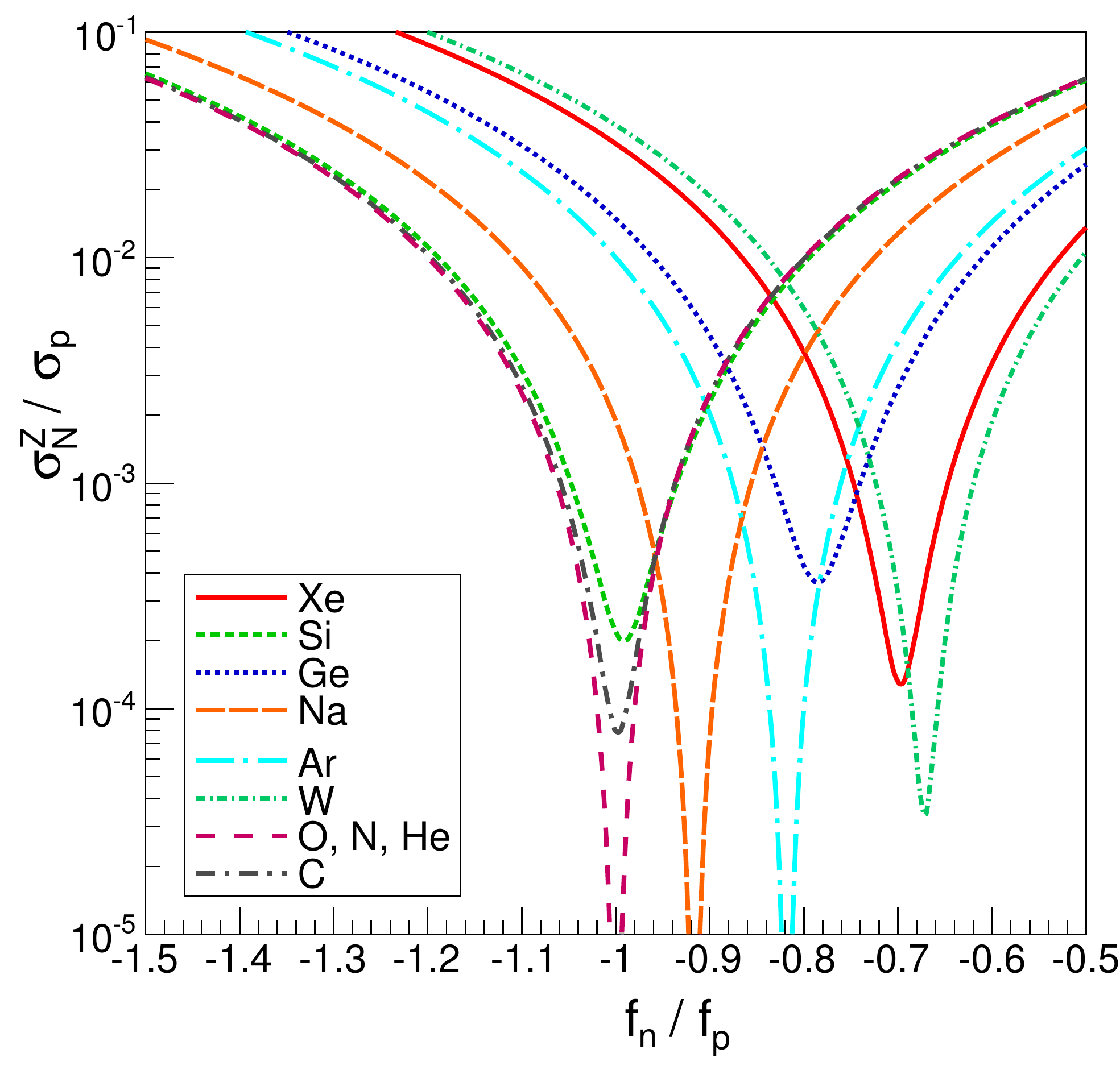}
\caption{\label{fig:fnfp}Ratio of the WIMP-neutron to WIMP-proton cross sections for different target nuclei used in direct detection experiments versus the ratio of couplings to neutrons and protons for isospin-violating models~\cite{feng}.}
\end{center}
\end{figure}

In addition to the WIMP mass and total cross section, it will be essential to learn as much as possible about the nature of the coupling to nuclei.  This may allow the disentanglement of competing particle physics models and would be needed for comparisons with collider data.  In the non-relativistic limit, WIMP-nucleon couplings are traditionally classified as ``spin-dependent'' (SD), when the sign of the scattering amplitude depends on the relative orientation of particle spins, or ``spin-independent'' (SI) when spin orientations do not affect the amplitude.  In the SI case, if all nucleons couple to WIMPs in the same way, the total nuclear cross section is enhanced by a factor $A^2$, with $A$ the atomic mass, due to coherent summation over all the scattering centers in the nucleus.  This greatly increases event rates on heavy nuclei relative to lighter nuclei.  However, in some models (so-called ``isospin violating dark matter''), the proton and neutron contributions can be different in magnitude or sign, breaking the simple $A^2$ scaling.  The resulting correction which translates the effective cross section relative to the standard $f_n$/$f_p$=1 case is shown in Fig.~\ref{fig:fnfp}.
One would like to separately measure the neutron and proton contribution through comparison of rates in different targets, although this can become impractical for very small cross sections due to high statistics required.  Since the ratio of neutrons to protons is rather similar for all heavy nuclei, using at least one light target nuclei is beneficial.

For SD interactions, the coupling is effectively to the net nuclear spin, due to cancellation between opposite spin pairs, and will differ depending on whether the net nuclear spin is carried primarily by a residual neutron or proton.  As a result, SD and SI searches are optimized by a different choice of target nuclei.  Because the spin-independent, proton-spin-dependent, and neutron-spin-dependent nuclear form factors are quite different for large momentum transfers in the dark matter nuclear-recoil interaction, it is possible, in principle, to distinguish the type of interaction by the shape of the recoil energy spectrum on a single target isotope.  However, for the low-momentum transfers typical of low mass ($<$10\,GeV) WIMP interactions, the differences are insignificant since all form factors are essentially unity.  Even for higher mass WIMPs that impart larger momentum transfers, detection with sufficiently high-statistics to allow deconvolution of the interaction cross section into the individual coupling terms using a single target would be quite challenging, especially since systematic uncertainties in the WIMP velocity distribution would similarly affect the expected recoil spectrum.

This is particularly exemplified in the general non-relativistic effective theory of dark matter direct detection \cite{fitzpatrick,haxton}.  Rather than making assumptions about couplings driven by specific models of the electroweak scale, or by appeals to minimalism, this theory considers the most general possible interactions with nucleons, and the corresponding nuclear response functions.  Instead of the traditional two (SI and SD) nuclear responses, the general effective theory includes five different couplings.~Since the relative amplitudes of these five couplings (and their possible interferences) are strongly dependent on the target material, large differences in the relative rates on different materials could be used to infer the relative strengths of the various WIMP-proton and WIMP-neutron couplings.  Combining targets that are sensitive in orthogonal directions can lead to improvements in the constraints on these couplings by up to several orders of magnitude.  Using multiple target materials to elucidate the particle interaction physics of dark matter is therefore an important consideration for selecting future direct detection experimental designs or a suite of experimental efforts.

%%%%%%%%%%%%%%%%%%%%%%%%%%%%%%%%%%%%%%%%%%%%%%%%%%%%%%%%%%%
\subsection{Post-Discovery Dark Matter Astrophysics}

Direct detection WIMP signals can also provide considerable information about the distribution of dark matter in our local neighborhood.  The standard assumption of a smooth Maxwellian distribution may be incorrect.  The local halo velocity distribution is subject to large modeling uncertainties, because of a lack of conventional observables giving information on dark matter halo substructure that could result from the particular formation history of our galaxy.  The energy spectrum of WIMPs can be distorted by the presence of tidal streams or other dark matter structure, such as a ``dark disk''~\cite{freese}.  By combining energy spectra on multiple targets, it will be possible to measure the WIMP velocity distribution independently of its particle properties, a possibility that has been referred to as ``WIMP Astronomy''.  Independent measurement of the WIMP mass - say from collider data - would greatly improve the ability of direct detection experiments to perform these measurements.

Measuring the rate modulation of dark matter induced nuclear recoils as a function of the Earth's direction through the galactic dark matter halo provides an additional method for understanding the velocity distribution.  For example, the magnitude of the modulation is a test of the Maxwellian velocity distribution hypothesis.  With sufficient statistics, measurements of annual modulation in the recoil spectrum shape can provide a constraint on WIMP mass, due to the fact that the mass determines the ``pivot energy'' where the modulation changes its sign.  Considerations of annual modulation measurement naturally lead to a desire for understanding the directionality of the WIMP scattering interactions, since a directional-specific modulation of the relative detector-WIMP relative velocity is the underlying mechanism for the modulation effect.  Directional information would therefore be a powerful tool for the investigation of halo substructure.

%%%%%%%%%%%%%%%%%%%%%%%%%%%%%%%%%%%%%%%%%%%%%%%%%%%%%%%%%%%
\subsection{A Dark Matter Roadmap}

\begin{figure}[h!]%Figure 26
\begin{center}
\includegraphics[width=\textwidth]{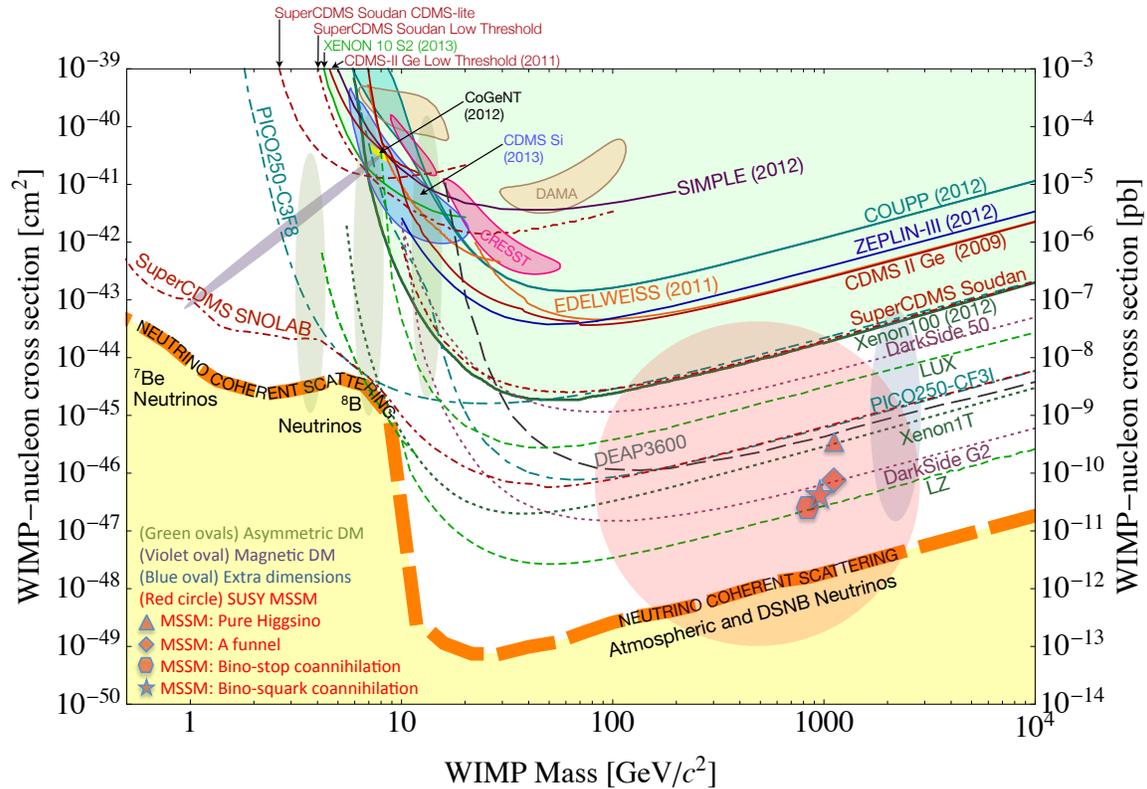}
\caption{\label{fig:SI-overview}A compilation of WIMP-nucleon spin-independent cross section limits (solid curves), hints for WIMP signals (shaded closed contours) and projections (dot and dot-dashed curves) for US-led direct detection experiments that are expected to operate over the next decade.  Also shown is an approximate band where coherent scattering of $^8$B solar neutrinos, atmospheric neutrinos and diffuse supernova neutrinos with nuclei will begin to limit the sensitivity of direct detection experiments to WIMPs. Finally, a suite of theoretical model predictions is indicated by the shaded regions, with model references included.}
\end{center}
\end{figure}

\begin{figure}[h!]%Figure 27
\begin{center}
\includegraphics[width=0.49\textwidth]{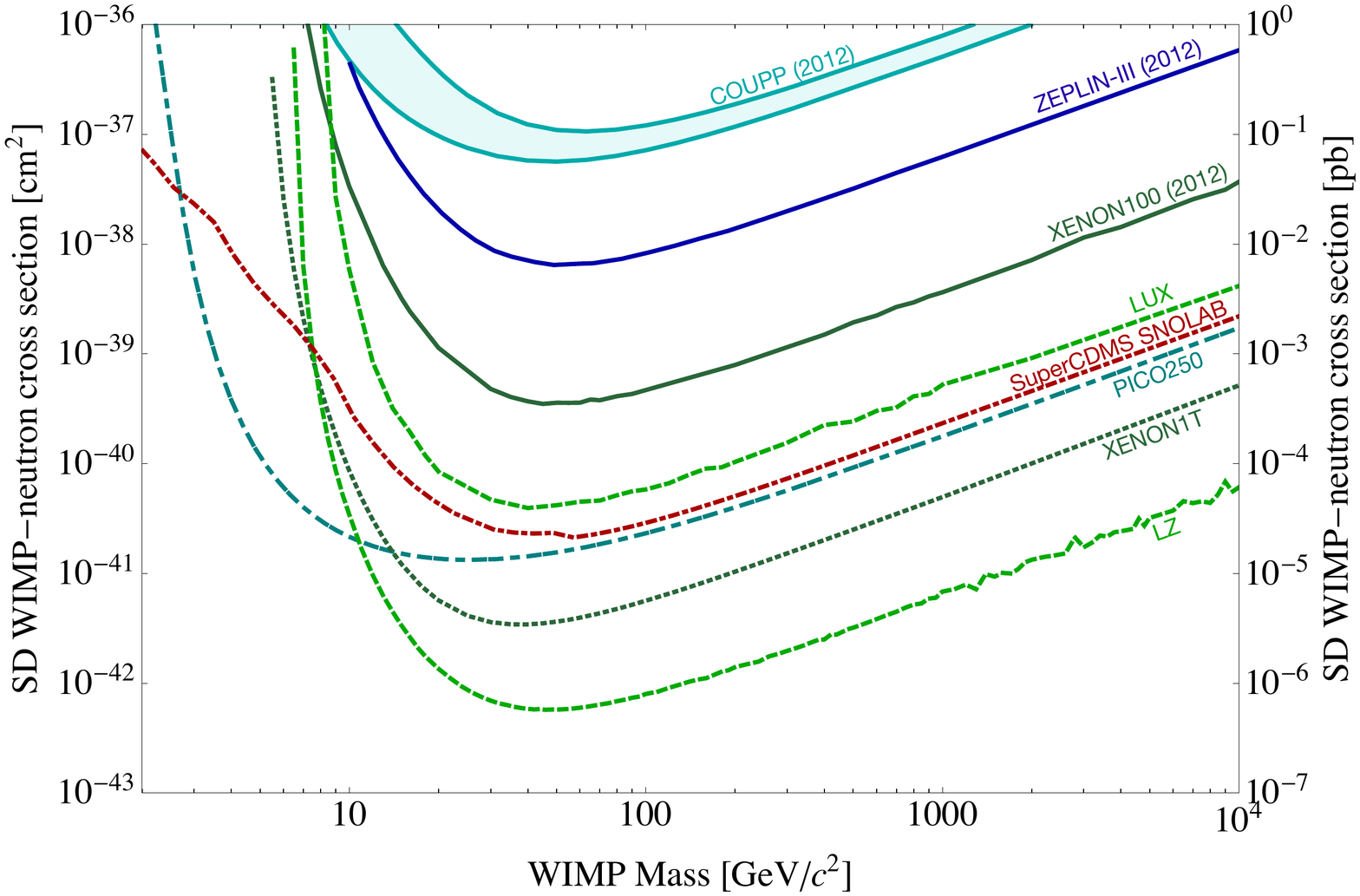}
\includegraphics[width=0.49\textwidth]{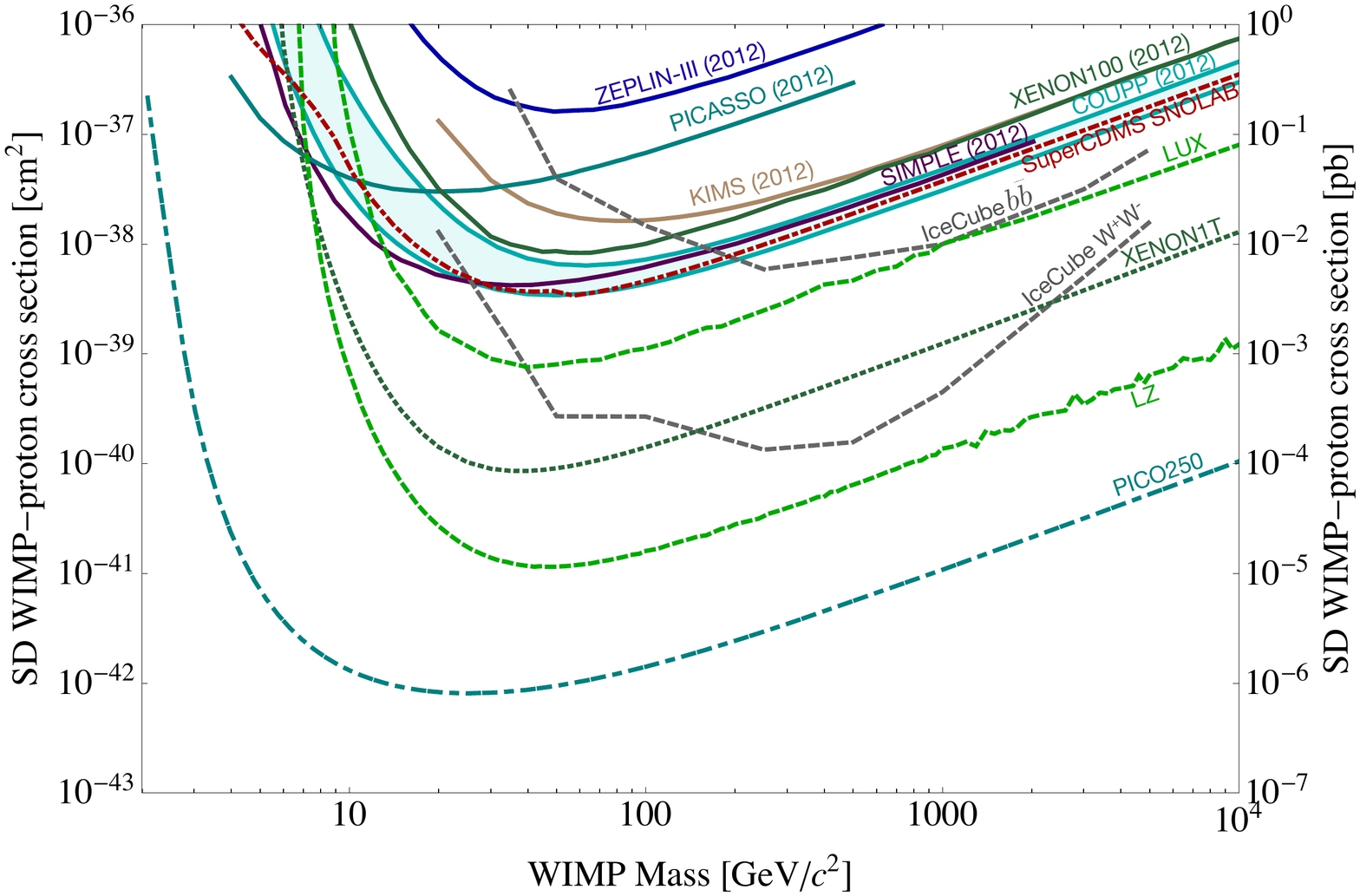}
\caption{\label{fig:SD-overview}A compilation of WIMP-nucleon spin-dependent cross section limits (solid curves) and projections (dot and dot-dashed curves) for US-led direct detection experiments that are expected to operate over the next decade.}
\end{center}
\end{figure}

Direct detection experiments provide a clear and compelling path towards an understanding of the cold dark matter that is widely believed to be composed of WIMPs.  These experiments explore significant portions of theory parameter space, some of which are completely inaccessible via collider or indirect detection techniques.  Fig.~\ref{fig:SI-overview}
and \ref{fig:SD-overview}
show the recent history and projected improvements in sensitivity for both spin-independent and spin-dependent couplings together with a range of theoretical benchmarks that can be probed over the next decade or so.

We believe that any proposed new direct detection experiment must demonstrate that it meets at least one of the following two criteria:

\begin{itemize}

\item Provide at least an order of magnitude improvement in cross section sensitivity for some range of WIMP masses and interaction types.

\item Demonstrate the capability to confirm or deny an indication of a WIMP signal from another experiment.

\end{itemize}

The US has a clear leadership role in the field of direct dark matter detection experiments, with most major collaborations having major involvement of US groups.  In order to maintain this leadership role, and to reduce the risk inherent in pushing novel technologies to their limits, a variety of US-led direct search experiments is required.  In addition, continuation of a robust detector R\&D program will ensure that new technologies can be brought to bear on WIMP signals as they appear.

\begin{figure}[h!]%Figure 28
\begin{center}
\includegraphics[width=1.0\textwidth]{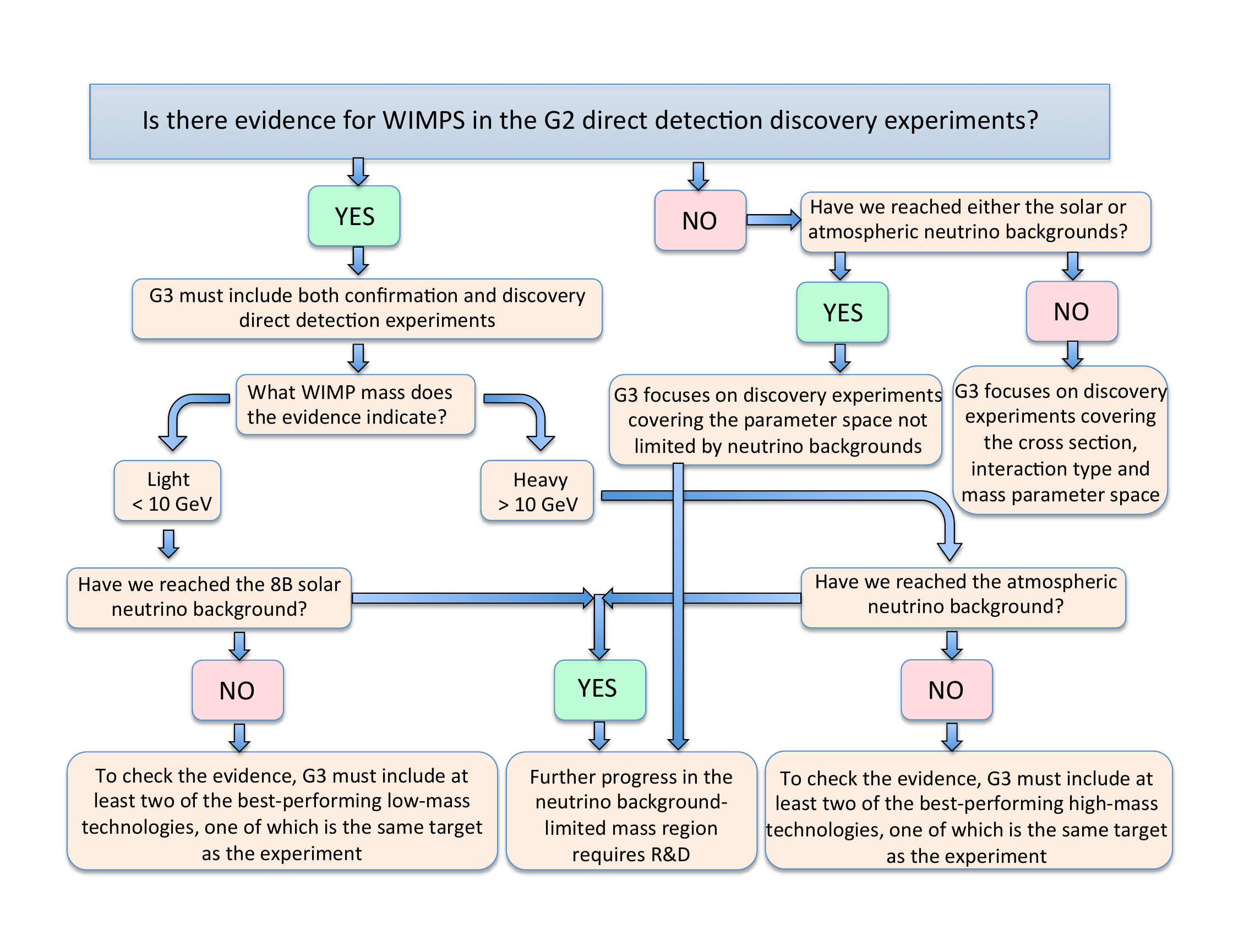}
\caption{\label{fig:roadmap}Decision tree for direct detection experiments from G2 to G3.}
\end{center}
\end{figure}

In a resource-limited environment, not every proposed direct detection experiment will be funded.  Information gleaned from past experiments, detector R\&D efforts and other types of dark matter searches has to be used to help inform funding agencies on how to choose a mix of experiments that will achieve the fundamental science goals of WIMP dark matter discovery and subsequent study.  Fig.~\ref{fig:roadmap}
shows how a ``decision tree'' for direct detection might utilize the information available from the current generation (G2) of experiments to make choices for the next generation (G3) experimental suite.  It is very important to keep in mind that, even for the simplest scenarios, the science goals are unlikely to be met with a single direct detection experiment, since confirmation from other experiments will be vital to convince the community that the particle nature of dark matter has finally been established.  The decision tree shown reflects our roadmap presented in Section~\ref{sec:cf1:exec} and summarized as the following three stages:

\begin{itemize}

\item[\bf A. Discovery:] Search broadly for WIMPS, with at least an order of magnitude improvement in sensitivity in each generation.

\item[\bf B. Confirmation:] Check any evidence for WIMP signals using complementary targets and the same target with enhanced sensitivity

\item[\bf C. Study:] If a signal is confirmed, extract maximal information about WIMP properties using multiple technologies.

\end{itemize}

%%%%%%%%%%%%%%%%%%%%%%%%%%%%%%%%%%%%%%%%%%%%%%%%%%%%%%%%%%%
\section{Summary}\label{sec:cf1:summary}

It is the consensus of the scientific community that identifying the particle nature of the dark matter in our universe is one of the most fundamental problems in particle physics today.  The solution to this problem may well lead the way to physics beyond the Standard Model.  Direct detection of dark matter particles in terrestrial laboratories would confirm their existence in our galaxy and begin to unravel their nature, especially with complementary information from production in colliders and indirect detection in our galaxy.

Over the last two decades, direct detection experiments have made tremendous progress in sensitivity to the Weakly Interacting Massive Particles (WIMPs) that are a favored candidate for particle dark matter.  This progress has been the result of technological innovations, a deeper understanding of backgrounds and an influx of scientists into the field.  We believe this rate of progress will continue, and perhaps even accelerate, over the last decade.  The next two generations of these experiments should probe another three orders of magnitude in WIMP-nucleon cross section sensitivity across a broad mass range from 1\,GeV to $>$10\,TeV, covering most of the currently-favored theory model regions.  If a WIMP signal is not seen, coherent neutrino scattering from solar, atmospheric and diffuse supernova neutrino sources will at least be detected. While this will be an interesting signal in its own right, it will become a difficult background in continuing the search for WIMPS.  However, if a WIMP is discovered and confirmed by multiple experiments, the field will shift towards a study of this new, Beyond the Standard Model, particle.  Direct detection experiments can provide information about the WIMP mass, and the form of its interactions with normal matter.

To conclude, we offer the following roadmap for the future of the dark matter direct detection experimental program worldwide.

\begin{center}

\textbf{A Roadmap for Direct Detection}

\textbf{\emph{Discovery}}

Search for WIMPS over a wide mass range (1 GeV to 100 TeV), with at least an order 
of magnitude improvement in sensitivity in each generation, until we encounter 
the coherent neutrino scattering signal that will arise from solar, atmospheric 
and supernova neutrinos

\textbf{\emph{Confirmation}}

Check any evidence for WIMP signals using experiments with complementary technologies, 
and also with an experiment using the original target material, but having better 
sensitivity

\textbf{\emph{Study}}

If a signal is confirmed, study it with multiple technologies in order to extract 
maximal information about WIMP properties

\textbf{\emph{R\&D}}

Maintain a robust detector R\&D program on technologies that can enable discovery, 
confirmation and study of WIMPs.

\end{center}

\bibliography{bibCF1v3}
\chapter*{Appendix:  Instrumentation Development for WIMP Direct Detection}
\renewcommand*\thesection{\arabic{section}}
\setcounter{section}{0}

\section{Abstract}
Dark matter direct detection experiments rely on a variety of technologies, often common across experiment types, and in many cases common with other HEP areas.  This appendix to the Cosmic Frontier Direct Detection of WIMP Dark Matter highlights the R\&D required to advance these technologies for the future direct detection experiments.

\section{Common technology development needs}

\subsection{Material Procurement and Assay}
The need for procurement and assay of radiopure materials is common to all direct detection experiments. Consequently, the entire direct detection field would benefit from several forms of material R\&D. This includes development of new types of materials and of devices using those materials, as well as development of new assay techniques that are faster, cheaper, and/or more sensitive.

A variety of common but radio-pure solid materials (such as stainless steel, titanium, copper, lead, PTFE teflon, poly etc) are needed by all experiments to build structures, shields, cryogenic components, cables, feed-throughs etc. Similarly, plastic and organic liquid scintillators are being considered for building muon and neutron vetoes, so finding appropriate radio-pure plastic and organic liquid scintillator materials could be if interest to all experiments. Improvements in purification of liquid materials would be of particular interest for noble liquids technologies. Given the substantial overlaps in needs of different technologies, stockpiling material for future use (such as xenon and depleted argon) may also be of broad interest.

Development of radioactively clean photodetectors is one of the highest-priority R\&D directions, as it would directly benefit noble liquid detectors, crystal arrays like NaI/CsI, as well as the active shields needed in all experiments. Noble liquids require large area coverage with low dark noise and are most interested in low-activity PMTs with high quantum efficiency, including new UV photocathodes.  Crystal arrays are likewise interested in such PMTs, but small modular arrays with light concentrators can make APDs and SiPMs attractive as well.  Large tanks or water or liquid scintillator providing active shielding have much the same PMT requirements as noble liquid targets.  Shields made of modular liquid or solid scintillator can concentrate the light via wavelength-shifting fibers, and may be in the market of APDs or SiPMs.  Such silicon alternatives are more radiopure in principle than PMTs, but vendors need to put better quality assurance practices in place and improve low-background packaging.  In the case of PMTs, the radioactive contamination can be reduced by improving the phototube glass, for example by using substitutes such as sapphire.  A mass-produced, large area, UV-sensitive, radiopure photo-sensitive panel would be a game-changer.

Regarding the material assay, development of a variety of new techniques and detectors should be pursued. This includes developing methods for neutron activation that work in new matrices and have better tolerance for trace contaminants, development of larger and more immersive hole-body screeners (CTF), improvements in the sensitivity of alpha/beta scanners and other techniques sensitive to the surface contaminants, further development of ICPMS and other mass spectroscopy techniques (applied to multiple materials) etc.

Radon mitigation is also of critical importance to all direct detection experiments. To suppress radon exposure, large enclosures with breathable radon-free air should be developed, as well as laboratory areas for assembly, handling, and storage of equipment. New ways of sealing detectors should also be investigated. To improve detection of radon plate-out, improved counting techniques should be pursued, such as alpha/beta counters and sequential ICPMS. Regarding radon plate-out removal, new cleaning and electropolishing methods should be investigated. Techniques that work at several stages of material production and use would be particularly useful for plate-out removal. Development and dissemination of improved techniques for emanation reduction (e.g. from welds), as well as of specialized radon filtration systems for liquids and gases would be valuable

\subsection{Neutron Detection and Shielding}

Since neutrons are one of the key backgrounds for direct detection experiments, neutron detection and shielding are of high priority and of critical importance for multiple technologies. 

To improve the neutron veto/monitor efficiency, several directions should be pursued: \\
(a) Increase the neutron capture cross section by loading the active medium with Gd, B, or Li. Studies should be conducted to compare different loading materials and optimize their concentration. \\
(b) Thermalize neutrons (high-A with hydrogenous, Pb-loaded scintillators).\\
(c) Pursue schemes for neutron identification, such as pulse-shape discrimination.\\
(d) Increase the light yield, including the development of photodetectors to read it out (see above).\\
(e) Develop modular solid alternatives to liquid scintillators. \\

Neutron survey (ÒbenchmarkingÓ) detectors should be developed, built, and used to survey experimental areas. They are imagined to be large and to operate over long periods of time, similar to reactor neutrino detectors. Bonner spheres could be used to measure the neutron spectrum across all energies.

Ceiling mounted ÒumbrellaÓ muon vetoes may be of interest, as a cheap way to cover a large area. Muons which never intersect the experiment can produce neutrons in the cavern walls and ceiling far from the detector at angles which intersect it.  Catching the parent muons reduces the need for active detection of the neutrons themselves. 

\subsection{Calibration}

Calibration of detector response (for both electron and nuclear recoils) is critical for all direct detection experiments. With ever increasing detector target masses, achieving sufficient calibration statistics and tracking calibration as a function of time becomes increasingly difficult. Improved calibration techniques would therefore be of broad interests. New ideas include:\\
(a) Internal spiking of liquids.\\
(b) Surface deposition of sources. Improvements in neutron sources are of particular importance, especially if developed to the point that experiments could directly deploy them with limited expertise. Multiple directions could be pursued, including:\\
(a) Beam-lines, howitzers, and deuterium neutron guns. \\
(b) Monochromatic, low-energy photoproduction sources.\\
(c) Broad spectrum sources.

\section{Specific Technology R\&D Directions}

With specific dark matter direct detection technology areas, such as noble liquid or cryogenic solid state, there are still specific R\&D topics that will be of value.  Although these are often pursued within specific projects, we list them here as many of them cut across HEP experimental programs. There are topics where common R\&D could pay off for both dark matter and some other HEP experimental program.

\subsection{Cryogenic Semiconductor Detectors}

The main challenges for future direct detection experiments based on cryogenic semiconductor detectors are associated with scaling to large target masses. Several R\&D directions are pursued.\\
(a) Increasing the mass of individual detectors. 150 mm-diameter Si crystals are commercially available, but fabrication process (polishing, deposition, lithography) is still to be tuned. Detector-grade Ge crystals are currently limited to 100 mm diameter. Dislocation-free Ge crystals hold the promise of enabling 150 mm diameter Ge detectors, but they are yet to be grown with sufficiently low impurity levels (industry ties may be particularly important). Semiconductor detectors of large diameter (and mass) are also of interest in coherent neutrino scattering experiments and in X-ray observations of astrophysical sources.\\
(b) Mass production of semiconductor detectors is necessary to reach 100-kg or 1-ton scale target mass. This R\&D direction may benefit from ties with industry or homeland security.\\
(c) High-density cryogenic cabling is needed to bring out the signals from many individual semiconductor detectors. Alternatively, the cabling density could be reduced using signal multiplexing, perhaps including digital signal processing and FPGA techniques.

Alternative readout techniques, for example based on the kinetic inductance approach, are also of interest as they can simplify the design of the detector payload, and potentially improve detector sensitivity. Related to this, further improvements in the first-stage amplification devices (SQUIDs, FETs) would also be valuable. Germanium detectors operated at 77 K that can achieve lower noise and capacitance, and better energy resolution, would be valuable for both dark matter and neutrino-less double beta decay experiments. Finally, optimizing cryogenic semiconductor detectors for measuring very small energy deposition (100 eV or less) is particularly important for the low WIMP-mass region, as well as for applications in other fields (such as for coherent neutrino scattering). 

\subsection{Inorganic Crystals}

Inorganic crystals such as NaI and CsI have played important roles in the direct dark matter detection field. Repeating the NaI measurement to test the DAMA/LIBRA annual modulation claim is important for the field, and requires improving the purification of materials and clean production of NaI (similarly for CsI). New crystal types should also be investigated.  Photodetectors are placed close to the array and need to be radiopure.

\subsection{Noble Liquid Detectors}

In the noble liquid area, R\&D directions of substantial interest include photodetector development, charge and light yield measurements, purification, and high voltage.  

Further development of large area, high-efficiency, low-background, and low-cost photodetectors would improve noble liquid detector performance. Such photodetectors should be sensitive to liquid Xe (170 nm) liquid Ar (125 nm), or tetraphenyl butadiene wavelength shifter (440 nm).

R\&D to further improve measurements of the liquid xenon and liquid argon nuclear and electron charge and light yields, as a function of energy and electric field, would have a significant impact on physics results. Light WIMP searches would benefit from light target R\&D, including charge and light yield measurements, bolometric readout of superfluid helium scintillation and long-lived triplet excimers, and single electron detection in gaseous or liquid helium.

Progress in liquid phase purification of noble liquids, including nitrogen removal, would be valuable. Compared to gas phase purification, liquid phase purification could allow much faster and more efficient removal of impurities, with lower heat load. It may also allow better removal of non-volatile impurities.

There are two broad areas of desirable R\&D work on high-voltage in liquid noble gas detectors: high voltage delivery into liquid-noble detectors, and suppression of electro-luminescence at high cathode voltage. As noble liquid detectors increase in size, much higher voltages are needed to maintain a given drift field. Noble gases can break down or glow under high-field conditions, preventing stable detector operation. A detailed understanding of the circumstances under which breakdown or light production occurs will allow higher drift fields, and therefore better electron recoil background discrimination and faster electron drift. 

\subsection{Superheated-Liquid Detectors}

Superheated-liquid detectors will benefit from R\&D in the areas of:\\
(a) The specific mechanism for maintaining the superheated liquid state and re-liquifying gas after boiling has occurred. The ideal scheme would be scalable to multi-ton targets, would allow for multiple target liquids, and would be highly stable against surface boiling. There are three schemes currently being pursued, Òbubble chambersÓ (COUPP), Òsuperheated dropletsÓ (PICASSO/SIMPLE) and ÒGeysersÓ (PICASSO/MOSCAB).\\
(b) Development of new liquid targets. Currently CF$_3$I (spin-dependent and spin-independent couplings) 
and C$_3$F$_8$ (spin-dependent only) are under active development. There is a need for at least one additional liquid with a high spin-independent cross section and perhaps for a liquid that has high sensitivity to neutrons, but low sensitivity to WIMPs by either spin-dependent or spin-independent couplings.\\
(c) Reliable, small, low-radioactivity acoustic sensors.\\
(d) New inner-vessel materials that are transparent, low in radioactivity, chemically compatible with the liquids and scalable to large volumes. \\
(e) Exploration of target liquids, thermodynamic conditions, backgrounds and shielding for low-threshold WIMP detectors.

\subsection{Directional Detection}

In the future, technology for directional detection may become a key need of the dark matter program.  Continued R\&D in this direction needs to be supported, including liquid/solid detectors and large, inexpensive gas detectors (while keeping the head-tail discrimination).  Once direct detection has indicated a target cross-section and particle mass, R\&D can be focused on directional measurement of that particle, but current priority should be boosting the energies of low-mass WIMPs with light target mass (e.g. HeCO$_2$).

\end{document}